\documentclass[11pt,a4paper]{article}
\pdfoutput=1
\usepackage{jheppub}
\usepackage{graphicx,booktabs,latexsym,amssymb,amsmath,bm,epsfig,psfrag,float}
\hypersetup{pdfnewwindow=true}

\usepackage{bbm}
\usepackage{mathtools}
\usepackage{hepunits}
\usepackage{color}
\usepackage{multirow}
\usepackage{sectsty}
\allsectionsfont{\boldmath}
\allowdisplaybreaks

\usepackage[utf8]{inputenc}

\newcommand{\Tr}{\mathrm{Tr}}

\newcommand{\Mtt}{$M_{t\bar{t}}$}

\preprint{
	\begin{flushright}	
		
		Cavendish-HEP-18/03\\
		IPPP/18/18 \\
		TTK-18-07 \\
		Nikhef 2018-009
	\end{flushright}}
	
\title{ Resummation for (boosted) top-quark pair production at NNLO+NNLL$'$ in QCD}	

\author[a]{Micha\l\ Czakon,}
\emailAdd{mczakon@physik.rwth-aachen.de}
\author[b,c]{Andrea Ferroglia,}
\emailAdd{aferroglia@citytech.cuny.edu}
\author[d]{David Heymes,}
\emailAdd{dheymes@hep.phy.cam.ac.uk}
\author[d]{Alexander Mitov,}
\emailAdd{adm74@cam.ac.uk}
\author[e]{Ben~D.~Pecjak,}
\emailAdd{ben.pecjak@durham.ac.uk}
\author[f,g]{Darren J. Scott,}
\emailAdd{d.j.scott@uva.nl}
\author[h]{Xing Wang,}
\emailAdd{x.wong@pku.edu.cn}
\author[h,i,j]{and Li Lin Yang}
\emailAdd{yanglilin@pku.edu.cn}

\affiliation[a]{Institut f\"{u}r Theoretische Teilchenphysik und Kosmologie,
	RWTH Aachen University
	D-52056 Aachen, Germany}
	
\affiliation[b]{New York City College of Technology, 
	Brooklyn, NY 11201, USA}
	
\affiliation[c]{The Graduate School and University Center,
	The City University of New York, 
	New York, NY 10016  USA}

\affiliation[d]{Cavendish Laboratory,
	University of Cambridge
	CB3 0HE Cambridge, UK}

\affiliation[e]{Institute for Particle Physics Phenomenology,
	University of Durham
	DH1 3LE Durham, UK}

\affiliation[h]{School of Physics and State Key Laboratory of Nuclear Physics and Technology
	Peking University, Beijing 100871, China}
\affiliation[f]{Institute for Theoretical Physics, University of Amsterdam, Science Park 904, 1098 XH Amsterdam, The Netherlands}
\affiliation[g]{Nikhef, Theory Group,
	Science Park 105, 1098 XG, Amsterdam, The Netherlands}

\affiliation[i]{Collaborative Innovation Center of Quantum Matter, Beijing, China}
\affiliation[j]{Center for High Energy Physics, Peking University, Beijing 100871, China}

\abstract{ We construct predictions for top quark pair differential
  distributions at hadron colliders that combine state-of-the-art NNLO
  QCD calculations with double resummation at NNLL$'$ accuracy of
  threshold logarithms arising from soft gluon emissions and of small
  mass logarithms. This is the first time a resummed calculation at
  full NNLO+NNLL$'$ accuracy in QCD for a process with non-trivial
  color structure has been completed at the differential level.
Of main interest to us is the stability of the $M_{t\bar{t}}$ and
top-quark $p_T$ distributions in the boosted regime where fixed order
calculations may become strongly dependent on the choice of dynamic
scales. With the help of numeric and analytic arguments we confirm
that the choice for the factorization and renormalization scales
advocated recently by some of the authors is indeed optimal. We
further derive a set of optimized kinematics-dependent scales for the
matching functions which appear in the resummed calculations.
Our NNLO+NNLL$'$ prediction for the top-pair invariant mass is
significantly less sensitive to the choice of factorization scale
than the fixed order prediction, even at NNLO. Notably, the
resummed and fixed order calculations are in nearly perfect agreement
with each other in the full $M_{t\bar{t}}$ range when the optimal
dynamic scale is used. For the top-quark $p_T$ distribution the
resummation performed here has less of an impact and instead we find
that upgrading the matching with fixed-order from NLO+NNLL$'$ to
NNLO+NNLL$'$ to be an important effect, a point to be kept in mind
when using NLO-based Monte Carlo event generators to calculate this
distribution.  }

\begin{document}

\maketitle

\newpage

\section{Introduction}
\label{sec:intro}

The top quark is the heaviest fundamental particle discovered so
far. Because of its large mass, the top quark is the Standard
Model particle which couples most strongly to the Higgs boson; as
such, it plays a pivotal role in the study of the electroweak symmetry
sector of the Standard Model. For this reason, and thanks to the large
number of top-quark pairs produced at the Large Hadron Collider (LHC),
accurate experimental measurements of several top-quark related
observables are either already available or will become available in
the next few years. 

An exciting feature of top physics at the LHC is that the large
collider energy enables the study of boosted top-quark production.  In
this context ``boosted'' refers to the kinematic regime where the
energies of the produced top quarks are much larger than the top-quark
mass. Boosted top quarks may appear in the study of many kinematic
distributions: top quarks with very high pair invariant mass, top
quarks with very large transverse momentum, or very forward top quarks
(large rapidity).  The LHC has already probed top quarks with
transverse momenta around 1~TeV \cite{Aaboud:2018eqg}, and will extend
the energy range up to a few TeV.  While these boosted energy top
quarks are more rarely produced than the low energy ones, and the high
energy regions do not contribute significantly to the total cross
section, they are phenomenologically important due to their potential
to probe directly physics beyond the electroweak scale.

On the theory side, state-of-the-art perturbative calculations for
on-shell top-quark pair production, in the boosted regime or
otherwise, have reached an accuracy beyond next-to-leading order (NLO)
in the strong coupling constant $\alpha_s$.  The most complete
Standard Model predictions for differential cross sections combine
next-to-next-leading-order (NNLO) QCD corrections with NLO electroweak
(EW) ones~\cite{Czakon:2017wor}. In this paper we will focus
exclusively on QCD corrections, with the understanding that the
results can eventually be combined with EW corrections which have been studied extensively in the literature \cite{Beenakker:1993yr,Kuhn:2005it,Bernreuther:2005is,Kuhn:2006vh,Bernreuther:2006vg,Hollik:2007sw,Bernreuther:2008md,Bernreuther:2010ny,Denner:2010jp,Bevilacqua:2010qb,Hollik:2011ps,Kuhn:2011ri,Manohar:2012rs,Bernreuther:2012sx,Denner:2012yc,Frederix:2013gra,Kuhn:2013zoa,Campbell:2016dks,Pagani:2016caq,
Denner:2016jyo}.

NNLO QCD calculations for differential cross sections in top-quark
pair production were  obtained in~\cite{Czakon:2015owf,
  Czakon:2016dgf}.  These calculations added to NNLO results for
more inclusive quantities such as the total cross section
\cite{Baernreuther:2012ws, Czakon:2012zr, Czakon:2012pz,
  Czakon:2013goa} and the forward-backward asymmetry at the Tevatron
\cite{Czakon:2014xsa}.  In \cite{Czakon:2015owf}, distributions such as the
$t\bar{t}$ (top-pair) invariant mass and the top-quark transverse
momentum distribution were evaluated at the LHC with center-of-mass
energies of $\sqrt{s}=8$~TeV and 13~TeV using renormalization and
factorization scales $\mu$ varied around the top mass $m_t$, and the
typical scale uncertainties were estimated to lie below $10\%$. In
\cite{Czakon:2016dgf}, on the other hand, dynamical scale choices were
investigated in order to determine which choice of scale is most
appropriate for fixed-order studies of multi-TeV differential cross
sections, based on the convergence of the fixed-order perturbative
series.  A notable result of that study is that the high-energy tails
of distributions can be quite sensitive to the parametric choice of
factorization scale, even at NNLO and especially in the case of the
top-pair invariant mass distribution.

The fact that the application of fixed-order perturbation theory in
the boosted regime is rather delicate is not entirely
surprising. Indeed, in the case of boosted top quarks one encounters
two potential difficulties.  The first is that in the boosted regime
the result of a fixed-order calculation contains mass logarithms of
the form $\ln(E_t/m_t)$ arising from quasi-collinear gluon emissions.
The second is that, due to the shape of the parton distribution
functions (PDFs), the effective partonic center-of-mass energies in
most events with boosted top quarks are not much larger than the
invariant masses of the top quark pairs.  This can lead to enhanced
corrections from so-called soft logarithms or threshold logarithms,
and indeed, in general the two types of logarithms multiply each other
at a given order in perturbation theory, due to emissions which are
simultaneously soft and collinear.  At the LHC, these logarithmic
corrections may become important numerically, to the point that higher
order corrections are not generically much smaller than lower order
ones, and while it may be possible to deal with this to some extent
through a judicious scale-setting procedure in a fixed-order
calculation, it is desirable to address the issue head-on by resumming
these two types of logarithms to all orders in perturbation theory and
adding them onto the fixed order result through a matching procedure.

A framework for the simultaneous resummation of threshold and
small-mass logarithms in differential cross sections in top-quark pair
production at the LHC was set up in \cite{Ferroglia:2012ku}, and
applied to the phenomenological study of $t\bar{t}$ invariant mass and
top-quark transverse momentum distributions in \cite{Pecjak:2016nee}.
The results are valid in the soft gluon emission limit, but contain an
extra layer of resummation for small-mass logarithms, so that
overlapping soft and small-mass logarithmic corrections are properly
taken into account.  The factorization formalism underlying the
resummation was derived using soft-collinear effective theory (SCET),
and the resummation was carried out in Mellin space.  While the focus
of that phenomenological study was the boosted regime, a matching
procedure with standard soft-gluon resummation results (i.e.  without
the simultaneous resummation of small-mass logarithms) at
next-to-next-to-leading logarithmic (NNLL) accuracy
\cite{Ahrens:2010zv} and to NLO fixed-order calculations was performed
in order to expand the validity of the results to the full phase
space.

In this paper we combine state-of-the-art results from analytic
resummation and fixed-order perturbation theory in QCD in order to
produce for the first time phenomenological predictions which match
NNLO results with resummation of soft and small-mass logarithms at the
level of differential $t\bar{t}$ invariant mass and top-quark
transverse momentum distributions.

The paper is organized as follows. In section~\ref{sec:prelim} we
immediately address the technical aspect which is new in this work,
namely the matching of the resummed calculation in the soft and
boosted limit to the NNLO calculations. In section~\ref{sec:MellinRes}
we explain the resummation procedures used in two distinct kinematic
limits. In particular, in section~\ref{sec:kinematics} we review the
kinematics for top-pair production and the form of the differential
cross section in Mellin space. In section~\ref{sec:MellinRes_soft} we
discuss the soft limit, where threshold logarithms are large but mass
logarithms are of generic size, while in
section~\ref{sec:MellinRes_boosted} we consider the boosted-soft
limit, where both soft and mass logarithms are considered large. With
all the analytic tools ready, in section~\ref{sec:systematic} we study
kinematic features of the top quark pair in the boosted region, and
use these insights to determine appropriate choices for the scales
appearing in the matching functions in the resummed formulas. In
section~\ref{sec:discussion} we present numerical results for the
top-pair invariant mass and top-quark $p_T$ distributions, paying
close attention to a comparison between pure NNLO results and those
supplemented by resummation, and then present conclusions in
section~\ref{sec:conclusions}. We perform more comparisons among
results at different orders in fixed-order and resummed perturbation
theory in appendix~\ref{sec:comparisons}, and relegate some lengthy
formulas for resummation exponents to appendix~\ref{sec:gis}.

\section{Matching fixed order and resummed calculations}
\label{sec:prelim}

We study the top-quark pair production process
\begin{align}
  N_1(P_1) + N_2(P_2) \rightarrow t(p_3) +\bar{t}(p_4) + X(p_X) \, ,
\end{align}
where $N_1$ and $N_2$ are the colliding hadrons (proton-proton for the
LHC and future colliders, proton-antiproton for the Tevatron), and $X$
is an inclusive hadronic final-state. In this work, the top quarks are
treated as on-shell particles.  QCD factorization allows us to write an arbitrary differential cross 
section in the schematic form
\begin{align}
\label{eq:schematic_fact}
d\sigma = \sum_{i,j} \, d\hat{\sigma}_{ij}\otimes \phi_{i/N_1} \otimes \phi_{j/N_2} \, .
\end{align}
The symbol $\otimes$ stands for a convolution over longitudinal
momentum fractions of the initial-state partons
$i,j\in\{q,\bar{q},g\}$ in the (anti-)proton, the $d\hat{\sigma}_{ij}$
are the differential partonic cross sections for the process $i+j\to
t\bar{t}+\hat{X}$, the $\hat{X}$ is a partonic final state, and
$\phi_{i/N}$ denotes the parton distribution function (PDF) of parton
$i$ in hadron $N$.  While the PDFs are non-perturbative objects to be
extracted from experiment, the differential partonic cross sections
can be calculated in perturbative QCD.
 
The aim of this paper is to provide state-of-the-art QCD calculations
for two specific differential cross sections: namely the invariant
mass of the top-quark pair, and the $p_T$ of the top-quark.
The baseline for the calculation is NNLO in fixed-order perturbation theory, to which we add the two types of resummation mentioned in the introduction and to be described in detail in section~\ref{sec:MellinRes}.  The first is performed in the
soft limit of the differential partonic cross sections, where the
top-quark pair carries almost all of the energy of the partonic
collision.  It can be obtained to NNLL$_m$ order using the results of
\cite{Ahrens:2010zv}, where we have labeled the logarithmic accuracy
of the resummation with a subscript $m$ to indicate that the result is
obtained for arbitrary values of $m_t$.  As the energy of the
top-quark pair becomes very large, this standard soft-gluon
resummation itself develops logarithms which become large in the limit
$m_t \to 0$.  We call this the ``boosted-soft limit", and perform a
joint resummation of overlapping soft and small-mass logarithms using
the formalism developed in~\cite{Ferroglia:2012ku}.  In this case it
is possible to increase the accuracy of the resummation to NNLL$'_b$
order, where now the subscript $b$ indicates that the results are
valid in the boosted-soft limit, and thus neglect corrections which
vanish in the limit $m_t\to 0$. The perturbative ingredients for these
two types of resummation (anomalous dimensions and matching
functions) and the order at which they need to be calculated to
achieve a given resummation accuracy is summarized in
table~\ref{tab:rg-counting} in section~\ref{sec:resAcc}.

We have just described three different calculational formalisms, each of which is tailored to a 
different kinematic situation.  The NNLO calculation is optimal in regions of phase space
where the top quarks are not highly boosted, and hard-gluon emissions are important.  The 
NNLL$_m$ result is applicable when soft-gluon radiation dominates, but small-mass logarithms
are unimportant.  When soft-gluon radiation dominates, and the top-quarks are highly boosted,
one would like to make use of the NNLL$'_b$ results.
To make optimal use of our results  we would like to have a unified description over the
whole phase space. For this purpose it is necessary to combine the
different formulas in such a way that no contribution is counted more than
once.  

To understand such a matching procedure, we first consider matching
NNLL$_m$ resummation with fixed-order results.  In that case, the
matching formula with (N)NLO reads
\begin{align}
\label{eq:soft-matched}
d\sigma^{\mathrm{(N)NLO+NNLL}_m} =d\sigma^{\mathrm{NNLL}_m}
+\bigg(d\sigma^{\mathrm{(N)NLO}}-
d\sigma^{\mathrm{NNLL}_m}\Big|_{\substack{\text{(N)NLO} \\ \text{expansion} }}\bigg) \, ,
\end{align}
where $d\sigma^{\mathrm{NNLL}_m}$ denotes the differential cross section evaluated 
to NNLL$_m$ accuracy.  The first term in the above equation contains the 
all-orders resummation result in the soft limit, and the difference of terms in parenthesis 
contains subleading terms in that limit, which are taken into account by the fixed-order 
calculation.  Explicit expressions for the fixed-order (N)NLO expansion of the resummation formulas 
needed in the matching are given in section~\ref{sec:Matching}.

We next match the resummation formulas in the soft and boosted-soft
limit with each other and with (N)NLO. To do so, we need to remove the overlap 
between the NNLL$'_b$  and NNLL$_m$ results to all orders in $\alpha_s$.  This is done
by exploiting the fact that the boosted-soft resummation formula is the
small-mass limit of the soft-gluon resummation formula at any fixed order in
$\alpha_s$, so we must subtract out the leading term in the limit $m_t\to0$ 
in order not to double count. The combined result, denoted $\mathrm{NNLL}'_{b+m}$, is thus given by
\begin{equation}
\label{eq:res_comb}
d \sigma^{\mathrm{NNLL}'_{b+m}} = d\sigma^{\mathrm{NNLL}'_b} + \left(d\sigma^{\mathrm{NNLL}_m} -
\left. d\sigma^{\mathrm{NNLL}_m}\right|_{m_t\to 0} \right) ,
\end{equation}
where the terms in the parenthesis account for
contributions which are suppressed by $m_t/M$ in the boosted-soft
limit and thus not included in the NNLL$'_b$ result. Matching with fixed order then proceeds in analogy to eq.~(\ref{eq:soft-matched}) resulting in
\begin{equation}
\label{eq:fully-matchedNNLO}
d \sigma^{\mathrm{(N)NLO+NNLL}'} = d\sigma^{\mathrm{NNLL}'_{b+m}}
+ \bigg(d\sigma^{\mathrm{(N)NLO}} -
 \left. d\sigma^{\mathrm{NNLL}'_{b+m}}\right|_{\substack{\mathrm{(N)NLO} \\
			\mathrm{expansion}}} \bigg) 		\,.
\end{equation} 
Again, the terms in the parenthesis account for subleading
terms in the soft limit, which are taken into account through a fixed-order
calculation, but are not accessible to either of the resummation formulas.
Calculating the subtraction term requires one to expand each term in eq.~(\ref{eq:res_comb}) to (N)NLO. 
The procedure for obtaining the different components of  the above equation is described in more 
detail in section~\ref{sec:Matching}.

Differential distributions obtained from the explicit evaluation of eq.~(\ref{eq:fully-matchedNNLO}) are a main result of this work, and can be found in section~\ref{sec:discussion}.  Before going into numerical studies,  we give details of the resummation   procedure, including recipes for obtaining the different pieces used in the matching procedure, in the next
section.  These details can safely be skipped by a reader interested in purely phenomenological results. In section~\ref{sec:systematic} we conduct a thorough analysis of the choice of scales for the matching functions which appear in the resummed results.

\section{Mellin-space resummation in the (boosted) soft limit}
\label{sec:MellinRes}

\subsection{Kinematics and differential cross sections}
\label{sec:kinematics}

In this section we review the kinematics involved in describing the limits in which resummation
is carried out.   At Born level, and to leading order in the soft limit considered below, two partonic channels contribute  to the partonic cross section:  the quark-antiquark annihilation channel
\begin{align} \label{eq:qq}
  q(p_1) + \bar{q}(p_2) \rightarrow t(p_3) +\bar{t}(p_4) \, ,
\end{align}
and the gluon fusion channel
\begin{align} \label{eq:gg}
  g(p_1) + g(p_2) \rightarrow t(p_3) +\bar{t}(p_4) \, .
\end{align}
The momenta of the incoming partons are related to the hadron momenta
according to ${p_i = x_i P_i}$ ($i=1,2$). The relevant Mandelstam
invariants are defined as
\begin{gather}
  s = (P_1+P_2)^2 \, , \quad \hat{s} = (p_1+p_2)^2 \, , \quad M_{t\bar{t}}^2 =
  (p_3+p_4)^2 \, , \nonumber
  \\
  t_1 = (p_1-p_3)^2 - m_t^2 \, , \quad u_1 = (p_2-p_3)^2 -m_t^2 \, .
\label{eq:mandel}
\end{gather}
In fixed order perturbation theory, starting from NLO accuracy, a new
$2 \to 3$ production channel, initiated by a quark and a gluon, opens
up. At NNLO one needs to account for the contribution of $2 \to 4$
processes (at tree level), as well as the contribution of $2 \to 3$
processes (up to one loop) and the $2 \to 2$ production channels in
eqs.~(\ref{eq:qq}) and~(\ref{eq:gg}) (up to two-loops).  All of these
channels are of course included in the NNLO results which we employ in
this work.

The soft emission region, which is of interest for the resummed
calculation, is defined by the limit $M_{t\bar{t}}^2 \to \hat{s}$ (sometimes also
referred to as the partonic threshold region).  In this limit, the
final state particles in addition to the top pair are soft.  In order
to describe the top-pair invariant mass distribution near the
partonic threshold, it is convenient to introduce the following
variables:
\begin{align}
  z = \frac{M_{t\bar{t}}^2}{\hat{s}} \, , \quad \tau = \frac{M_{t\bar{t}}^2}{s} \, , \quad
  \beta_t = \sqrt{1-\frac{4m_t^2}{M_{t\bar{t}}^2}} \, , \quad \beta=
  \sqrt{1-\frac{4m_t^2}{\hat{s}}} \, .
\end{align}
The quantity $\beta_t$ is the 3-velocity of the (anti-)top quark in
the $t\bar{t}$ rest frame, while $\beta$ is often invoked to describe
the partonic threshold for the total cross section
\cite{Bonciani:1998vc, Czakon:2009zw, Beneke:2011mq,
  Beneke:2010da}. In the soft limit $z \to 1$, one has $\beta \to
\beta_t$. Moreover, in that limit the scattering angle $\theta$ is
related to the Mandelstam variables according to
\begin{align}
  \label{eq:tu}
  t_1 = -\frac{M_{t\bar{t}}^2}{2} ( 1 - \beta_t \cos\theta ) \, , \quad u_1 =
  -\frac{M_{t\bar{t}}^2}{2} ( 1 + \beta_t \cos\theta ) \, ,
\end{align}
from which one can easily verify the usual relation $M_{t\bar{t}}^2+t_1+u_1=0$.

We will perform resummation on the double differential cross section in the top-pair invariant mass and 
the scattering angle $\theta$.  Applying the generic QCD factorization formula eq.~(\ref{eq:schematic_fact})  allows one to write this differential cross section as
\begin{equation}
	\frac{d^2\sigma(\tau)}{dM_{t\bar{t}}\,d\cos\theta}=\frac{8\pi\beta_t}{3sM_{t\bar{t}}}\sum\limits_{ij}\int^1_\tau \frac{dz}{z} \, \mathcal{L}_{ij}(\tau/z,\mu_f) \, C_{ij}(z,M_{t\bar{t}},m_t,\cos\theta,\mu_f) \, ,
\label{eq:x-sec}
\end{equation}
where $M_{t\bar{t}}$ and $\cos\theta$ are in the ranges
\begin{align}
  2m_t \leq M_{t\bar{t}} \leq s \, , \quad  |\cos\theta| \leq 1 \,.
\end{align}
The hard-scattering kernels $C_{ij}$ are proportional to the partonic
cross sections and can be calculated in perturbation theory at the
factorization scale $\mu_f$, while $\mathcal{L}_{ij}$ are
non-perturbative parton luminosity functions, defined as
\begin{equation}
\mathcal{L}_{ij}(y,\mu_f) = \int_y^1 \frac{dx}{x} \, \phi_{i/N_1}(x,\mu_f) \, \phi_{j/N_2}(y/x,\mu_f) \, .
\end{equation}

In this work, we study the resummation in two limits where soft and collinear gluon
emissions dynamically generate scales much lower than the scattering
energy:
\begin{align}
\label{eq:softlimit}
\text{soft limit:} & \quad \hat{s},|t_1|,m_t^2 \gg \hat{s}(1-z)^2 \, ,
\\
\label{eq:boostedlimit}
\text{boosted-soft limit:} & \quad \hat{s},|t_1|
\gg m_t^2 \gg \hat{s}(1-z)^2  \gg m_t^2(1-z)^2 \,.
\end{align}
In each of these limits the perturbative expansions of the
hard-scattering kernels contain large logarithms of scale ratios,
which can be resummed to all orders in $\alpha_s$.  For the discussion
of resummation that follows, it is convenient to study the cross
section in Laplace or Mellin space.  The Mellin transform and its
inverse are defined by
\begin{align}
\tilde{f}(N) = \mathcal{M}[f](N) = \int_0^1 dx \,x^{N-1}{f}(x) \, , \quad f(x) = \mathcal{M}^{-1}[\tilde{f}](x) = \frac{1}{2 \pi i}\int_{c-i \infty}^{c+i \infty}dN \, x^{-N} \tilde{f}(N),
\label{mellint}
\end{align}
where in the inverse transform the real part of the contour $c$ is
chosen such that it lies to the right of all singularities in the
function $\tilde{f}(N)$.  Convolutions such as the differential cross
section in eq.~(\ref{eq:x-sec}) become simple products in Mellin
space.  Indeed, by performing the Mellin transform of
eq.~(\ref{eq:x-sec}) with respect to $\tau$, we find the differential
cross section in Mellin space, which reads
\begin{align}
  \frac{d^2\widetilde{\sigma}(N)}{dM_{t\bar{t}} \, d\cos\theta}=
  \frac{8\pi\beta_t}{3sM_{t\bar{t}}}\sum_{ij}
  \widetilde{\mathcal{L}}_{ij}(N,\mu_f) \, \widetilde{c}_{ij}(N,M_{t\bar{t}},m_t,\cos\theta,\mu_f)
  \, .
\label{eq:x-sec-mellin}
\end{align}
The soft limit $z\to 1$ corresponds to $N\to \infty$, as can be seen by
taking the Mellin transform of the plus distributions appearing in the
partonic cross section, which are generated by soft gluon emissions:
\begin{align}
\mathcal{M}[P_0](N) &= -\ln{\bar{N}} + \mathcal{O}\left(\frac{1}{N}\right) \ , \nonumber
 \\
 \mathcal{M}[P_1](N) &= \frac{1}{2}\left(\ln^2{\bar{N}} + \frac{\pi^2}{6}\right)+ \mathcal{O}\left(\frac{1}{N}\right) \ , \nonumber
  \\
\mathcal{M}[P_2](N) &=  -\frac{1}{3}\left(\ln^3{\bar{N}} + \frac{\pi^2}{2} \ln{\bar{N}} + 2 \zeta(3)\right)+ \mathcal{O}\left(\frac{1}{N}\right) \ , \nonumber
 \\
\mathcal{M}[P_3](N) &=  \frac{1}{4}\left(\ln^4{\bar{N}} + \pi^2 \ln^2{\bar{N}} + 8 \zeta(3)\ln{\bar{N}} + \frac{3 \pi^4}{20}\right) + \mathcal{O}\left(\frac{1}{N}\right)\ ,
\label{eq:MtranfofP}
\end{align}
where
\begin{align}
\label{eq:plusDist}
P_n(z) = \left[\frac{\ln^n (1-z)}{1-z} \right]_+ \, ,
\end{align}
and we have introduced the variable $\bar{N} = Ne^{\gamma_E}$ in order
to simplify the expressions, with $\gamma_E$ denoting the Euler
constant. We note that the partonic cross section contains terms of
the form $\alpha_s^n P_k(z)$ where $0 \leq k \leq 2n-1$ at N$^n$LO in
its perturbative expansion. In Mellin space this becomes $\alpha_s^n
L^k$ where $L=\ln \bar{N}$ and $0 \leq k \leq 2n$, which will be
important when we consider the tower of logarithms which can be
resummed in section~\ref{sec:resAcc}. In Mellin moment space, the soft
and boosted-soft limits in eqs.~(\ref{eq:softlimit}) and
(\ref{eq:boostedlimit}) become
\begin{align}
\label{eq:softlimit-mellin}
\text{Mellin-space  soft limit:} & \quad \hat{s},|t_1|,m_t^2 \gg \frac{\hat{s}}{N^2} \, ,
\\
\label{eq:boostedlimit-mellin}
\text{Mellin-space boosted-soft limit:} & \quad \hat{s},|t_1|
\gg m_t^2 \gg \frac{\hat{s}}{N^2} \gg \frac{m_t^2}{N^2} \,.
\end{align}

Besides the invariant mass distribution, the distribution in the
transverse momentum $p_T$ of the top quark is also interesting. In
\cite{Ferroglia:2013awa}, a different formulation, dubbed ``1PI'', was
employed to deal with such ``single-particle-inclusive'' observables
in the boosted regime.
In contrast, the formulation used in this paper was called ``pair-invariant-mass'' (PIM) kinematics.
Near the partonic threshold, the PIM and 1PI formulations differ only by power-suppressed
contributions. We can exploit this fact to express the differential
cross section with respect to $p_T$ in terms of the double
differential cross section in eq.~(\ref{eq:x-sec}), thereby avoiding
the introduction of the 1PI formulation. In the threshold limit, the
transverse momentum $p_T$ and the rapidity $\hat{y}$ of the top-quark
in the partonic center-of-mass frame are given by
\begin{align}
p_T = \frac{M_{t\bar{t}}\beta_t}{2}\sin\theta \,, \quad \hat{y} = \frac{1}{2} \ln \frac{1+\beta_t\cos\theta}{1-\beta_t\cos\theta} \, ,
\end{align}
and we can write the differential cross section with respect to $p_T$
and $\hat{y}$ in Mellin space as
\begin{align}
 \frac{d^2\widetilde{\sigma}(N)}{dp_T \, d\hat{y}} = 2\sin\theta \, \frac{d^2\widetilde{\sigma}(N)}{dM_{t\bar{t}} \, d\cos\theta} = \frac{16\pi\beta_t\sin\theta}{3sM_{t\bar{t}}} \sum_{ij} \widetilde{\mathcal{L}}_{ij}(N,\mu_f) \, \widetilde{c}_{ij}(N,M_{t\bar{t}},m_t,\cos\theta,\mu_f) \, .
\label{eq:x-sec-pt}
\end{align}
In the above equation, it is understood that $M_{t\bar{t}}$ and
$\cos\theta$ should be expressed in terms of the integration variables
according to
\begin{align}
\label{eq:transmassDef}
M_{t\bar{t}} = 2m_T \cosh(\hat{y}) \equiv 2\sqrt{p_T^2+m_t^2} \cosh(\hat{y}) \, , \quad \cos\theta = \frac{1}{\beta_t} \, \tanh(\hat{y}) \, ,
\end{align}
where we have defined the transverse mass $m_T$.
The transverse momentum distribution can be obtained by integrating
over $\hat{y}$ in the range
\begin{align}
 |\hat{y}| \leq \mathrm{arccosh} \left( \frac{\sqrt{s}}{2m_T} \right) ,
\end{align}
while the range of $p_T$ is
\begin{align}
0 \leq p_T \leq \sqrt{\frac{s}{4} - m_t^2} \, .
\end{align}

\subsection{Resummation in the soft limit}
\label{sec:MellinRes_soft}

Resummation of top-quark hadroproduction cross sections in the
partonic threshold limit $z \to 1$ was considered in
\cite{Kidonakis:1997gm}.  More recently, top-pair production in the
soft limit was re-analyzed by means of SCET methods\footnote{For a
  didactic introduction to SCET, see \cite{Becher:2014oda}.}. With
this technique, it was possible to study the resummation of soft gluon
emission corrections both in PIM and 1PI kinematics
\cite{Ahrens:2010zv,Ahrens:2011mw} in momentum space. Here we follow
closely the discussion and notation of the NNLL calculation of
\cite{Ahrens:2010zv}, although we perform the resummation in Mellin
rather than momentum space. This approach was adopted in
\cite{Pecjak:2016nee} for top-pair production.  Subsequently, it was
also applied to the evaluation of the soft emission corrections to $t
\bar{t} W^{\pm}$ \cite{Broggio:2016zgg}, $t \bar{t} H$
\cite{Broggio:2016lfj}, and $t \bar{t} Z$ \cite{Broggio:2017kzi} to
NNLL accuracy.  The starting point is the factorization of the
partonic cross section in the soft limit. In Mellin space, this takes
the form
\begin{multline}
\label{eq:soft-fact}
\widetilde{c}_{ij}(N,M_{t\bar{t}},m_t,\cos\theta,\mu_f) =
\Tr \Bigg[ \bm{H}^m_{ij}(M_{t\bar{t}},m_t,\cos\theta,\mu_f) \\*
\times \widetilde{\bm{s}}^m_{ij}\left(\ln\frac{M_{t\bar{t}}^2}{\bar{N}^2 \mu_f^2},M_{t\bar{t}},m_t,\cos\theta,\mu_f \right)\Bigg]
+ \mathcal{O}\left(\frac{1}{N}\right) .
\end{multline}
The hard functions $\bm{H}^m_{ij}$ and the soft functions
$\widetilde{\bm{s}}^m_{ij}$, referred to generically as matching
functions, are matrices in color space --- explicit results for the
two-by-two matrices in the $q\bar{q}$ channel and three-by-three
matrices in the $gg$ channel up to NLO in $\alpha_s$ can be extracted
from \cite{Ahrens:2010zv}, where the derivation of a momentum-space
factorization formula analogous to eq.~(\ref{eq:soft-fact}) is given
in detail.
Comparing to the notation employed in \cite{Ahrens:2010zv}, here we have introduced a superscript $m$ on these functions, indicating that they contain full dependence on the top quark mass $m_t$. This is to distinguish them from the corresponding functions with $m_t=0$ used later in the boosted-soft limit.

Given the factorized form of the partonic cross section in Mellin
space, one can derive and solve renormalization group (RG) equations
for the component functions. This allows one to evaluate the hard and
soft functions at an arbitrary hard scale $\mu_h$ and soft scale
$\mu_s$, where large logarithms are absent. One then uses RG evolution
to obtain the hard scattering kernels at the factorization scale
$\mu_f$. These RG-improved hard-scattering kernels can be written as
\begin{multline}
\label{eq:soft-resummed}
\widetilde{c}_{ij}(N,\mu_f) =
\Tr \Bigg[\widetilde{\bm{U}}^m_{ij}(\!\bar{N},\mu_f,\mu_h,\mu_s) \, \bm{H}^m_{ij}(\mu_h) \, \widetilde{\bm{U}}_{ij}^{m\dagger}(\!\bar{N},\mu_f,\mu_h,\mu_s)
 \widetilde{\bm{s}}^m_{ij}\left(\ln\frac{M_{t\bar{t}}^2}{\bar{N}^2 \mu_s^2},\mu_s \right)\!\!\Bigg] \\ + \mathcal{O}\left(\!\frac{1}{N}\!\right) ,
\end{multline}
where we have suppressed the dependence of all functions in
eq.~(\ref{eq:soft-resummed}) on the variables $M_{t\bar{t}}$, $m_t$ and
$\cos\theta$.  The evolution matrices $\widetilde{\bm{U}}^m_{ij}$ contain all large logarithms
in an exponentiated form, and thereby resum them to all orders in
$\alpha_s$. The explicit form of the evolution matrices was derived in
\cite{Ahrens:2010zv}, and in Mellin space reads (suppressing for the moment the
subscript $ij$ and the argument $\bar{N}$)
\begin{multline}
\label{eq:Umat}
\widetilde{\bm{U}}^m(\mu_f,\mu_h,\mu_s) =
\exp \bigg\{2 S_{\Gamma_{\text{cusp}}}(\mu_h, \mu_s)- a_{\Gamma_{\text{cusp}}}(\mu_h, \mu_s) \left(\ln \frac{M_{t\bar{t}}^2}{\mu_h^2}-i \pi \right) \\
+ a_{\Gamma_{\text{cusp}}}(\mu_f, \mu_s) \ln \bar{N}^2
+ 2 a_{\gamma^\phi}(\mu_s,\mu_f)\bigg\} \\ \times \bm{u}^m(M_{t\bar{t}},m_t,\cos\theta,\mu_h,\mu_s)
\, .
\end{multline}
The explicit exponential in the above equation is color-diagonal, and contains
the evolution functions
\begin{align}
\label{eq:RGexps}
S_\gamma(\nu,\mu)=-\int^{\alpha_s(\mu)}_{\alpha_s(\nu)}d\alpha\,\frac{\gamma(\alpha)}{\beta(\alpha)}\int^\alpha_{\alpha_s(\nu)}\frac{d\alpha '}{\beta(\alpha ')} \, , \quad a_\gamma(\nu,\mu)=-\int^{\alpha_s(\mu)}_{\alpha_s(\nu)}d\alpha\frac{\gamma(\alpha)}{\beta(\alpha)} \, ,
\end{align}
where $\gamma$ stands for an anomalous dimension such as $\Gamma_{\rm cusp}$ or
$\gamma^\phi$, and $\beta(\alpha_s) = d\alpha_s(\mu)/d \ln \mu$ is the
QCD $\beta$-function. The matrix $\bm{u}^m$ in the second line of eq.~(\ref{eq:Umat}) is given by
\begin{equation}
 \bm{u}^m(M_{t\bar{t}},m_t,\cos\theta,\mu_h,\mu_s) = \mathcal{P} \exp \int^{\alpha_s(\mu_s)}_{\alpha_s(\mu_h)} \frac{d \alpha}{\beta(\alpha)} \,
\bm{\gamma}^{h,m}(M_{t\bar{t}},m_t,\cos\theta,\alpha) \, , \label{eq:utm}
\end{equation}
where $\bm{\gamma}^{h,m}$ is the color non-diagonal part of hard
anomalous dimension with full mass dependence, and $\mathcal{P}$
denotes path-ordering. The definition and the explicit expressions for
the various anomalous dimensions can be readily found in the appendix
of \cite{Ahrens:2010zv}. Although we have dropped indices indicating
the partonic channel in eqs.~(\ref{eq:Umat}) and~(\ref{eq:utm}), one
should keep in mind that the anomalous dimension $\gamma^\phi$,
(related to the PDFs), the cusp anomalous dimensions $\Gamma_{\rm cusp}$, and the non-color diagonal anomalous dimension
$\bm{\gamma}^{h,m}$ are different in the quark-annihilation and gluon
fusion channels.

The above integrals appearing in the evolution matrices can be
evaluated and truncated to a given logarithmic order. The results can
be expressed in terms of the strong coupling constant evaluated at the
various scales, $\alpha_s(\mu_f)$, $\alpha_s(\mu_h)$ and
$\alpha_s(\mu_s)$, as was done in \cite{Ahrens:2010zv}. This is
convenient if the soft scale $\mu_s$ is chosen directly in momentum
space, and is thus a real number. However, in the current paper, we
employ a more conventional choice of the soft scale in Mellin space,
$\mu_s \sim \Lambda/\bar{N}$ for some mass scale $\Lambda$, which is now a complex number. It is therefore
more convenient to re-express $\alpha_s(\mu_s)$ in terms of
$\alpha_s(\mu_h)$, keeping in mind that $\ln(\mu_h/\mu_s)$ is a large
logarithm. Actually, it is conventional to express all
$\alpha_s(\mu_i)$ in terms of $\alpha_s(\mu_h)$, where $\mu_i$ is any
scale other than $\mu_h$ appearing in the formula. This can be done by
using the perturbative evolution of the strong coupling, which up to
3-loop order is given by (see, e.g. \cite{Moch:2005ba})
\begin{multline}
\label{eq:as-exp}
\alpha_s(\mu) =
\frac{\alpha_s(\mu_h)}{X} \Bigg[ 1 - \frac{\alpha_s(\mu_h)}{4\pi} \, \frac{\beta_1}{\beta_0} \, \frac{\ln X}{X}
\\*
+ \left(\frac{\alpha_s(\mu_h)}{4\pi}\right)^2 \frac{1}{X^2}
\left[\frac{\beta_1^2}{\beta_0^2} \left( \ln^2 X - \ln X -1 + X \right)
+\frac{\beta_2}{\beta_0}(1-X)    \right]  + {\cal O}(\alpha_s^3(\mu_h)) \Bigg] \, ,
\end{multline}
where
\begin{align}
X = 1 - \frac{\alpha_s(\mu_h)}{2\pi} \beta_0 \ln \frac{\mu_h}{\mu} \,.
\end{align}
Logarithmic orders are then determined according to powers of $\alpha_s(\mu_h)$ with $\ln(\mu_h/\mu_i)$ counted as order $1/\alpha_s(\mu_h)$. For this reason we introduce two ${\cal O}(1)$ parameters: $\lambda_s$ and $\lambda_f$ defined by
\begin{align}
\label{eq:lambda}
\lambda_i = \frac{\alpha_s(\mu_h)}{2 \pi} \beta_0 \ln \frac{\mu_h}{\mu_i} \, ,
\end{align}
for $i=\{s,f\}$.
The part of the exponent proportional to the identity matrix in eq.~(\ref{eq:Umat}) can now be expanded as a series in $\alpha_s(\mu_h)$, and the resulting evolution matrix can be written in the form
\begin{multline}
\label{eq:Umatgs}
\widetilde{\bm{U}}^m(\mu_f,\mu_h,\mu_s) = \exp \left\{\frac{4 \pi}{\alpha_s(\mu_h)} \, g^m_1(\lambda_s,\lambda_f)+g^m_2(\lambda_s,\lambda_f)+\frac{\alpha_s(\mu_h)}{4 \pi} \, g^m_3(\lambda_s,\lambda_f) + \cdots \right\}
\\*
\times \bm{u}^m(M_{t\bar{t}},m_t,\cos\theta,\mu_h,\mu_s) \,.
\end{multline}
Explicit expressions for the RG exponents $g_i^m$ are given in
appendix~\ref{app:gfn_massive}, while a method for evaluating $\bm{u}^m$ is
detailed in~\cite{Ahrens:2010zv} using techniques
from~\cite{Buras:1991jm}.

The all-order resummed hard scattering kernel
eq.~(\ref{eq:soft-resummed}) is formally independent of the matching
scales $\mu_s$ and $\mu_h$. However, the truncation of the resummed
formula to a given logarithmic order introduces residual dependence on
these scales. In order to perform the resummation, one must choose
these scales in such a way that the fixed-order expansion of the hard and soft
functions are free of large logarithms. The explicit form of the
one-loop hard and soft functions reveals that the leading logarithmic
terms are $\alpha_s \ln^2(M_{t\bar{t}}^2/\mu_h^2)$ and $\alpha_s
\ln^2(M_{t\bar{t}}^2/\bar{N}^2/\mu_s^2)$, respectively, which motivates the
na\"ive choices $\mu_h \sim M_{t\bar{t}}$ and $\mu_s \sim M_{t\bar{t}}/\bar{N}$. However, we will study the analytic form of the hard and soft functions in greater detail in section~\ref{sec:systematic} in order to make a more informed choice of appropriate scales. It should be noted however, that in picking the soft scale $\mu_s$ directly in Mellin space (as we do in this work) the resummed hard-scattering kernel
contains a branch cut at large $N$ due to the Landau pole in the
running of $\alpha_s$. This leads to ambiguities in the choice of
contour for the inverse Mellin transform in eq.~(\ref{mellint}). We come
back to this issue when discussing the numerical implementation of our
results in section~\ref{sec:MellinInv}. As mentioned earlier, one
could also choose the soft scale directly in momentum space, which is
then independent of $N$, as was done in \cite{Ahrens:2010zv}. With
this choice the inverse Mellin (or Laplace) transform for the soft
function can be carried out analytically and is free from the Landau
pole problem. On the other hand, this comes at the price of resumming
a different tower of logarithms compared to the pure partonic
threshold ones, as discussed in \cite{Ahrens:2011mw} and explored in
more detail in, e.g. \cite{Bonvini:2012az,Bonvini:2014qga}.

\subsection{Resummation in the boosted-soft limit}
\label{sec:MellinRes_boosted}

Resummation of top-quark hadroproduction cross sections in the boosted-soft limit eq.~(\ref{eq:boostedlimit}) was considered in
\cite{Ferroglia:2012ku}.  We first collect the main results from that
paper in the absence of perturbative corrections involving closed
top-quark loops. In that case, the functions $\bm{H}^m_{ij}$ and
$\widetilde{\bm s}_{ij}^m$ in eq.~(\ref{eq:soft-fact}) can be
subfactorized in the $m_t\to 0$ limit as
\begin{align}
\label{eq:smallMassFac}
 \bm{H}^m_{ij}(M_{t\bar{t}},m_t,\cos\theta,\mu_f) &= \bm{H}_{ij}(M_{t\bar{t}},\cos\theta,\mu_f) \, C_D^2(m_t,\mu_f) + \mathcal{O} \left( \frac{m_t}{M_{t\bar{t}}} \right) ,
 \\  \label{eq:smallMassFac2}
 \widetilde{\bm{s}}^m_{ij} \left( \ln\frac{M_{t\bar{t}}^2}{\bar{N}^2 \mu_f^2},M_{t\bar{t}},m_t,\cos\theta,\mu_f \right) &= \widetilde{\bm{s}}_{ij} \left( \ln\frac{M_{t\bar{t}}^2}{\bar{N}^2 \mu_f^2},M_{t\bar{t}},\cos\theta,\mu_f \right) \, \widetilde{s}_D^2 \left( \ln\frac{m_t}{\bar{N}\mu_f}, \mu_f \right) \nonumber
\\
&\hspace{14em} + \mathcal{O} \left( \frac{m_t}{M_{t\bar{t}}} \right) .
\end{align}
In the above formulas, the hard functions $\bm{H}_{ij}$ and soft
functions $\widetilde{\bm{s}}_{ij}$ without the superscript $m$ are
independent of the top-quark mass $m_t$. They were calculated to NNLO
in \cite{Broggio:2014hoa} and \cite{Ferroglia:2012uy}, respectively,
and can also be applied to di-jet production. All $m_t$-dependence is
factorized into the two functions $C_D$ and $\widetilde{s}_D$, which
are related to the perturbative heavy-quark fragmentation function \cite{Melnikov:2004bm}
and were extracted at NNLO in \cite{Ferroglia:2012ku}.  After this
refactorization, the result for the partonic cross section in the
boosted-soft limit reads
\begin{multline}
\label{eq:boostedFac}
  \widetilde{c}_{ij}(N,M_{t\bar{t}},m_t,\cos\theta,\mu_f) =  \Tr
  \left[ \bm{H}_{ij}(M_{t\bar{t}},\cos\theta,\mu_f) \, \widetilde{\bm{s}}_{ij} \Biggl(
    \ln\frac{M_{t\bar{t}}^2}{\bar{N}^2\mu_f^2}, M_{t\bar{t}},\cos\theta, \mu_f \Biggr) \right]
  \\
  \times C_D^2(m_t,\mu_f)\, \widetilde{s}_D^2 \Biggl(
  \ln\frac{m_t}{\bar{N}\mu_f}, \mu_f \Biggr) + \mathcal{O} \left(
    \frac{1}{N} \right) + \mathcal{O} \left( \frac{m_t}{M_{t\bar{t}}} \right) .
\end{multline}

As for the soft limit, the resummed hard-scattering kernel can be
obtained by deriving and solving RG equations for the component
functions in the above factorization formula.  We write the result in
the form
\begin{multline}
\label{eq:boosted-resummed}
\widetilde{c}_{ij}(N,\mu_f) =
\Tr \Bigg[\widetilde{\bm{U}}_{ij}(\bar{N},\mu_f,\mu_h,\mu_s) \, \bm{H}_{ij}(\mu_h) \, \widetilde{\bm{U}}_{ij}^{\dagger}(\bar{N},\mu_f,\mu_h,\mu_s) \widetilde{\bm{s}}_{ij}\left(\ln\frac{M_{t\bar{t}}^2}{\bar{N}^2 \mu_s^2},\mu_s \right)\Bigg]\\
\times
\, \widetilde{U}_D^2(\bar{N},\mu_f,\mu_{dh},\mu_{ds})
\, C_D^2(m_t,\mu_{dh})\, \widetilde{s}_D^2 \Biggl(
\ln\frac{m_t}{\bar{N}\mu_{ds}}, \mu_{ds} \Biggr)
+ \mathcal{O} \left( \frac{1}{N} \right) + \mathcal{O} \left( \frac{m_t}{M_{t\bar{t}}} \right) .
\end{multline}
In the l.h.s.\ of eq.~(\ref{eq:boosted-resummed}) we dropped the dependence on the arguments
$M_{t\bar{t}}$, $\cos\theta$ and $m_t$. Similarly, $\widetilde{\bm{U}}_{ij}$, $\bm{H}_{ij}$ and $\widetilde{\bm{s}}_{ij}$ also depend on $M_{t\bar{t}}$ and $\cos\theta$.

As mentioned with the massive hard and soft functions, we postpone discussion of the most appropriate scale choices for the matching functions $\bm{H}_{ij}$ and $\widetilde{\bm{s}}_{ij}$ until section~\ref{sec:systematic}. However, since the functions $C_D$ and $\widetilde{s}_D$ depend only on the scales $m_t$ and $m_t/\bar{N}$, we can immediately make the assignment $\mu_{dh}=m_t$ and $\mu_{ds}=m_t/\bar{N}$. The
evolution of the hard and soft functions is encoded in
the functions $\widetilde{\bm{U}}_{ij}$, which are matrices in color
space, while the evolution of $C_D$ and $\widetilde{s}_D$ is in the
color-diagonal function $\widetilde{U}_D$.
Suppressing for the moment the channel labels $ij$, the explicit expressions for the evolution matrices are given by
\begin{align}
\label{eq:Umat-massless}
\widetilde{\bm{U}}(\mu_f,\mu_h,\mu_s) &= \exp \bigg\{2 S_A(\mu_h, \mu_s)- a_A(\mu_h, \mu_s) \left(\ln \frac{M_{t\bar{t}}^2}{\mu_h^2}-i\pi\right) + a_A(\mu_f, \mu_s) \ln \bar{N}^2 \nonumber \\*
&\hspace{6em} + 2 a_{\gamma^\phi}(\mu_s,\mu_f) + 2 a_{\gamma^{\phi_q}}(\mu_s,\mu_f)\bigg\} \, \bm{u}(M_{t\bar{t}},\cos\theta,\mu_h,\mu_s) \, ,
\\
\label{eq:Umat-fragFn}
\widetilde{U}_D(\mu_f,\mu_{dh},\mu_{ds}) &= \exp \bigg\{ -2S_{\Gamma^q}(\mu_{dh},\mu_{ds})
+a_{\Gamma^q}(\mu_{dh},\mu_{ds})\ln \frac{m_t^2}{\mu_{dh}^2}
-a_{\Gamma^q}(\mu_f,\mu_{ds})\ln \bar{N}^2
\nonumber \\*
&\hspace{6em} -2a_{\gamma^S}(\mu_{dh},\mu_{ds})-2 a_{\gamma^{\phi_q}}(\mu_{dh},\mu_f)\bigg\}
\, .
\end{align}
The functions $S_A$, $a_\gamma$, and $\bm{u}$ are defined analogously to those in eqs.~(\ref{eq:RGexps}) and~(\ref{eq:utm}). Here $A$ is given by $2\Gamma_{\rm cusp}^q$ in the $q\bar{q}$ channel and $\Gamma_{\rm cusp}^q + \Gamma_{\rm cusp}^g$ in the $gg$ channel.
Definitions of the various anomalous dimensions and their explicit
perturbative expansions can be found in \cite{Ferroglia:2012ku}. In
order to write the perturbative expansion of the evolution functions
in the same form as eq.~(\ref{eq:Umatgs}) we introduce two additional
$\mathcal{O}(1)$ parameters, $\lambda_{dh}$ and $\lambda_{ds}$ in complete analogy to $\lambda_s$ and $\lambda_f$, as defined earlier in
eq.~(\ref{eq:lambda}). Following the same procedure as in the soft
limit, the evolution matrices can then be written as
\begin{align}
\label{eq:Umatg-boosted}
\widetilde{\bm{U}}(\mu_f,\mu_h,\mu_s) = & \exp \left\{\frac{4 \pi}{\alpha_s(\mu_h)} \, g_1(\lambda_s,\lambda_f)+g_2(\lambda_s,\lambda_f)+\frac{\alpha_s(\mu_h)}{4 \pi} \, g_3(\lambda_s,\lambda_f)\right\} \nonumber \\*
& \times \bm{u}(M_{t\bar{t}},\cos\theta,\mu_h,\mu_s) \, ,
\end{align}
and
\begin{align}
\label{eq:Umatg-boosted-d}
U_D(\mu_f, \mu_{dh},\mu_{ds}) =  \exp \bigg\{ & \frac{4 \pi}{\alpha_s(\mu_h)} \, g^D_1(\lambda_{dh},\lambda_{ds},\lambda_f)+g^D_2(\lambda_{dh},\lambda_{ds},\lambda_f ) \nonumber \\
& +\frac{\alpha_s(\mu_h)}{4 \pi} \, g^D_3(\lambda_{dh},\lambda_{ds},\lambda_f)\bigg\} \, .
\end{align}
We list the (lengthy) expressions for the RG-exponents $g_i$ and
$g^D_i$ in appendix~\ref{app:gfn_massless}.

In the presence of heavy-quark loops the factorization of the partonic
cross section in the boosted-soft limit is more involved.  It is
necessary to introduce additional coefficients related to matching
six-flavor PDFs, heavy-quark fragmentation functions, and $\alpha_s$
onto five-flavor ones. To ease notation we cluster these contributions
from heavy quarks together in coefficients $\tilde{c}^{ij}_t$.  Such
corrections are proportional to powers of $n_h$, the number of heavy
flavors, and introduce additional $m_t$ dependence into the formula
via logarithms of the form $\ln^n(m_t/M_{t\bar{t}})$.  Our factorized hard
scattering kernel is then written as
\begin{multline}
\label{eq:boosted-resummed-nf}
  \widetilde{c}_{ij}(N,M_{t\bar{t}},m_t,\cos\theta,\mu_f) =  \Tr
  \left[ \bm{H}_{ij}(M_{t\bar{t}},\cos\theta,\mu_f) \, \widetilde{\bm{s}}_{ij} \Biggl(
    \ln\frac{M_{t\bar{t}}^2}{\bar{N}^2\mu_f^2}, M_{t\bar{t}},\cos\theta, \mu_f \Biggr) \right]
  \\
  \times C_D^2(m_t,\mu_f)\, \widetilde{s}_D^2 \Biggl(
  \ln\frac{m_t}{\bar{N}\mu_f}, \mu_f \Biggr)\tilde{c}^{ij}_t\Biggl(\ln \frac{1}{\bar{N}^2},m_t,\mu_f\Biggr) + \mathcal{O} \left(
    \frac{1}{N} \right) + \mathcal{O} \left( \frac{m_t}{M_{t\bar{t}}} \right) .
\end{multline}

It is not entirely clear whether logarithms of the form $\ln^n(m_t/M_{t\bar{t}})$ appearing through heavy-quark loops\footnote{\label{fn:toploop}Our definition of the boosted-soft limit, eq.~(\ref{eq:boostedlimit}), is such that production of additional on-shell top-quark pairs through soft radiation is kinematically forbidden. This is reasonable phenomenologically since the production of four top quarks is usually considered as a different process than top quark pair production. As a result, we need only consider contributions proportional to $n_h$ related to virtual corrections.} can be systematically resummed. Therefore, we add them onto the resummation formula eq.~(\ref{eq:boosted-resummed}) using fixed-order perturbation theory. At NNLO, these mass logarithms come from two sources: those from the interference of two-loop amplitudes with tree-level ones, and those from one-loop amplitudes squared. The one-loop squared contributions can be extracted from the results in \cite{Korner:2008bn, Kniehl:2008fd}, while the two-loop terms were calculated in \cite{Czakon:2007ej, Czakon:2007wk}. Numerically, we have found that the contributions of these $n_h$ terms to the differential cross sections are almost negligible.

\subsection{Resummation accuracy}
\label{sec:resAcc}

Having obtained resummed hard scattering kernels in the soft
eq.~(\ref{eq:soft-resummed}) and boosted-soft
eq.~(\ref{eq:boosted-resummed}) limits, we now examine what level of
resummation can be achieved given the current status of perturbative
calculations.
At this point, it should be pointed out that there exist two naming schemes for the logarithmic accuracies of resummed results, as discussed in \cite{Bonvini:2012az} and summarized in table~1 of \cite{Bonvini:2014joa}. While they are purely conventions and one is free to choose either, it is important to have internal consistency with the earlier works \cite{Ferroglia:2012ku, Pecjak:2016nee, Ahrens:2010zv, Ferroglia:2013awa, Ahrens:2011mw}. We therefore adopt the so-called ``Notation$'$'' outlined in table~1 of \cite{Bonvini:2014joa} to denote the accuracies of our resummed results. 
In table~\ref{tab:rg-counting}, we list
the perturbative orders at which the matching functions and anomalous
dimensions need to be evaluated in order to achieve resummation at a
given logarithmic accuracy. 

As highlighted in section~\ref{sec:kinematics} in the discussion
following eq.~(\ref{eq:plusDist}), in Mellin space the perturbative
expansion of the resummed cross section gives corrections of the form
$\alpha_s^n L^k$ where $L = \ln \bar{N}$. The power $k$ of logarithms included in the expansion of the resummed result at a given logarithmic accuracy is also indicated in the last column of table~\ref{tab:rg-counting}. As can be seen there, the difference between the NNLL and NNLL$'$ accuracies amounts to a single logarithm at each order in perturbation theory.

\begin{table}[t!]
\centering
\begin{tabular}{|l|l|l|l|c|}
\hline
        & $\Gamma^i_{\rm cusp}$, $\beta$ & $\bm{\gamma}^h$, $\gamma^{S}$, $\gamma^{\phi}$ & $\bm{H}^{(m)}$, $\widetilde{\bm{s}}^{(m)}$, $C_D$, $\widetilde{s}_D$ & $\alpha_s^n L^k$
\\ \hline
NLL     & NLO    & LO      & LO    & $2n-1 \leq k \leq 2n$
\\
NNLL    & NNLO   & NLO     & NLO   & $2n-3 \leq k \leq 2n$
\\
NNLL$'$ & NNLO   & NLO     & NNLO  & $2n-4 \leq k \leq 2n$
\\ \hline
\end{tabular}
\caption{\label{tab:rg-counting}Our naming scheme for the logarithmic accuracies. We list the perturbative orders at which the cusp anomalous dimension, the QCD $\beta$-function, all other anomalous dimensions and matching functions need to be evaluated in order to obtain resummation at a given logarithmic order.}
\end{table}

The cusp anomalous dimension is fully known to three-loop
order~\cite{Moch:2004pa}, results for the other anomalous dimensions
to NLO can be found
in~\cite{Becher:2009kw,Becher:2009qa,Ferroglia:2009ep,Ferroglia:2009ii,Kidonakis:2009ev,Korchemsky:1987wg,Korchemsky:1991zp},
and the massive hard $\bm{H}^m_{ij}$ and soft functions
$\widetilde{\bm{s}}^m_{ij}$ have been extracted to NLO
\cite{Ahrens:2010zv}. We can therefore perform resummation in the soft
limit up to the NNLL accuracy\footnote{The NNLO massive soft function has recently been calculated in \cite{nnlosoft}. The NNLO massive hard function could be extracted from the virtual amplitude in \cite{Baernreuther:2013caa}. It should therefore be possible to push the resummation accuracy to include NNLL$_m'$ in the future.}. On the other hand, the matching functions in the massless limit
$\bm{H}_{ij}$, $\widetilde{\bm{s}}_{ij}$, $C_D$ and $\widetilde{s}_D$ are
all known to NNLO
\cite{Ferroglia:2012ku,Broggio:2014hoa,Ferroglia:2012uy}, enabling
resummation to NNLL$'$ accuracy in the boosted-soft limit. In terms of the evolution factors in
eqs.~(\ref{eq:Umatgs}),~(\ref{eq:Umatg-boosted}), and (\ref{eq:Umatg-boosted-d}), we need to keep the first three $g$-functions and compute the evolution matrices $\bm{u}^m$ and $\bm{u}$ to second order to obtain NNLL or NNLL$'$ accuracy. Keeping only the first two $g$-functions and the LO $\bm{u}^m$ and $\bm{u}$ matrices results in NLL
resummation as can be seen by the lower perturbative order of the anomalous dimensions in the first line of table~\ref{tab:rg-counting}.

\subsection{Mellin inversion}
\label{sec:MellinInv}

Given the results for Mellin-space resummed hard-scattering kernels in eqs.~(\ref{eq:soft-resummed}) and~(\ref{eq:boosted-resummed}), we must perform the inverse Mellin transform of eq.~(\ref{mellint}) in order to get the differential cross section in momentum space. For the invariant mass distribution we have
\begin{align}
  \frac{d^2\sigma}{dM_{t\bar{t}} \, d\cos\theta} &= \frac{8\pi\beta_t}{3sM_{t\bar{t}}}\sum_{ij} \frac{1}{2\pi i} \int_{c-i\infty}^{c+i\infty} dN \, \tau^{-N} \, \widetilde{\mathcal{L}}_{ij}(N,\mu_f) \, \widetilde{c}_{ij}(N,M_{t\bar{t}},m_t,\cos\theta,\mu_f) \nonumber
  \\
  &= \frac{8\pi\beta_t}{3sM_{t\bar{t}}}\sum_{ij} \!\int_\tau^\infty \! \frac{dz}{z} \, \mathcal{L}_{ij}(\tau/z,\mu_f)\frac{1}{2\pi i}\int_{c-i\infty}^{c+i\infty}\!\!\!\!  dN \, z^{-N} \, \widetilde{c}_{ij}(N,M_{t\bar{t}},m_t,\cos\theta,\mu_f) \, ,
  \label{eq:xs-inverse}
\end{align}
and the transverse momentum distribution follows similarly. 
As indicated in eq.~(\ref{eq:xs-inverse}), there are two ways of
carrying out the inverse transform. The first line utilizes the parton
luminosity functions in Mellin space ($N$-space), while the second
line uses the PDFs in momentum space ($x$-space). The $x$-space PDFs
are easier to obtain, but the $z$-integration is numerically unstable
due to the singular behaviour of $\mathcal{L}(y)$ around $y \sim 0$,
and one needs to use some tricks to improve the convergence
\cite{Kulesza:2002rh}. The use of the $N$-space PDFs, on the other
hand, avoids these instabilities and leads to a fast numerical
implementation. However, public libraries such as LHAPDF
\cite{Buckley:2014ana} only give $x$-space PDFs and one needs to
construct the $N$-space ones by performing the Mellin transform and
analytically continuing to complex $N$. To this end we employ the
methods from \cite{Bonvini:2012sh, Bonvini:2014joa}, namely, we
approximate the $x$-space luminosity functions in terms of Chebyshev
polynomials, from which the Mellin transform can be carried out
analytically.

At fixed
order in $\alpha_s$, the inverse Mellin transform in eq.~(\ref{eq:xs-inverse}) is
well-defined (in the sense of the delta function and plus
distributions) and independent of the integration contour as long as
it lies to the right of all (physical) singularities of the
hard-scattering kernels $\widetilde{c}_{ij}$. However, after
resummation the $\widetilde{c}_{ij}$ develop dependence on the running
coupling $\alpha_s(\mu)$ at the soft scale $\mu_s$ with the canonical
choice $\mu_s \sim M_{t\bar{t}}/\bar{N}$ (and likewise with the soft-collinear
scale $\mu_{ds} \sim m_t/\bar{N}$ in the boosted-soft case). This
introduces an unphysical Landau pole singularity at large $N$ in
$\widetilde{c}_{ij}$, and the inverse transform is ambiguous against
the choice of contour. In this paper we adopt the so-called Minimal
Prescription (MP) \cite{Catani:1996yz}, in which the contour is chosen
to be to the left of the Landau pole but to the right of all other
singularities. It is because of this prescription that the first
integration in $z$ runs from $\tau$ to $\infty$, not $\tau$ to
$1$. Our hard scattering kernel $\tilde{c}_{ij}$ no longer vanishes
for $z>1$, but does asymptote to zero quickly enough so that parton
model assumptions are not violated. This point is addressed in detail
in appendix B of \cite{Catani:1996yz}. Other prescriptions such as the
Borel Prescription exist in the literature \cite{Forte:2006mi}, and we
have checked that in our case there is no large numerical difference
between the two.

\subsection{Ingredients for matching with fixed-order}
\label{sec:Matching}

We outlined the procedure for matching fixed-order calculations with soft (and boosted-soft) 
gluon resummation formulas in section~\ref{sec:prelim}. In the following, we make more precise some of the definitions introduced in the matching equations.  

In eq.~(\ref{eq:soft-matched}), we need the NLO and the NNLO expansions of the NNLL$_m$ resummed result. For the NLO expansion, we can make use of the following result:
\begin{align}
\label{eq:NLOexp}
d\sigma^{\mathrm{NNLL}_m}\Big|_{\substack{\text{NLO} \\ \text{expansion} }}= 
d\sigma^{\mathrm{NNLL}_m}\Big|_{\mu_h=\mu_s=\mu_f} \, .
\end{align}
The reason that the NLO expansion of the NNLL formula is so simple is
that $\mu_h$ and $\mu_s$ dependence in the NLO matching functions
cancels against factors that come from expanding the RG evolution
factors in eq.~(\ref{eq:soft-resummed}).  The end result can thus be
obtained directly by setting these scales to $\mu_f$ at the beginning
as indicated in eq.~(\ref{eq:NLOexp}), which turns off the RG
evolution and leaves the NLO matching functions evaluated at those
scales.  A similar result holds for the NNLO expansion of the NNLL$'_b$ result, namely
\begin{align}
\label{eq:NNLObexp}
d\sigma^{\mathrm{NNLL}'_b}\Big|_{\substack{\text{NNLO} \\ \text{expansion} }}= 
d\sigma^{\mathrm{NNLL}'_b}\Big|_{\mu_i = \mu_f} \, .
\end{align}
where $\mu_i\in\{\mu_h,\mu_s,\mu_{ds},\mu_{dh}\}$.

The NNLO expansion of the NNLL$_m$ resummed result, which we write as 
\begin{align}
\label{eq:NNLL-NNLO}
d\sigma^{\mathrm{NNLL}_m}\Big|_{\substack{\text{NNLO} \\ \text{expansion} }}& = 
d\sigma^{\mathrm{NNLL}_m}\Big|_{\substack{\text{NLO} \\ \text{expansion} }}
+ d\sigma^{\mathrm{NNLL}_m,(2)}  \, ,
\end{align}
is not as simple, because the NNLO matching functions are absent. As a result, the $\mu_h$ and $\mu_s$ dependence does not completely cancel at NNLO. We can express $d\sigma^{\mathrm{NNLL}_m,(2)}$ by inserting the following $\widetilde{c}^{(2)}$ into eq.~(\ref{eq:x-sec-mellin}) or (\ref{eq:x-sec-pt}) in place of $\widetilde{c}_{ij}$:
\begin{align}
\label{eq:c2}
\widetilde{c}^{(2)} &= \Tr \left[ \bm{H}_m^{(2)}(\mu_f) \, \widetilde{\bm{s}}_m^{(0)}(\mu_f) + \bm{H}_m^{(1)}(\mu_f) \, \widetilde{\bm{s}}_m^{(1)}(\mu_f) + \bm{H}_m^{(0)}(\mu_f) \, \widetilde{\bm{s}}_m^{(2)}(\mu_f) \right]  \nonumber
\\
&- \Tr \left[ \bm{H}_m^{(2)}(\mu_h) \, \widetilde{\bm{s}}_m^{(0)}(\mu_s) + \bm{H}_m^{(1)}(\mu_h) \, \widetilde{\bm{s}}_m^{(1)}(\mu_s) + \bm{H}_m^{(0)}(\mu_h) \, \widetilde{\bm{s}}_m^{(2)}(\mu_s)\right] ,
\end{align}
where we have suppressed all arguments in the expansion coefficients of the hard and soft functions with the exception of the scale at which they should be evaluated. The expansion coefficients are defined through
\begin{align}
\label{eq:expCoeffs}
\bm{H}^m&= \alpha_s^2\left[\bm{H}_m^{(0)}+\left(\frac{\alpha_s}{4\pi}\right) \bm{H}_m^{(1)}
+\left(\frac{\alpha_s}{4\pi}\right)^2 \bm{H}_m^{(2)} +\cdots \right]\, , 
\nonumber \\
\widetilde{\bm{s}}^m&=\widetilde{\bm{s}}_m^{(0)} +\left(\frac{\alpha_s}{4\pi}\right)\widetilde{\bm{s}}_m^{(1)} 
+ \left(\frac{\alpha_s}{4\pi}\right)^2\widetilde{\bm{s}}_m^{(2)}  + \cdots \, .
\end{align}

The form of $\widetilde{c}^{(2)}$ in eq.~(\ref{eq:c2}) requires a bit of explanation.
First of all, eq.~(\ref{eq:NLOexp}) tells us that $\widetilde{c}^{(2)}$ should vanish if we set $\mu_h=\mu_s=\mu_f$, since the NNLL$_m$ calculation includes only the NLO matching
coefficients.  Secondly, note the fact that the NNLL$_m$ resummed result $d\sigma^{\mathrm{NNLL}_m}$ \emph{would} be independent of $\mu_s$ and $\mu_h$ up to NNLO if the NNLO contributions from the hard and soft functions were known and included (which would upgrade the resummation accuracy to NNLL$'_m$). 
Therefore, in a fixed order expansion of the resummed result to NNLO, one should recover the first line of eq.~(\ref{eq:c2}) by adding the NNLO contributions from the hard function (evaluated at $\mu_h$) and the soft function (evaluated at $\mu_s$). The form of eq.~(\ref{eq:c2}) follows directly from these two facts. 
Evidently, we must also explain how to evaluate eq.~(\ref{eq:c2}) without knowing the massive two-loop hard and soft functions $\bm{H}_m^{(2)}$ and $\bm{S}_m^{(2)}$. The logarithmic terms of these two functions can be determined from their RG equations and in fact this is all one needs. The non-logarithmic ``constant'' terms which could also appear in $\bm{H}_m^{(2)}$ and $\bm{S}_m^{(2)}$ (and which are not determined by the RG equation) cancel between the first and second lines in eq.~(\ref{eq:c2}) and therefore do not contribute to $\widetilde{c}^{(2)}$. This concludes the matching of the NNLL$_m$ results with fixed-order calculations.

We now turn to eqs.~(\ref{eq:res_comb}) and (\ref{eq:fully-matchedNNLO}), which describe the matching between (N)NLO, NNLL$_m$, and NNLL$'_b$ results. For this we need the NNLO expansion of $d\sigma^{\text{NNLL}'_b}$ given in eq.~(\ref{eq:NNLObexp}), as well as the NNLO expansion of the NNLL resummation formulas given in eq.~(\ref{eq:NNLL-NNLO}). A further ingredient is the $m_t\to0$ limit of the NNLL$_m$ formula appearing in eq.~(\ref{eq:res_comb}). To evaluate that we exploit the fact that the boosted-soft resummation formula is the small-mass limit of the soft resummation formula at any fixed order in $\alpha_s$. This leads to the result
\begin{align}
d\sigma^{\mathrm{NNLL}_m}\big|_{m_t\to 0} &=
d\sigma^{\mathrm{NNLL}_b}\big|_{\substack{\mu_{ds}=\mu_s \\ \mu_{dh}=\mu_h}} \, .
\end{align}
This follows because setting $\mu_{dh}=\mu_h$ and $\mu_{ds}=\mu_s$ in the boosted-soft result removes RG evolution between the functions $\bm{H}$ and $c_D$
in eq.~(\ref{eq:smallMassFac}), and $\widetilde{\bm{s}}$ and
$\widetilde{s}_D$ in eq.~(\ref{eq:smallMassFac2}), thus leaving behind
the leading contributions from threshold resummation in the limit
$m_t/M_{t\bar{t}}\to 0$.

It is worth mentioning some additional subtleties concerning the virtual top-quark loops at NNLO (we mentioned in the footnote on page \pageref{fn:toploop} that we do not consider real top-quark pair emissions). As discussed in the last paragraph of section~\ref{sec:MellinRes_boosted}, it is not entirely clear how to resum these contributions to all orders in $\alpha_s$, and we choose to add them in fixed order. For the NNLO+NNLL$'$ result this is automatically taken into account by the matching formula while for the NLO+NNLL$'$ result we have to add them manually. A complication arises from the fact that the soft resummed result, $d\sigma^{\mathrm{NNLL}_m}$, generates some (but not all) of the $\alpha_s^2n_h$ terms through RG running. We need to subtract these terms out before adding back the full NNLO heavy quark contributions introduced at the end of section~\ref{sec:MellinRes_boosted} in order to avoid double counting. Again, we only consider these contributions for completeness, practically their numerical impact is negligible.

\section{Choosing the kinematics-dependent matching scales}
\label{sec:systematic}

The (massive) hard and soft functions depend on the Mandelstam
variables $t_1,u_1$ and $\hat{s}$, and the appropriate choice of the matching
scales $\mu_h$ and $\mu_s$ appearing in the resummation formalism
depends on the kinematic regime probed by the differential cross section under study.   
The focus of this section will be to identify a well-motivated set of these 
kinematics-dependent matching scales 
for the top-pair invariant mass and $p_T$ distributions, 
which can be used all the way from the low-energy regime where
all kinematic invariants are on the order of the top-quark mass, up to
the boosted regime where $m_t$ is small compared to $M_{t\bar{t}}$ or $p_T$. 
We begin our analysis by highlighting and comparing some important kinematic features of the
low-energy and high-energy regions of the $M_{t\bar{t}}$ and $p_T$
distributions in section~\ref{sec:KinFeatures}.  We then devote section~\ref{sec:scalesMtt} to scale
choices for the top-pair invariant mass distribution and section~\ref{sec:scalesPT} to
those for top-quark $p_T$ distribution.

Throughout the rest of the paper we use the following numerical inputs. We fix the LHC collider energy to $\sqrt{s}=13$~TeV,
take $m_t=\unit{173.3}{\GeV}$, and use the NNPDF3.0 PDF sets with
${\alpha_s(M_Z)=0.118}$ \cite{Ball:2014uwa} in conjunction with LHAPDF6 \cite{Buckley:2014ana}. We use NNLO PDFs for all
predictions unless otherwise indicated. In the numerical evaluation of the
resummed formulas we have made use of the \texttt{CUBA} integration
library \cite{Hahn:2004fe}.

\subsection{\boldmath Some kinematic considerations that underpin the choice of scales}
\label{sec:KinFeatures}

In this section we point out some important kinematic
features of the $M_{t\bar{t}}$ and top-quark $p_T$ distributions which are instrumental in determining appropriate values of the matching scales $\mu_h$ and $\mu_s$. 
In particular, we study the high-energy and low-energy
regimes for both distributions, and explain the differences between them that impact their
description in fixed-order and resummed perturbation theory.

The main idea is that soft gluon resummation works in the limit where
the kinematic features of a given observable resemble those of the LO
process. This can be studied by introducing kinematic variables sensitive to
higher-order hard emissions.  A kinematic variable we find particularly 
useful is
\begin{align}
\label{eq:HTdef}
R_T \equiv \frac{H_T}{M_{t\bar{t}}} \equiv \frac{1}{M_{t\bar{t}}} \bigg[ \sqrt{m_t^2+p^2_{T,t}} + \sqrt{m_t^2+p^2_{T,\bar{t}}}  \bigg] \, .
\end{align}
At LO 
\begin{align}
\label{eq:LOrt}
R_T^2 = \frac{4t_1 u_1}{M_{t\bar{ t}}^4} = 1- \left(1-\frac{4m_t^2}{M_{t\bar{t}}^2}\right) \cos^2\theta \, ,
\end{align}
and $R_T\leq 1$, with $R_T=1$ corresponding to $\theta=\pi/2$ (central
scattering). Moreover, one finds that the Jacobian factors arising from rewriting
the double differential cross sections in terms of $R_T$ and either $M_{t\bar{t}}$ or $p_T$ have 
integrable singularities proportional  to $1/\sqrt{1-R_T^2}$.  For instance, 
\begin{align}
\label{eq:RTjac}
dM_{t\bar{t}}\,  d\cos\theta = \frac{R_T}{\beta_t \sqrt{1-R_T^2}} \,dM_{t\bar{t}}\, dR_T   \, .
\end{align}
Beyond LO, however, $H_T$ is only constrained to be
smaller than $\sqrt{\hat{s}}$. Hard emissions generate
non-vanishing cross section in the region $R_T>1$, in addition to
contributing to the $R_T<1$ region, and the singularity at $R_T=1$ is resolved into 
a Jacobian peak.  The quantity $R_T$ thus offers a
useful kinematic discriminant: the more sensitive an observable to the
region $R_T \leq 1$, the greater the potential for an improved
prediction through resummation.  Conversely, observables characterized
by $R_T >1$ are inaccessible to soft kinematics and dominated by hard emissions.

\begin{figure}[t!]
\centering
\includegraphics[width=0.495\textwidth]{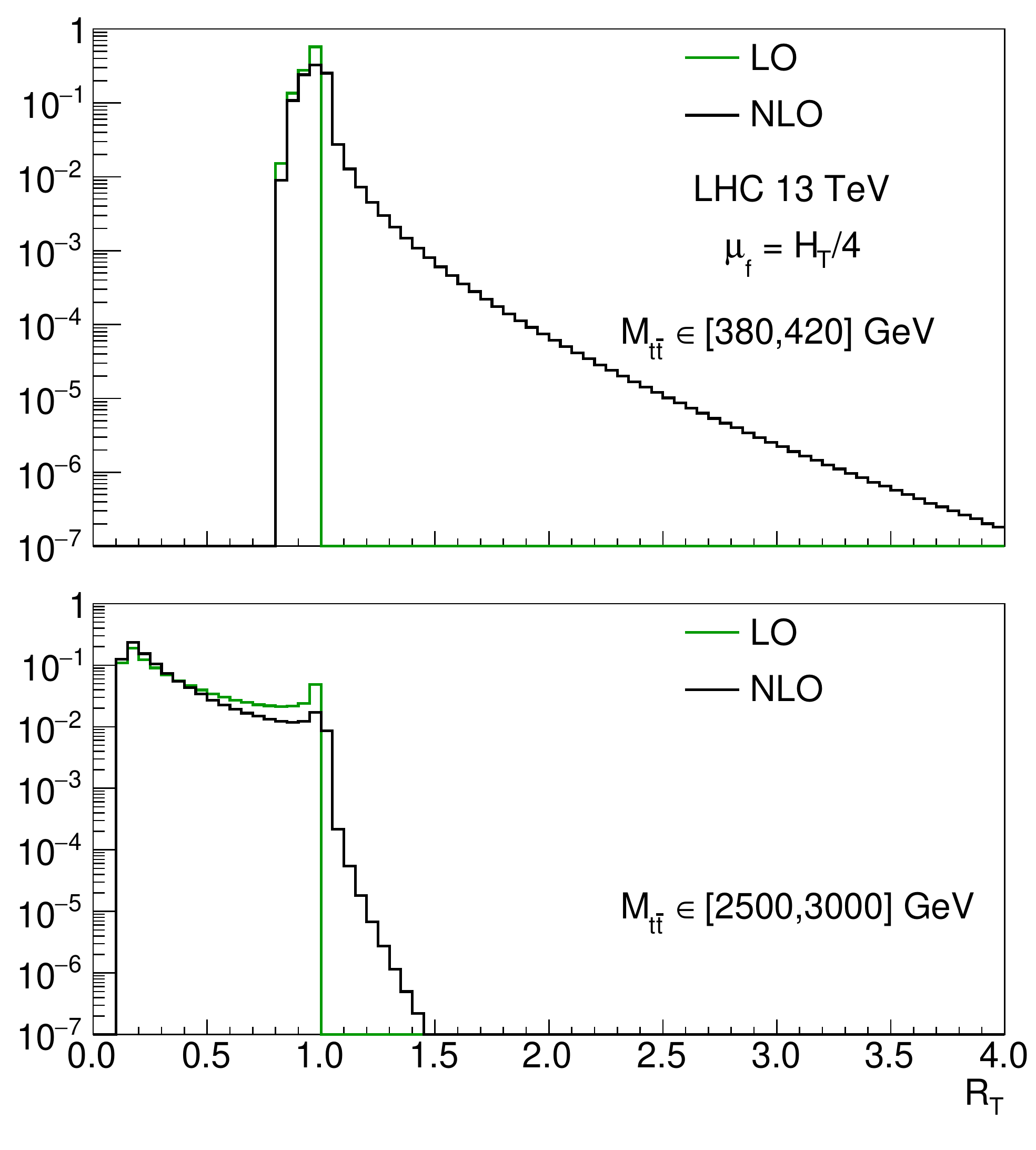}
\includegraphics[width=0.495\textwidth]{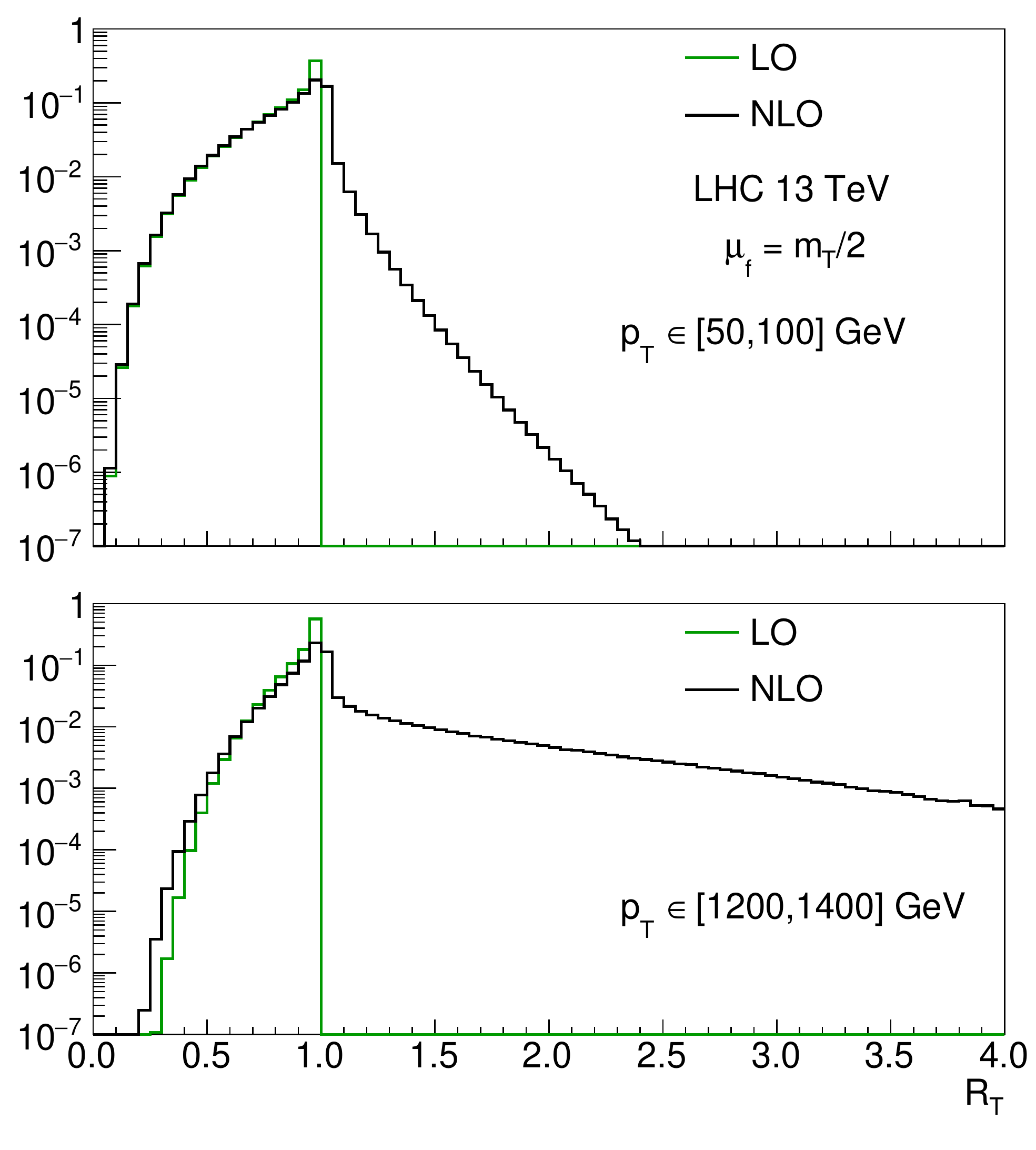}
\vspace{-5ex}
\caption{\label{fig:HTovMdist} The distributions of $R_T=H_T/M_{t\bar{t}}$ at LO (green) and NLO (black), in the four regions $M_{t\bar{t}} \in [380, 420]$~GeV (top-left), $M_{t\bar{t}} \in [2500, 3000]$~GeV (bottom-left), $p_T \in [50, 100]$~GeV (top-right), and $p_T \in [1200, 1400]$~GeV (bottom-right). The distributions are normalized to the integrated cross section in each of the four regions.}
\end{figure}

In figure~\ref{fig:HTovMdist}, we show the distributions of $R_T$ in four
kinematic regions, normalized to the integrated cross section in each
region. The first is $M_{t\bar{t}}\in [380,420]$~GeV, representative
of ``low $M_{t\bar{t}}$'', $M_{t\bar{t}} \gtrsim 2m_t$; the second is
$M_{t\bar{t}}\in [2500,3000]$~GeV, representative of ``high
$M_{t\bar{t}}$'', $M_{t\bar{t}} \gg 2m_t$; the third and fourth
regions are $p_T \in [50,100]$~GeV and $p_T \in [1200, 1400]$~GeV,
representative of ``low $p_T$'' and ``high $p_T$'' respectively. We
will also refer to these as ``low-energy" and ``high-energy" bins of
$M_{t\bar{t}}$ or $p_T$.  An interesting observation is that the
differences between the LO and NLO distributions are
quite distinctive in the four cases.  It is the task of this subsection
to explore the explanations and implications of these facts.

We first discuss the low-$M_{t\bar{t}}$ bin of the top-pair invariant mass distribution, whose $R_T$ distribution is shown in the top-left plot of figure~\ref{fig:HTovMdist}.
At both LO (the green curve) and NLO (the black curve), the distribution is peaked at $R_T \sim 1$, a fact following from eq.~(\ref{eq:RTjac}). However, at NLO, hard emissions generate a non-negligible fraction of the distribution in the $R_T>1$ region. In terms of a perturbative description beyond NLO, this indicates that while soft-gluon resummation can potentially improve the description in this bin, matching with NNLO in order to describe the $R_T>1$ region is needed for a precise prediction.

The $R_T$ distribution in the high-$M_{t\bar{t}}$ bin is displayed in the bottom-left plot of figure~\ref{fig:HTovMdist}.
A remarkable feature of this bin is that the $R_T$ distribution is peaked at values close to $R_T \sim 0.2$,
both at LO and NLO. This feature can be traced to a property of the partonic cross section for $gg\to t\bar{t}$ scattering, namely that the Born diagrams possess $t$- and $u$-channel singularities in the limit $m_t\to 0$. 
These singularities are cut off by the finite  value of the top-quark mass.  However, we show below with analytic arguments that their presence  implies that in the boosted regime where $m_t \ll M_{t\bar{t}}$ the integral over $\cos\theta$ used to calculate the differential cross section is dominated by the regions of small $R_T$. One implication of this feature is that  the high-invariant mass region is especially amenable to soft-gluon 
resummation, as we shall see in section~\ref{sec:discussion}.  Another is that, in spite of what one might expect,
the most relevant perturbative scale in the high-energy tail of the top-pair invariant mass distribution
is $H_T$ rather than $M_{t\bar{t}}$ itself.  As a result, the partonic cross 
section, as well as the hard and soft functions appearing in the factorized form of the cross section in the 
soft limit, are better evaluated at scales $\mu \sim H_T$ in this region.  This was indeed a main result 
of the fixed-order study of the top-pair invariant mass distribution to NNLO  \cite{Czakon:2016dgf}, which 
identified $\mu_f = H_T/4$ as the choice at which fixed-order perturbation theory converges 
best.

We now turn to the $R_T$ distributions in the two bins of top-quark
$p_T$. The low and high $p_T$ regions are shown in the top-right plot
and the bottom-right plot of figure~\ref{fig:HTovMdist},
respectively.
The Jacobian peaks at $R_T \sim 1$ are still present, similar to the $M_{t\bar{t}}$ case. The distributions exhibit long tails toward lower values of $R_T$. These come from the $t$- and $u$-channel singularities, the same effect at work in the high $M_{t\bar{t}}$ bin discussed in the last paragraph. 
On the other hand, the behavior of the NLO distribution in this case is rather different from the $M_{t\bar{t}}$ case.
The measured value of the top quark $p_T$ can only constrain the combined $p_T$ of the anti-top-quark and the extra parton, but puts no constraints on their separate transverse momenta. It is therefore kinematically allowed to have the $t\bar{t}$ pair recoiled against a separate hard parton, such that one ends up with a large $H_T$ and a small $M_{t\bar{t}}$. This explains the long tail at the NLO towards $R_T > 1$, particularly in the high $p_T$ bin. In fact, at $\mu_f = m_T/2$, one finds that $47$\% of the distribution lies at $R_T>1$.
From these facts, we expect that it is important to incorporate the effects of hard emissions for the $p_T$ distribution, especially in the high-$p_T$ region.

\subsection{Scales for the $M_{t\bar{t}}$ distribution}
\label{sec:scalesMtt}

The main goal of this and the next subsection is to identify the
optimal choices of the hard scale $\mu_h$ and the soft scale $\mu_s$,
based on the kinematic features of the hard and soft functions. Since
the hard and soft functions may be evaluated analytically, this will
also help to understand the kinematic features of the $R_T$ distribution
given in the previous subsection. 

The philosophy of RG-improved perturbation theory is to choose the
matching scales such that the fixed-order expansion of the hard and
soft functions is well behaved.  The massless hard and soft functions
depend on the kinematic invariants $M_{t\bar{t}}^2$, $t_1$ and $u_1$.
As long as all of these scales are of the same size numerically, the
choices $\mu_h \sim M_{t\bar{t}}$ and $\mu_s \sim
M_{t\bar{t}}/\bar{N}$ free these functions of potentially large
logarithmic corrections and thus ensure good perturbative convergence.
The situation becomes more subtle in the boosted regime, where $m_t
\ll M_{t\bar{t}}$. In that case, when the top quark is produced in the
very forward region, $\cos\theta \to 1$, the kinematic invariants
develop a hierarchy $|t_1|\ll M_{t\bar{t}}^2\sim |u_1|$.  An analogous
hierarchy develops for very backward production, with
$|t_1|\leftrightarrow |u_1|$.  Both of these situations correspond to
the region $H_T\ll M_{t\bar{t}}$, since for Born
kinematics
\begin{align}
\label{eq:t1Limit}
-t_1 \big|_{m_t\to 0} &\approx \frac{M_{t\bar{t}}^2}{2} (1-\cos\theta) + m_t^2 \cos\theta \xrightarrow{\cos\theta \to 1} p_T^2+m_t^2 \equiv m_T^2 = H_T^2/4\, ,
\\
\label{eq:u1Limit}
-u_1 \big|_{m_t\to 0} &\approx \frac{M_{t\bar{t}}^2}{2} (1+\cos\theta) - m_t^2 \cos\theta \xrightarrow{\cos\theta \to -1} m_T^2 =H_T^2/4 \, .
\end{align}
On the other hand, the region where $m_t \sim M_{t\bar{t}}$
corresponds to $H_T \sim M_{t\bar{t}}$, irrespective of the value of
$\cos\theta$.

The top-pair invariant mass distribution is calculated from the double
differential cross section in eq.~(\ref{eq:x-sec}) by integrating over the
scattering angle in the region $-1<\cos\theta<1$. At large
$M_{t\bar{t}}$, the results for the $R_T$ distribution shown in the
bottom-left panel of figure~\ref{fig:HTovMdist} and discussed in the
previous subsection make clear that the integral is dominated by the
region where $|\cos\theta|\sim 1$ and $R_T$ is significantly smaller
than unity.  This kinematic feature is explained by the fact that at Born level the
$gg$ partonic cross section has $t$- and $u$-channel singularities, related to the hard function. 
For example,  in the limit $t_1\to 0$, the LO hard function reads
\begin{align}
  \bm{H}_{gg}^{(0)}\big|_{t_1\to 0} = \frac{1}{2x_t}
 \begin{pmatrix}
    \frac{1}{N_c^2} & \frac{1}{N_c}  & \frac{1}{N_c}
    \\
    \frac{1}{N_c} & 1 &1
    \\
    \frac{1}{N_c} &1& 1
  \end{pmatrix}
 \, ,
\end{align}
where $x_t\equiv -t_1/M_{t\bar{t}}^2$ and $N_c=3$ is the number of
colors in QCD.  The expression in the limit $u_1\to 0$ is obtained
by replacing $x_t\to 1-x_t$.  Therefore, at fixed $M_{t\bar{t}} \gg m_t$, 
the cross section gets large contributions from the region 
$H_T \ll M_{t\bar{t}}$, due to $t$- and $u$- channel enhancements in the $gg$ channel.
The $q\bar{q}$ channel is free of such $t$- and $u$- channel enhancements.

The dynamical enhancement of the forward and backward scattering
regions at large $M_{t\bar{t}}$ has important implications for the 
choice of the matching scales $\mu_h$ and $\mu_s$.  We can study this
issue analytically by expanding the higher-order corrections to the 
hard and soft functions in the limit $R_T \to 0$.  The $gg$ partonic cross section is symmetric under the 
exchange of  $t_1$ and $u_1$, so the $R_T \to 0$ limit can be easily obtained from the $x_t \to 0$ 
results, which we focus on for concreteness.  While the soft function itself has no $1/x_t$ singularities in this 
limit, it enters the factorization formula in a matrix product with the hard function, so we must deal
with both functions at once in order to take the correct limit of the differential cross section.  This 
leads us to study the higher-order perturbative corrections at the level of the objects 
\begin{align}
\label{eq:HardC}
\mathcal{H}^{\rm LO}_{ij}(\mu_h) &= \alpha_s^2 (\mu_h) \, {\rm Tr}\left[\bm{H}_{ij}^{(0)}\widetilde{\bm{s}}_{ij}^{(0)}\right] \, , \nonumber
\\
\mathcal{H}^{\rm NLO}_{ij}(\mu_h) &=  \mathcal{H}^{\rm LO}_{ij}(\mu_h) + \frac{\alpha_s^3(\mu_h)}{4\pi} \, {\rm Tr}\left[\bm{H}_{ij}^{(1)}(\mu_h)\widetilde{\bm{s}}_{ij}^{(0)}\right] \,, \nonumber
\\
\mathcal{H}^{\rm NNLO}_{ij}(\mu_h) & =  \mathcal{H}^{\rm NLO}_{ij}(\mu_h) + \frac{\alpha_s^4(\mu_h)}{(4\pi)^2} \, {\rm Tr}\left[\bm{H}_{ij}^{(2)}(\mu_h)\widetilde{\bm{s}}_{ij}^{(0)}\right] \,,
\end{align}
and
\begin{align}
\label{eq:SoftC}
\mathcal{S}^{\rm LO}_{ij} &= {\rm Tr}\left[\bm{H}_{ij}^{(0)}\widetilde{\bm{s}}_{ij}^{(0)}\right] \, , \nonumber
\\
\mathcal{S}^{\rm NLO}_{ij}(\mu_s) & =  \mathcal{S}^{\rm LO}_{ij} +
\frac{\alpha_s(\mu_s)}{4\pi} {\rm Tr}\left[\bm{H}_{ij}^{(0)}\widetilde{\bm{s}}_{ij}^{(1)}(\mu_s)\right] \,, \nonumber\\
\mathcal{S}^{\rm NNLO}_{ij}(\mu_s) & =  \mathcal{S}^{\rm NLO}_{ij}(\mu_s)+
\left(\frac{\alpha_s(\mu_s)}{4\pi} \right)^2{\rm Tr}\left[\bm{H}_{ij}^{(0)}\widetilde{\bm{s}}_{ij}^{(2)}(\mu_s)\right] \,,
\end{align}
where we have suppressed the dependence of the matching functions on
all parameters and scales other than $\mu_h$ and $\mu_s$. 
Explicit results for the NLO corrections can be written as 
\begin{align}
\frac{\mathcal{H}^{\rm NLO}_{gg}(\mu_h)}{\mathcal{H}^{\rm LO}_{gg}(\mu_h)} 
\bigg|_{t_1\to 0} &=  1 \!+\! 
 \frac{\alpha_s(\mu_h)}{36\pi}\left[-78\ln^2 \left(\frac{-t_1}{\mu_h^2}  \right)
 \!+\!24\ln \left(\frac{-t_1}{\mu_h^2}\right)(3\!+\!2\ln x_t)\!+\!37\pi^2- 168  \right] 
\, , \nonumber \\
\frac{\mathcal{S}^{\rm NLO}_{gg}(\mu_s)}{\mathcal{S}^{\rm LO}_{gg}} 
\bigg|_{t_1\to 0} &= 1 \!+\! \frac{\alpha_s(\mu_s)}{36\pi}\bigg[ 78 \ln^2\left(\frac{-t_1}{\mu_s^2\bar{N}^2}  \right)
 -48\ln \left(\frac{-t_1}{\mu_s^2\bar{N}^2}\right)\ln x_t+24 \ln^2 x_t \!+\!31\pi^2  \bigg] \,,
\label{eq:HScorNLO}
\end{align}
and those for the  NNLO corrections as  
\begin{align}
\frac{ \mathcal{H}^{\text{NNLO}}_{gg}(\mu_h) }{ \mathcal{H}^{\text{NLO}}_{gg}(\mu_h) } \bigg|_{t_1 \to 0} &= 1 \!+\! \left( \frac{\alpha_s(\mu_h)}{4\pi} \right)^2 \bigg[ 37.6 \ln^4 \left( \frac{-t_1}{\mu_h^2} \right) - \big( 46.2 \ln x_t + 47.2 \big) \ln^3 \left( \frac{-t_1}{\mu_h^2}  \right) \nonumber
\\
&+ \big( 14.2 \ln^2 x_t + 22.2 \ln x_t - 248 \big) \ln^2 \left( \frac{-t_1}{\mu_h^2} \right) \nonumber 
\\
& + \big( 154 \ln x_t + 102 \big) \ln \left( \frac{-t_1}{\mu_h^2} \right) + 12.7 \ln x_t + 577 \bigg] + \mathcal{O}(\alpha_s^3) \, , \nonumber
\\
\frac{ \mathcal{S}^{\text{NNLO}}_{gg}(\mu_s) }{ \mathcal{S}^{\text{NLO}}_{gg}} \bigg|_{t_1\to 0} &= 1 \!+\! \left( \frac{\alpha_s(\mu_s)}{4\pi} \right)^2 \bigg[ 37.6 \ln^4 \left( \frac{-t_1}{\mu_s^2\bar{N}^2} \right) - \big( 46.2 \ln x_t + 22.1 \big) \ln^3 \left( \frac{-t_1}{\mu_s^2\bar{N}^2} \right) \nonumber
\\
&+ \big(37.3 \ln^2 x_t + 20.4 \ln x_t + 354 \big) \ln^2 \left( \frac{-t_1}{\mu_s^2\bar{N}^2} \right) \nonumber
\\
&- \big( 14.2 \ln^3 x_t + 20.4 \ln^2 x_t + 218 \ln x_t + 12.9 \big) \ln \left( \frac{-t_1}{\mu_s^2\bar{N}^2} \right) \nonumber
\\
&+ 3.56 \ln^4 x_t + 6.81 \ln^3 x_t + 109 \ln^2 x_t - 42.6 \ln x_t + 356 \bigg] + \mathcal{O}(\alpha_s^3)\, ,
\label{eq:HScorNNLO}
\end{align}
where we have set $N_c=3$ and the number of light quarks to $N_l=5$ in the above 
equations.\footnote{Although not immediately apparent from the results above, 
one finds that the real parts of ${\bm H}_{gg}^{(1,2)}$ are proportional to ${\bm H}_{gg}^{(0)}$ in the 
$x_t\to 0 $ limit.}

An important feature of eqs.~(\ref{eq:HScorNLO})
and~(\ref{eq:HScorNNLO}) is that both the NLO and the NNLO corrections
depend on the two physical scales $-t_1$ and $M_{t\bar{t}}$ (through
the ratio $x_t$). Therefore, any choice of $\mu_h$ and $\mu_s$ will
lead to corrections of the form $\alpha_s^n\ln^{m}(x_t)/x_t$ in the
$x_t\to 0$ limit.  However, the structure of such corrections is
rather different for the hard and soft functions.  For the hard
function, the choice $\mu_h^2=-t_1$ frees the NLO corrections of such
logarithmic corrections in $x_t \rightarrow 0$ limit, and reduces the
logarithmic terms in the NNLO corrections to a single power of $\ln
x_t$.  On the other hand, the choice $\mu_h=M_{t\bar{t}}$ generates a
double logarithmic series whose corrections have the form
$\alpha_s^n\ln^{2n}(x_t)/x_t$.  Using that  $\sqrt{-t_1}=H_T/2$ in the $x_t \to 0$ limit along
with the symmetry of the $gg$ channel under $t_1\leftrightarrow u_1$, one thus
expects the perturbative corrections from the hard function to the $M_{t\bar{t}}$ distribution
to be well behaved across phase space with the choice $\mu_h = H_T/2$.

For the soft corrections $\mathcal{S}_{gg}$, double logarithmic
corrections of the form $\alpha_s^n \ln^{2n}(x_t)/x_t$ are generated
even when $\mu_s = \sqrt{-t_1}/\bar{N}$ is chosen. To understand this result,
we note that NLO corrections to the massless soft function can be
written in the form \cite{Ferroglia:2012ku,Ferroglia:2012uy}
\begin{align}
\widetilde{\bm{s}}^{(1)} = -\sum_{(I, J)=1}^4 \bm{w}_{IJ} 
\left[\ln^2\left(\frac{s_{IJ}}{\mu_s^2\bar{N}^2} \right)+ 
\frac{\pi^2}{6}+2 {\rm Li}_2\left(1-\frac{s_{IJ}}{M_{t\bar{t}}^2}
\right) \right] \, ,
\end{align}
where $s_{12}=s_{34}=M_{t\bar{t}}^2$, $s_{13}=s_{24}= -t_1$,
$s_{14}=s_{23}=-u_1$, and the sum over $(I,J)$ excludes the terms
where $I=J$. The matrices ${\bm w}_{IJ}$ differ for the $q\bar{q}$ and
$gg$ channels, and can be found in, for instance,
\cite{Ferroglia:2012uy}.  The NLO corrections to the soft functions
are thus characterized by the three different scales $s_{IJ}/\bar{N}$,
the relative importance of each scale being determined by 
the properties of matrix elements ${\bm w}_{IJ}$, which are pure color factors.
Taking the $x_t\to 0$ limit of $\mathcal{S}^{\rm(N)NLO}_{gg}$ then leads to the 
double logarithmic series mentioned above, irrespective of the choice of $\mu_s$.

The analytic results above give very useful insight into the nature of
perturbative corrections arising from the hard and soft
functions.
In particular, they hint at the use of a $H_T$-based scale for the hard function, which is also a reasonable choice for the soft function.
However, they are derived in the formal limit $R_T\to 0$,
which for the top-pair invariant mass distribution is relevant because
of a dynamical enhancement from $t$- and $u$- channel singularities in
the high $M_{t\bar{t}}$ region.  It is therefore useful and necessary
to supplement these analytic arguments with a numerical study.  To do
this, we define NLO and NNLO $K$ factors for the hard and soft
functions in the following way.  For the hard functions, we evaluate
the differential cross section with respect to $M_{t\bar t}$ using
$\mathcal{H}^{\rm (N)NLO}_{ij}$ for the hard-scattering kernel in
eq.~(\ref{eq:x-sec-mellin}), and define NLO and NNLO $K$ factors by
dividing these results by those found using $\mathcal{H}^{\rm
  LO}_{ij}$.  These ratios are indicated by $K_{ij}^{\rm H,
  (N)NLO}(M_{t\bar{t}},\mu_h)$.  The parton luminosity cancels in the
ratio since the soft function is independent of the Mellin parameter $\bar{N}$ at LO, so we can
write
\begin{align}
\label{eq:HardK}
K^{\rm H, NLO}_{ij}(M_{t\bar t},\mu_h) &= \int_{-1}^1 d\cos\theta \, \mathcal{H}^{\rm NLO}_{ij}(\mu_h) \Bigg/ \int_{-1}^1 d\cos\theta \, \mathcal{H}^{\rm LO}_{ij}(\mu_h) \, , \nonumber \\
\\
K^{\rm H, NNLO}_{ij}(M_{t\bar t},\mu_h) &= \int_{-1}^1 d\cos\theta \, \mathcal{H}^{\rm NNLO}_{ij}(\mu_h) \Bigg/ \int_{-1}^1 d\cos\theta \, \mathcal{H}^{\rm LO}_{ij}(\mu_h) \, . \nonumber
\end{align}
Ratios $K^{\rm S, (N)NLO}(M_{t\bar{t}},\mu_s)$, which take into account corrections from the 
soft function, are defined similarly.   The important difference in the case of the soft function
is that it depends on $\bar{N}$ and one must take the product with the
Mellin-transformed parton luminosities before performing the inverse
Mellin transform in order to get the contribution to the differential
cross section.  For simplicity, we use these luminosities evaluated at
the scale $\mu_f = M_{t\bar t}$. We have checked that the luminosity dependence
nearly completely cancels in the ratio defining the soft $K$ factors, so that the
results depend very little on the exact choice of $\mu_f$ and also the collider energy.

\begin{figure}[t!]
\centering
\begin{align}
\includegraphics[width=0.495\textwidth]{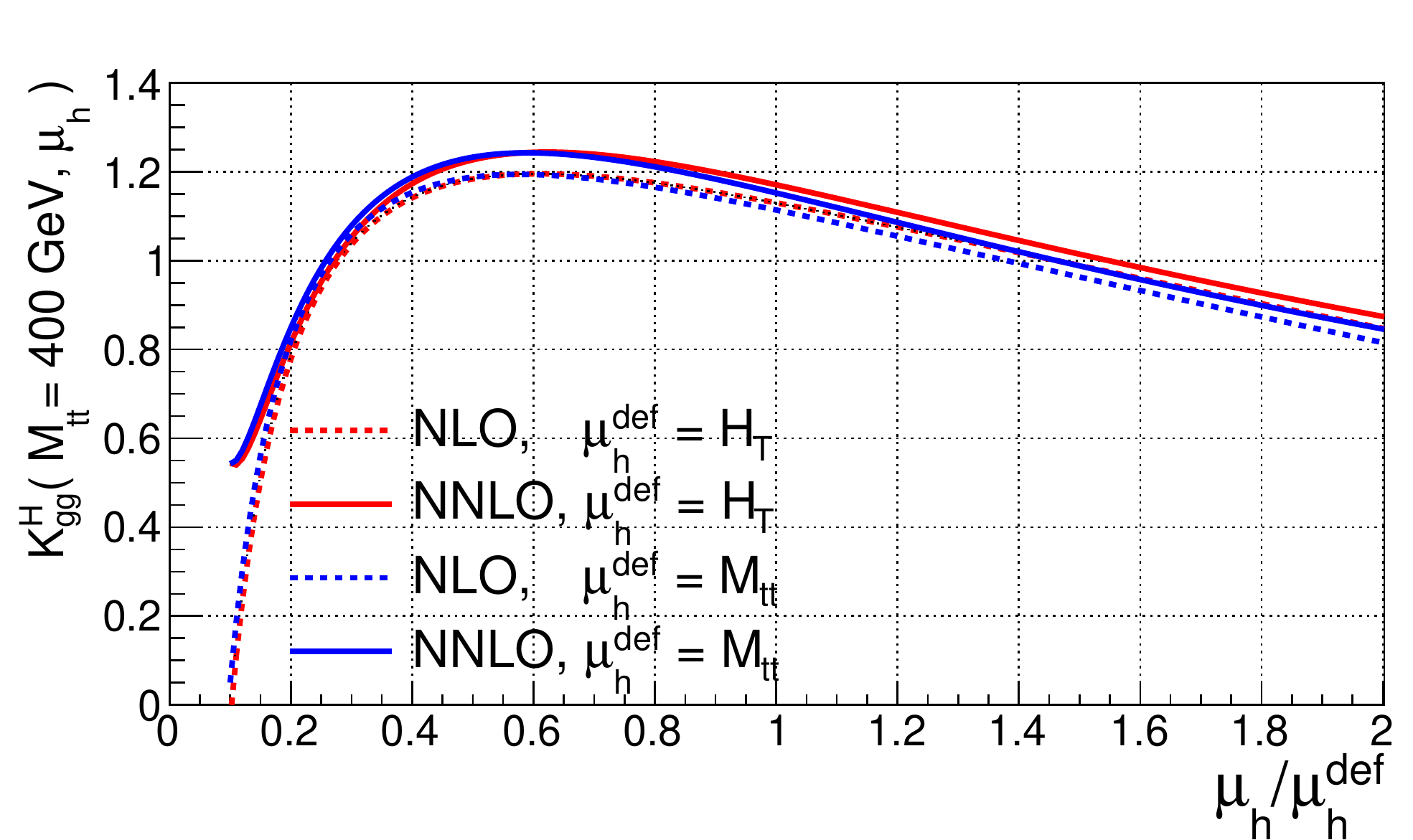}&
\includegraphics[width=0.495\textwidth]{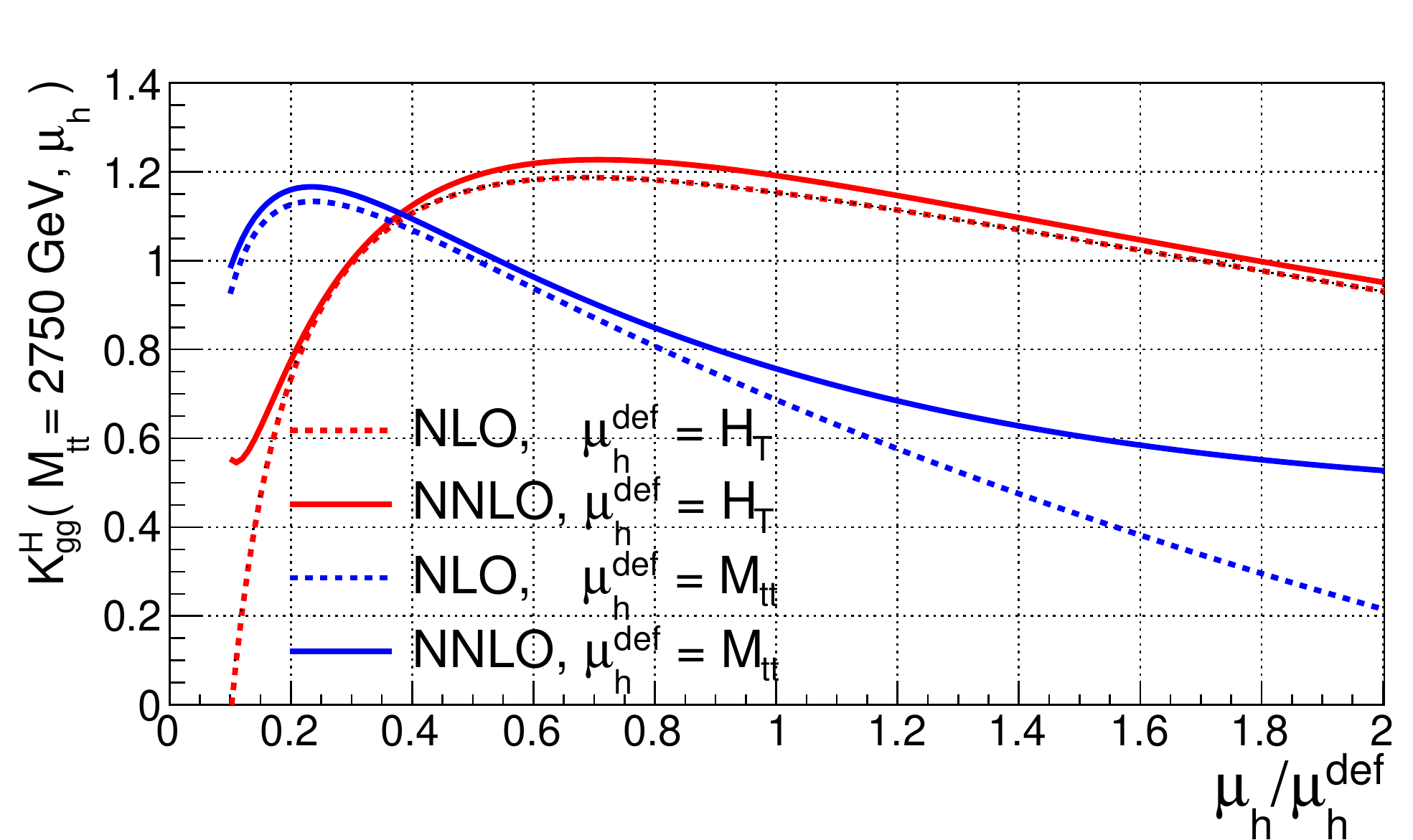}\nonumber \\
\includegraphics[width=0.495\textwidth]{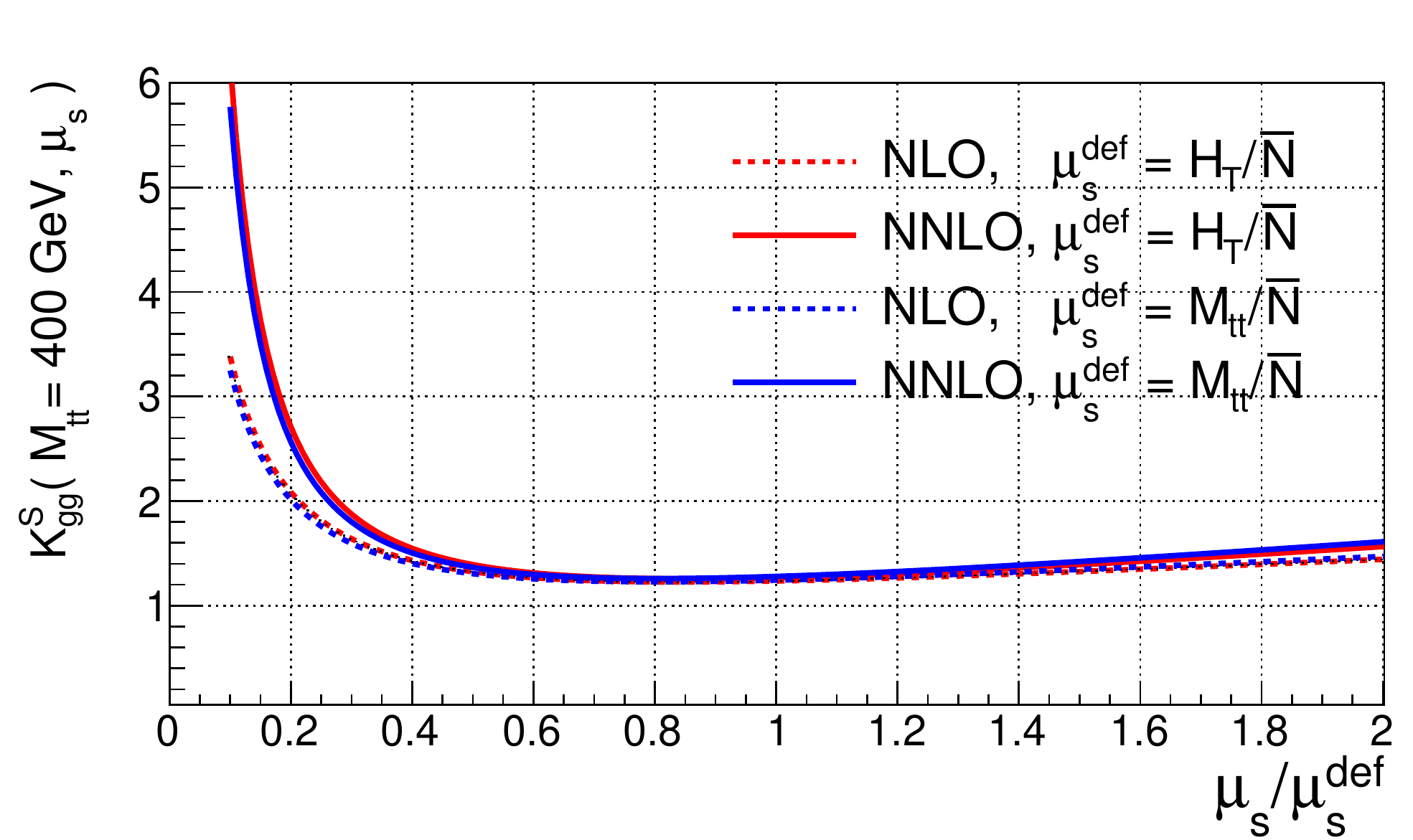}&
\includegraphics[width=0.495\textwidth]{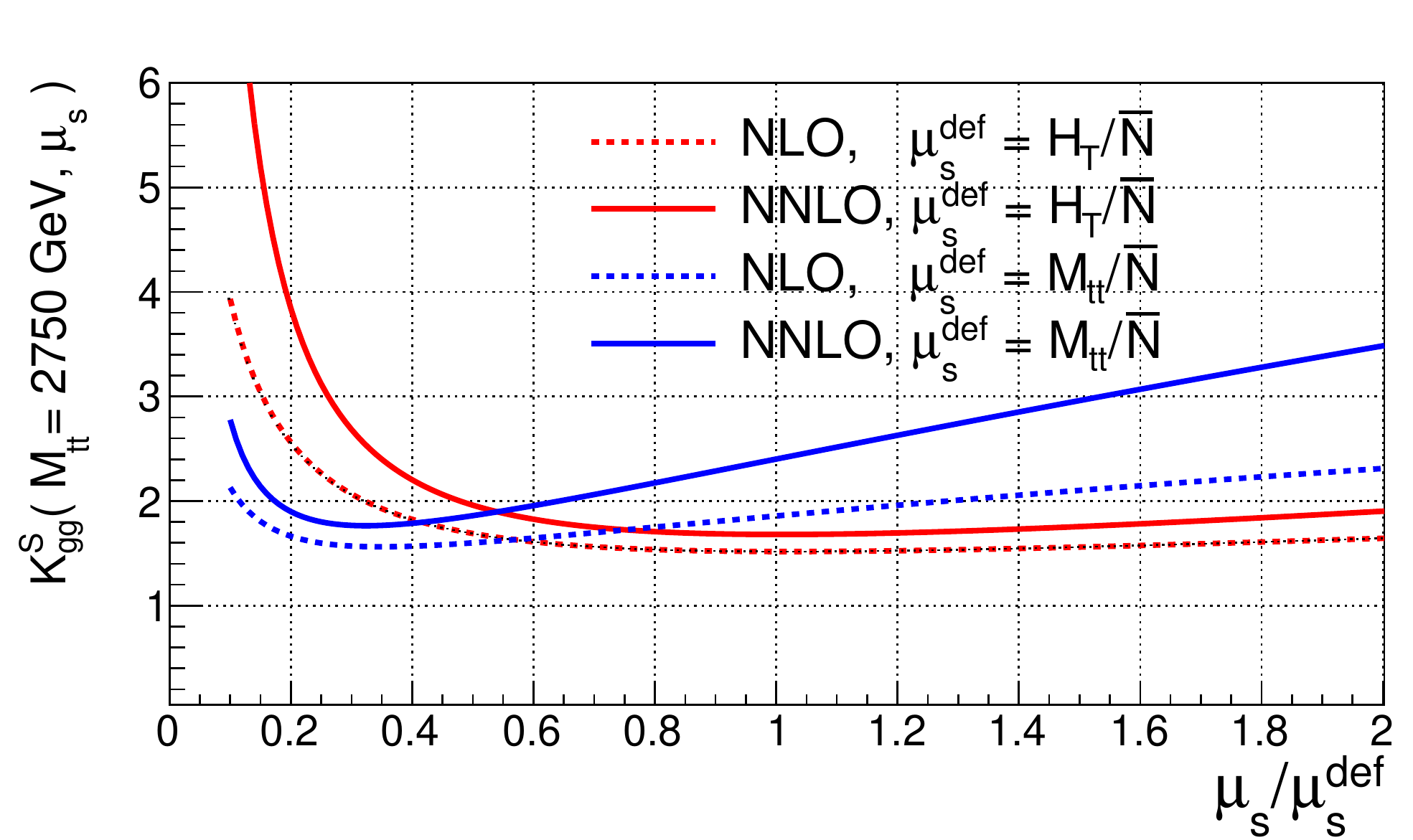}\nonumber
\end{align}
\caption{\label{fig:KHSmtt}
The $K$ factors for the corrections to the hard (top) and soft (bottom) functions in the $gg$ channel 
as defined  in the text, at $M_{t\bar{t}}=400$~GeV (left)
 and $M_{t\bar{t}}=2750$~GeV (right).}
\end{figure}

Numerical results for the hard and soft $K$ factors in the $gg$
channel are shown in figure~\ref{fig:KHSmtt}, for two values of
$M_{t\bar{t}}$.  The first is $M_{t\bar{t}}=400$~GeV (left-hand
plots), indicative of $m_t \sim M_{t\bar t}$, and the second is
$M_{t\bar{t}}=2750$~GeV (right-hand plots), indicative of the high
invariant-mass region where $m_t\ll M_{t\bar{t}}$.  Results for the
hard function, shown in the top two panels of the figure, are given for the
two different parametric choices $\mu_h = r H_T$ (red lines) and
$\mu_h = r M_{t\bar{t}}$ (blue lines), where $r$ is a numerical factor.
Results for the soft function, shown in the bottom two panels of the
figure, are given instead for $\mu_s = r H_T/\bar{N}$ (red lines) and
$\mu_s = r M_{t\bar{t}}/\bar{N}$ (blue lines). At
$M_{t\bar{t}}=400$~GeV there is little difference between the $K$
factors with the two different parametric choices, and both the hard
and the soft $K$ factors are moderate as long as the proportionality
factor $r$ is not too small.  At the higher value of $M_{t\bar{t}}$,
the corrections with the two different parametric scale choices differ
by quite a large amount. For the hard function, the corrections remain
moderate for $\mu_h \sim H_T/2$, as anticipated from the analysis
above.  For the soft function, the NLO corrections cannot be made
smaller than about 50\%.  This happens, for instance, at $\mu_s \sim
H_T/\bar{N}$.   The $K$ factor at this scale is flat with respect to changes of
$\mu_s$ around this value, at both the low and high values of \Mtt\, and
the NNLO corrections are also moderate. 

The above analysis leads us to identify $\mu_h=H_T/2$ and $\mu_s =
H_T/\bar{N}$ as a well-motivated choice of matching scales across the full
range of $M_{t\bar{t}}$ in the $gg$ channel.  We have checked that the
soft and hard $K$ factors in the $q\bar{q}$ channel (which gives a
considerably smaller contribution to the cross section at both large
and small $M_{t\bar{t}}$) are also well behaved for these choices.
Therefore, we will use these choices by default in all further
numerical analysis of the $M_{t\bar{t}}$ distribution.  
Of course, at the level of differential cross sections the dependence on these
scales cancels against that in the RG evolution factors, so that
resummed results are independent of the exact choice at the order at
which one is working.  The unphysical, residual dependence can be
reduced by calculating more orders in the logarithmic series, but with
a proper choice of matching scales these higher-order terms are
expected to be small corrections.

It is worth noting that \cite{Pecjak:2016nee}, as well as all other works on soft-gluon resummation 
with PIM kinematics, used $M_{t\bar{t}}$-based matching scales instead of $H_T$-based ones.
For moderate $M_{t\bar{t}}$ there is little difference, but at higher $M_{t\bar{t}}$ the perturbative
uncertainties estimated from $\mu_h$ and $\mu_s$ variations are larger with the 
$M_{t\bar{t}}$-based choice. Some numerical results with $M_{t\bar{t}}$-based
choices are given at the end in appendix~\ref{sec:comparisons}.

\subsection{Scales for the $p_T$ distribution}
 \label{sec:scalesPT}    

\begin{figure}[t!]
\centering
\begin{align}
\includegraphics[width=0.495\textwidth]{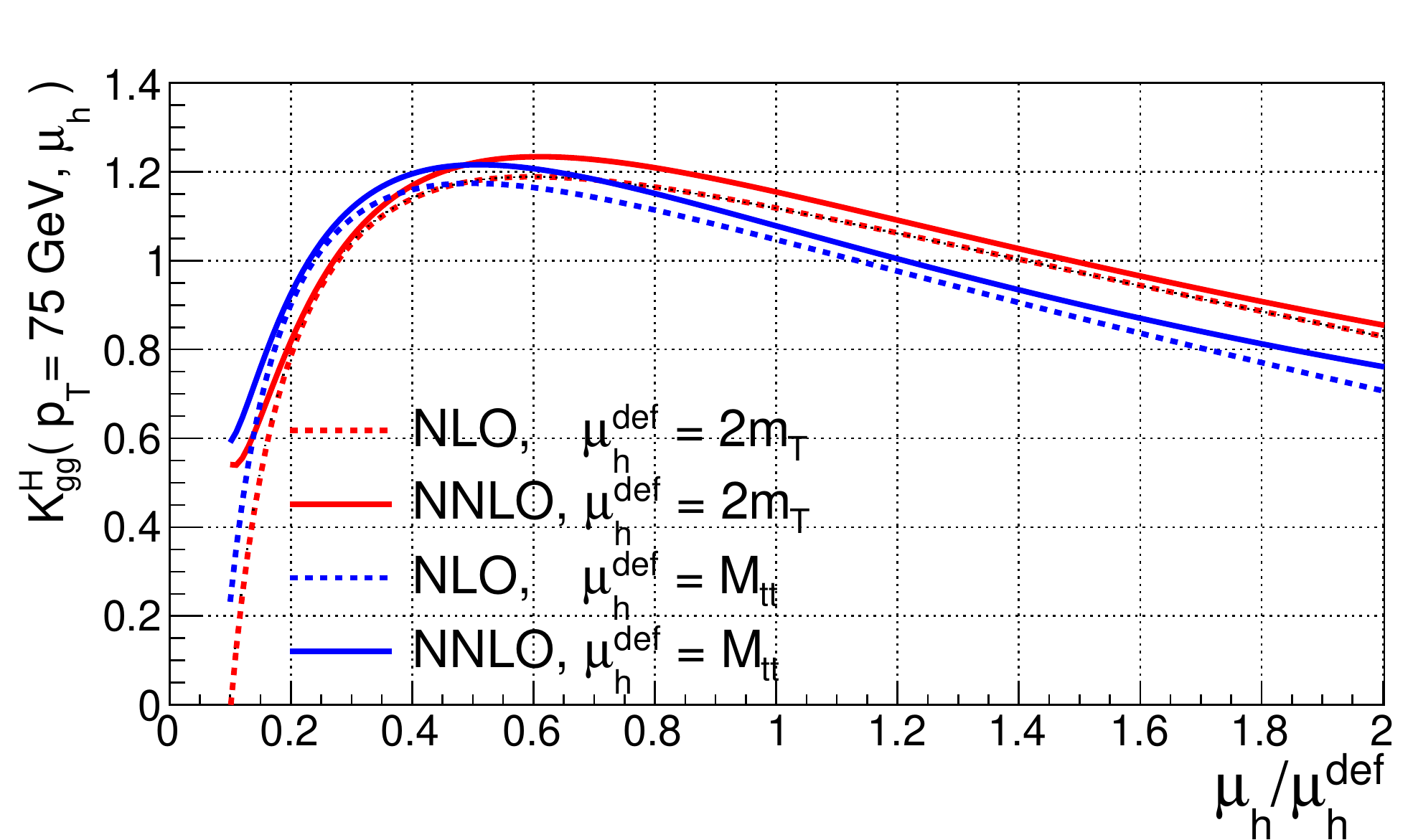}&           
\includegraphics[width=0.495\textwidth]{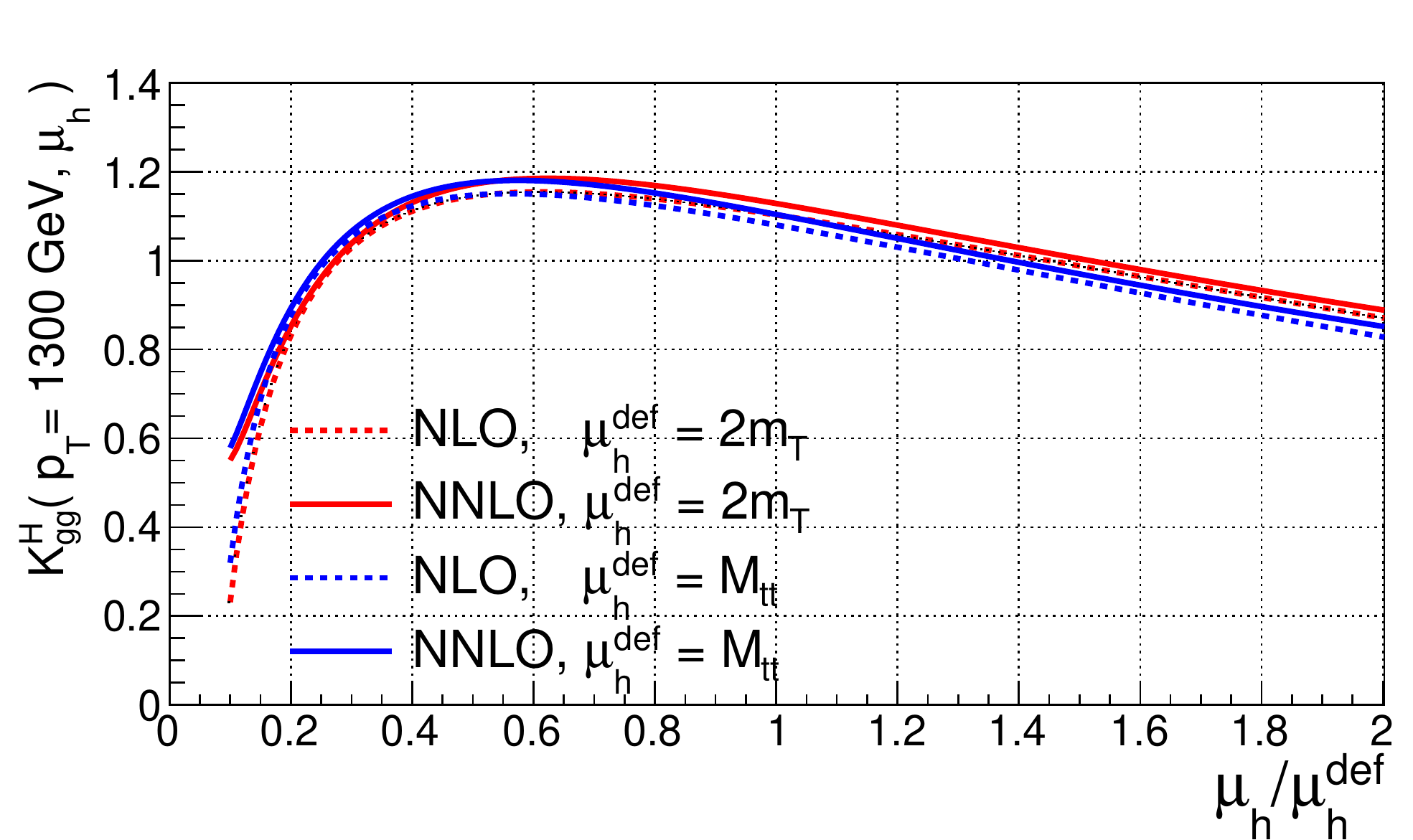}\nonumber \\
\includegraphics[width=0.495\textwidth]{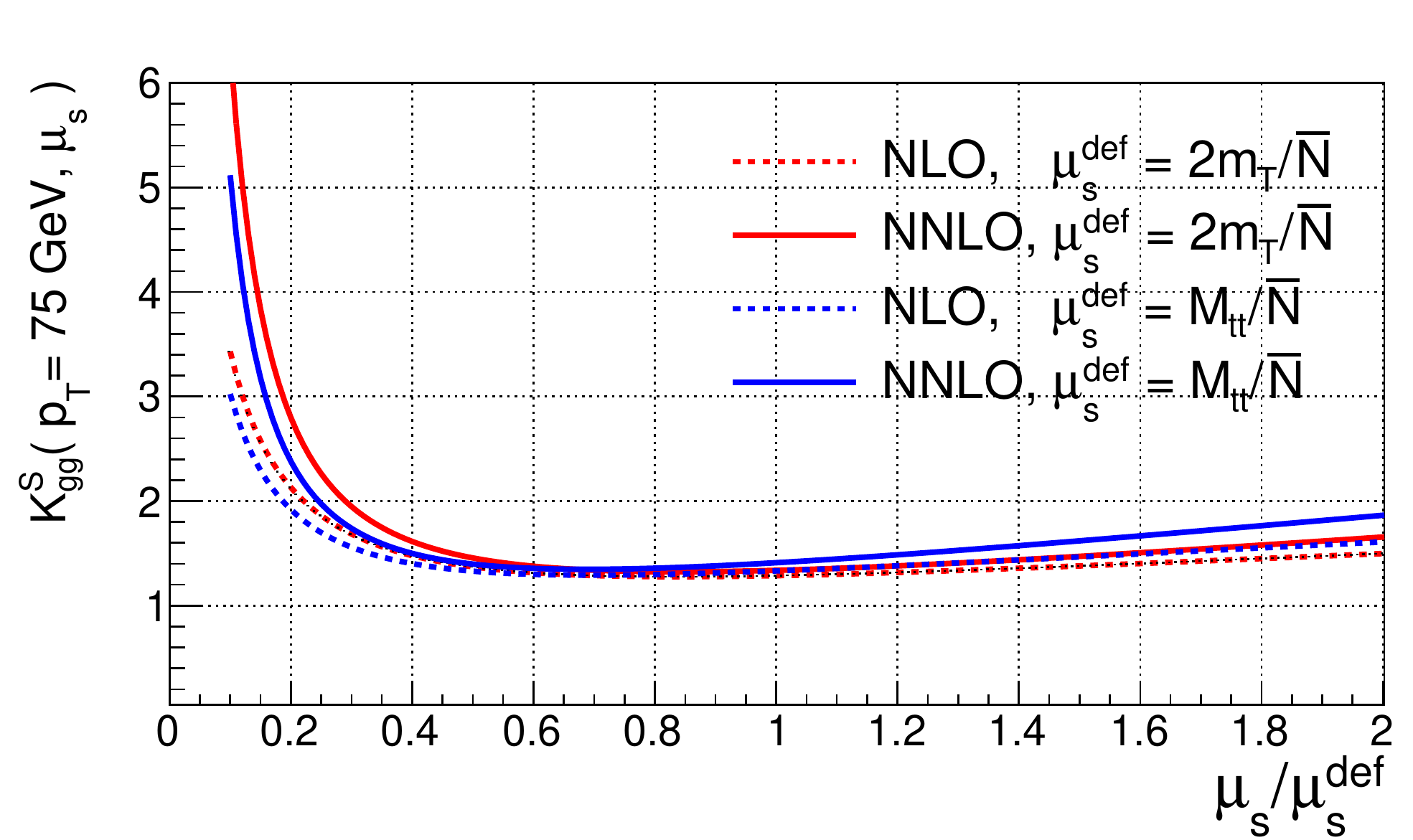}&
\includegraphics[width=0.495\textwidth]{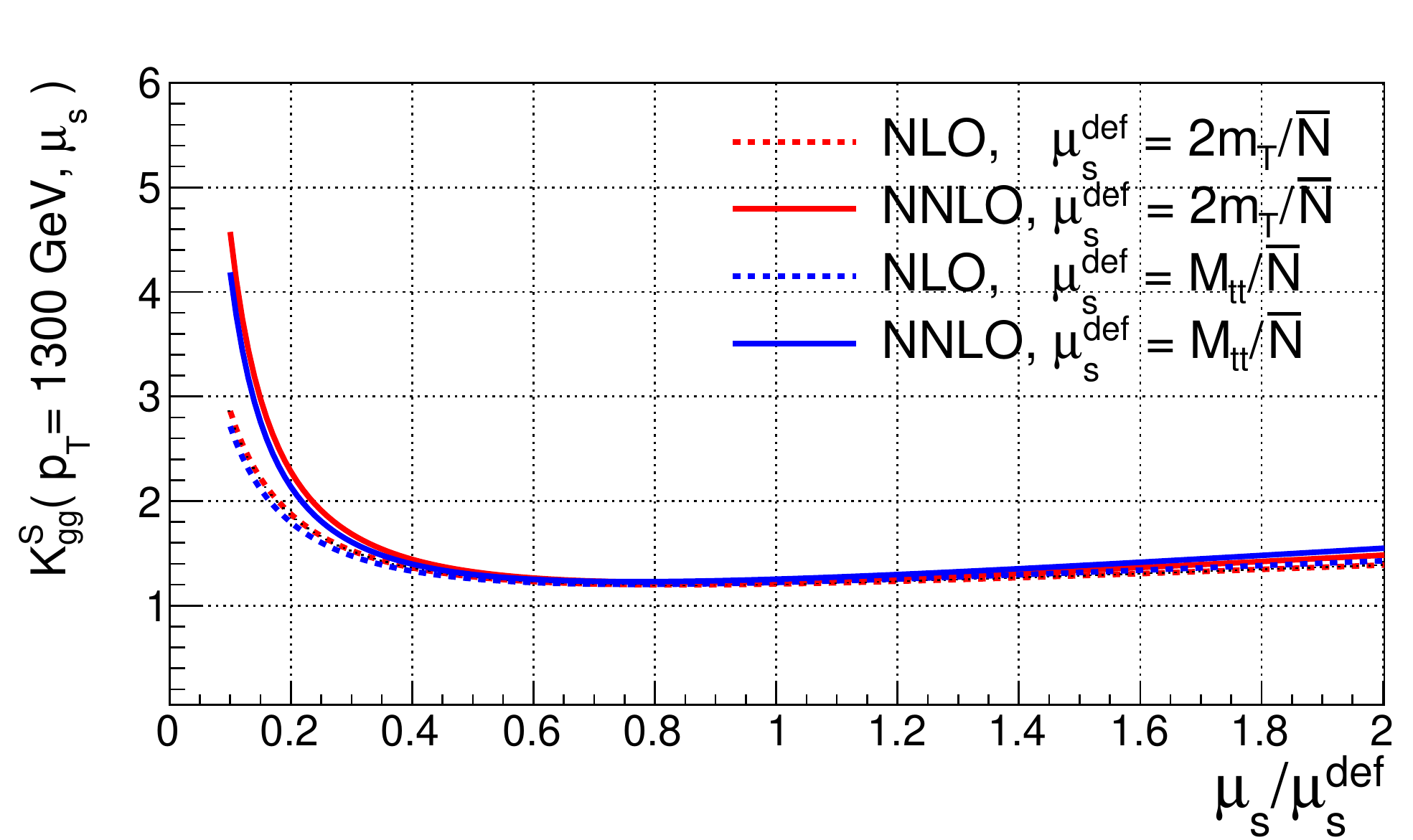}\nonumber
\end{align}
\caption{\label{fig:KHSdpt}
The $K$ factors for the corrections to the hard (top) and soft (bottom) functions in the $gg$ channel 
as defined  in the text, at $p_T=75$~GeV (left) and $p_T=1300$~GeV (right).}
\end{figure}

We now turn our attention to the top-quark $p_T$ distribution.  We
again motivate a suitable choice of $\mu_h$ and $\mu_s$ by studying
$K$ factors for the hard and soft functions analogous to those for the
$M_{t\bar{t}}$ distribution, but this time obtained by substituting
the hard-scattering kernel in eq.~(\ref{eq:x-sec-pt}) with  the appropriate
terms in eqs.~(\ref{eq:HardC}) and~(\ref{eq:SoftC}) before inverting the Mellin transform to
compute the cross section as a function of $p_T$.  We show the results
in figure~\ref{fig:KHSdpt} for a high and a low value of $p_T$,
examining both $M_{t\bar{t}}$ and $m_T$ based scales.  Note that while
$K^{H,(N)NLO}(M_{t\bar{t}},\mu_h)$ displayed explicitly in
eq.~(\ref{eq:HardK}) can be calculated without reference to the parton
luminosities, this is not the case for $K^{H,(N)NLO}(p_T, \mu_h)$. We used
$\mu_f = m_T$ in the luminosities in calculating both the hard and
soft $K$ factors, and have checked that varying $\mu_f$ to other
values produces only a small effect in the $K$ factor ratios.  

The soft and hard $K$ factors exhibit only a mild hierarchy between the
two types of scale choices in the low-$p_T$ region, which is even smaller
in the high-$p_T$ region.  This can be understood from  
figure~\ref{fig:HTovMdist}, which shows that in both $p_T$ regions 
the cross section sits mainly at $2m_T\sim M_{t\bar{t}}$ due to the 
Jacobian peak at $R_T=1$.  The $K$ factors are moderate when $\mu_h = m_T$ and $\mu_s =
2m_T/\bar{N}$, and we shall use these as the default choices in our
resummed calculations. Recall that in the soft limit the $p_T$ of the
top is equal to that of the anti-top, so that $m_T=H_T/2$ and these
are the exact same choices as for the $M_{t\bar{t}}$ distribution
after a trivial renaming.

\section{Results and discussion}
\label{sec:discussion}

In this section we give our main results for the top-pair invariant mass and 
(anti) top-quark $p_T$ distributions, as well as the total cross section, with 
a focus on comparing NNLO results with NNLO+NNLL$'$ ones.   
Some further comparisons across different perturbative orders and between standard and threshold resummed PDFs are presented in
appendices~\ref{sec:comparisons} and~\ref{sec:resPDFs} respectively.  Although we  present only a limited
set of results for the LHC operating at a center-of-mass energy of $\unit{13}{\TeV}$, distributions with alternate binning and at different collider energies can be produced on request from the authors.

\begin{figure}[t!]
\centering
\includegraphics[width=0.495\textwidth]{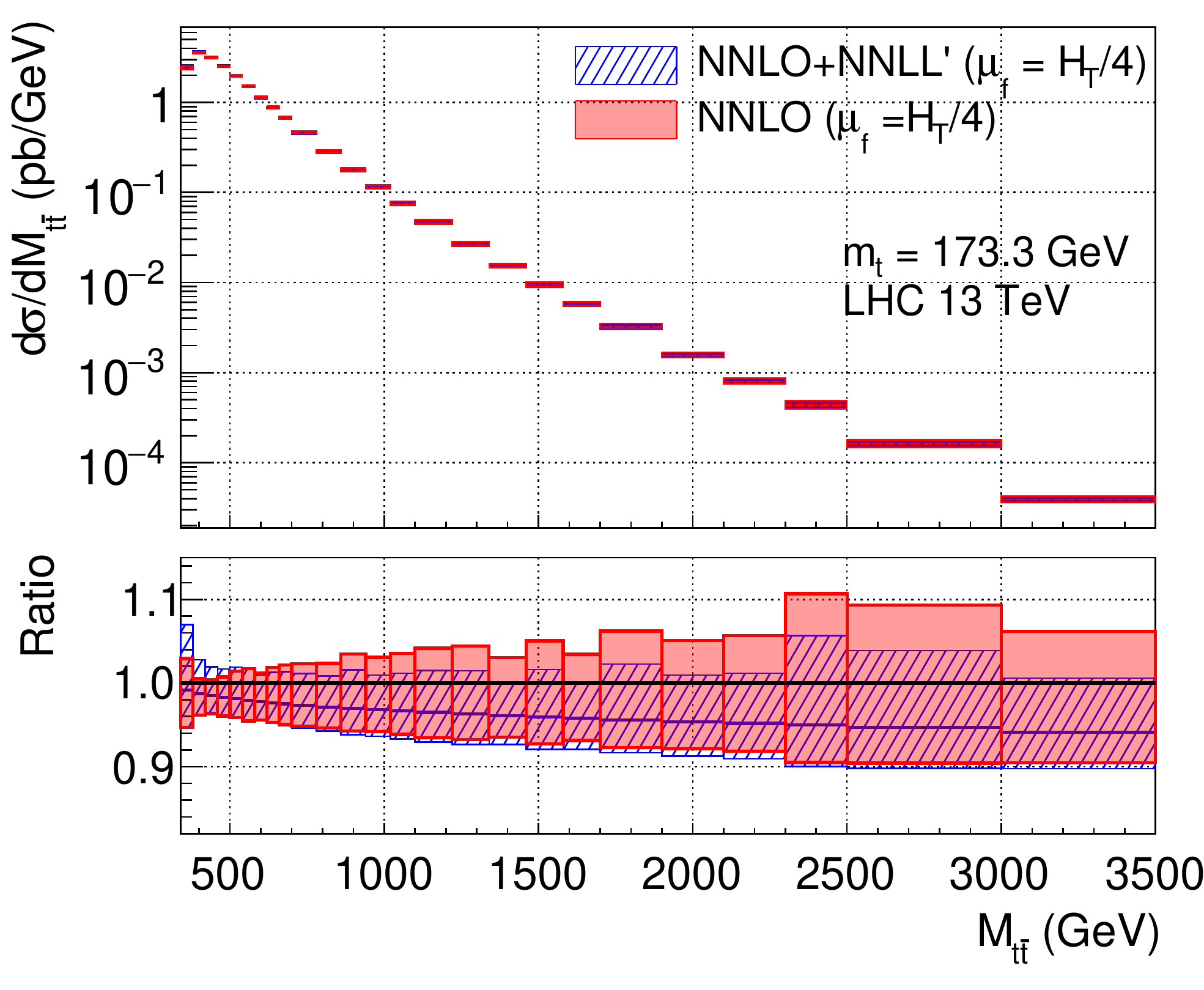}
\includegraphics[width=0.495\textwidth]{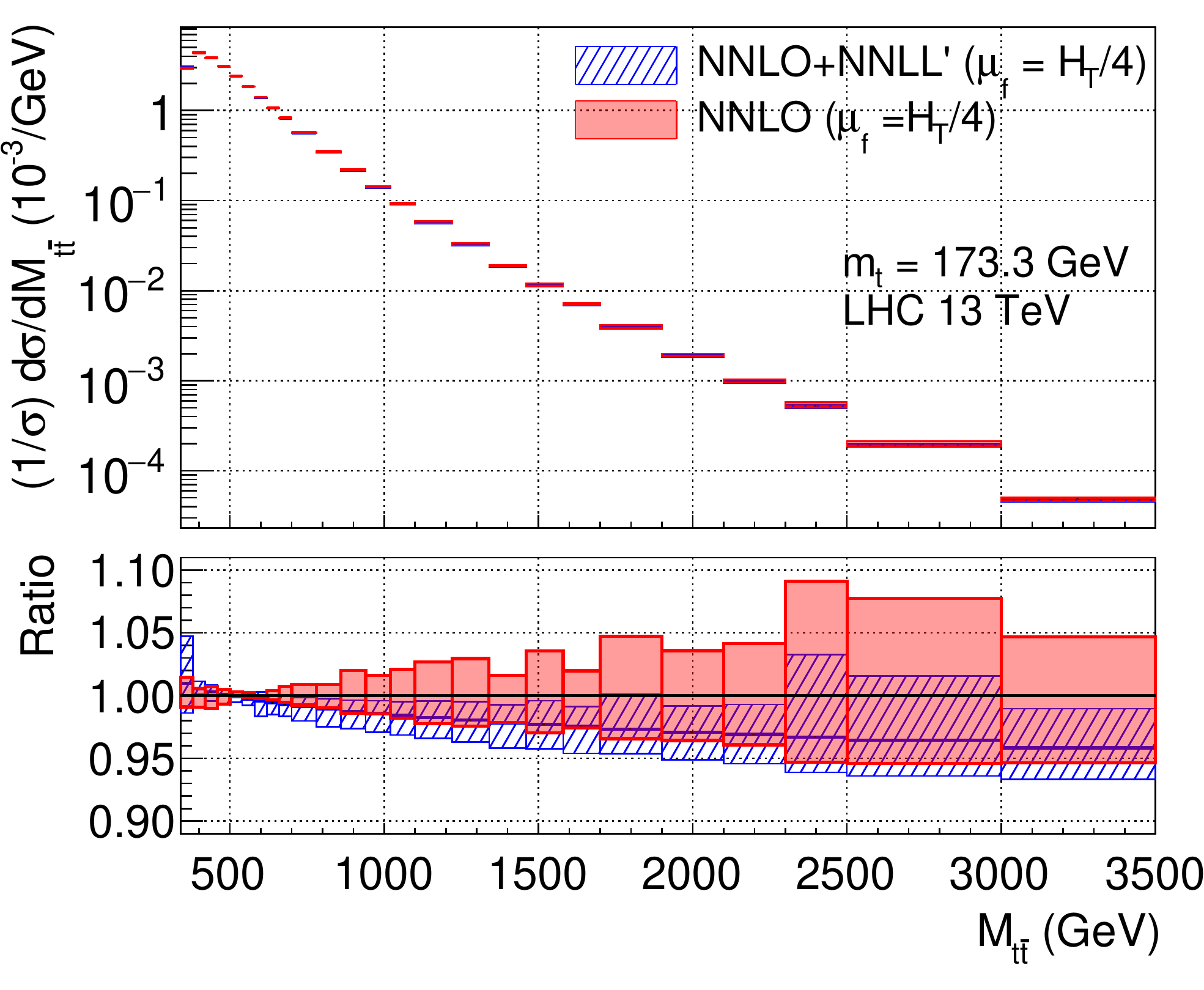}
\caption{\label{fig:mttDists_final} Results for the absolute (left) and normalized (right) 
top-pair invariant mass distribution at the LHC with $\sqrt{s}=13$~TeV.
In all cases the ratio is to the NNLO result with $\mu_f = H_T/4$. The uncertainty bands are obtained through scale 
variations as described at the beginning of section~\ref{sec:discussion} and in eqs.~(\ref{eq:unc_ref1}) and~(\ref{eq:unc_ref2}).}
\end{figure}

Results for the absolute (normalized) $M_{t\bar{t}}$ distribution are
shown in left (right) panel of figure~\ref{fig:mttDists_final}.  The
NNLO results use $\mu_f = H_T/4$ by default (we shall always set the
renormalization scale appearing in the NNLO calculation to $\mu_r
=\mu_f$ unless otherwise specified), which is the scale favored by the
analysis of perturbative convergence of the fixed-order series
performed in \cite{Czakon:2016dgf}. The NNLO+NNLL$'$ results are
obtained from the matching relation eq.~(\ref{eq:fully-matchedNNLO}).
All pieces of that equation must be evaluated at a common $\mu_f$,
which is also chosen as $\mu_f = H_T/4$ by default.  In addition, we
draw on the analysis of the previous section and use $\mu_h= H_T/2$
and $\mu_s = H_T/\bar{N}$ by default, as well as $\mu_{dh}=m_t$ and
$\mu_{ds}=m_t/\bar{N}$.  In both the NNLO and the NNLO+NNLL$'$
results, the bands in figure~\ref{fig:mttDists_final} represent
perturbative uncertainties estimated through scale variations.  For
the NNLO calculation, we obtain the bands by keeping the factorization
and renormalization scales equal and varying them up and down by a
factor of two.  For the NNLO+NNLL$'$ calculation, both the
factorization scales and the resummation scales are independently
varied in the interval $\left[\mu_{i,0}/2, 2\mu_{i,0}\right]$, where
$i \in \{f,h,s,dh,ds\}$ and the subscript ``$0$'' denotes the default
value of that scale as previously specified. To determine the upper
and lower uncertainties $\Delta O^+$ and $\Delta O^-$ for the cross
section $O$ in a given bin, one first evaluates
\begin{align}
\label{eq:unc_ref1}
\Delta O_i^+ = \text{max}\{O(\kappa_i=1/2,\kappa_i=1,\kappa_i=2)\} - \bar{O}  \, ,\nonumber \\
\Delta O_i^- = \text{min}\{O(\kappa_i=1/2,\kappa_i=1,\kappa_i=2)\} - \bar{O} \, ,
\end{align}
for each scale $i$, where $\kappa_i=\mu_{i}/\mu_{i,0}$ and $\bar{O}$
denotes the value of the cross section as given by
  eq.~(\ref{eq:fully-matchedNNLO}) in that bin using the default
scale choices. For example, $O(\kappa_f=2)$ means each term in
eq.~(\ref{eq:fully-matchedNNLO}) is evaluated at $\mu_f = 2
\mu_{f,0}$, with all other scales set to their default value.  
The upper (lower) uncertainty bands are then given by $\bar{O}+\Delta O^{+}$ ($\bar{O}-\Delta O^{-}$), 
where
\begin{equation}
\label{eq:unc_ref2}
\Delta O^{\pm} = \sqrt{\sum_i \left(\Delta O_i^\pm\right)^2} \, ,
\end{equation}
so that this method amounts to adding the uncertainties from
independent scale variations in quadrature.\footnote{While we have
  used correlated $\mu_r=\mu_f$ variations in the NNLO piece of the
  calculation, we have checked that the uncertainties estimated this
  way differ little from those obtained by varying $\mu_f$ and $\mu_r$
  with the 7-point method. The NNLO+NNLL$'$ calculation varies four
  resummation scales and also $\mu_f = \mu_r$ independently and adds
  the uncertainties in quadrature, so a direct numerical comparison
  with the 7-point method is not straightforward. However, we have
  experimented with a 7-point scan over $\mu_f$ and $\mu_h$, and found
  that the uncertainty estimates change only marginally compared to
  the quadrature method.}

\begin{figure}[t!]
\centering
\includegraphics[width=0.495\textwidth]{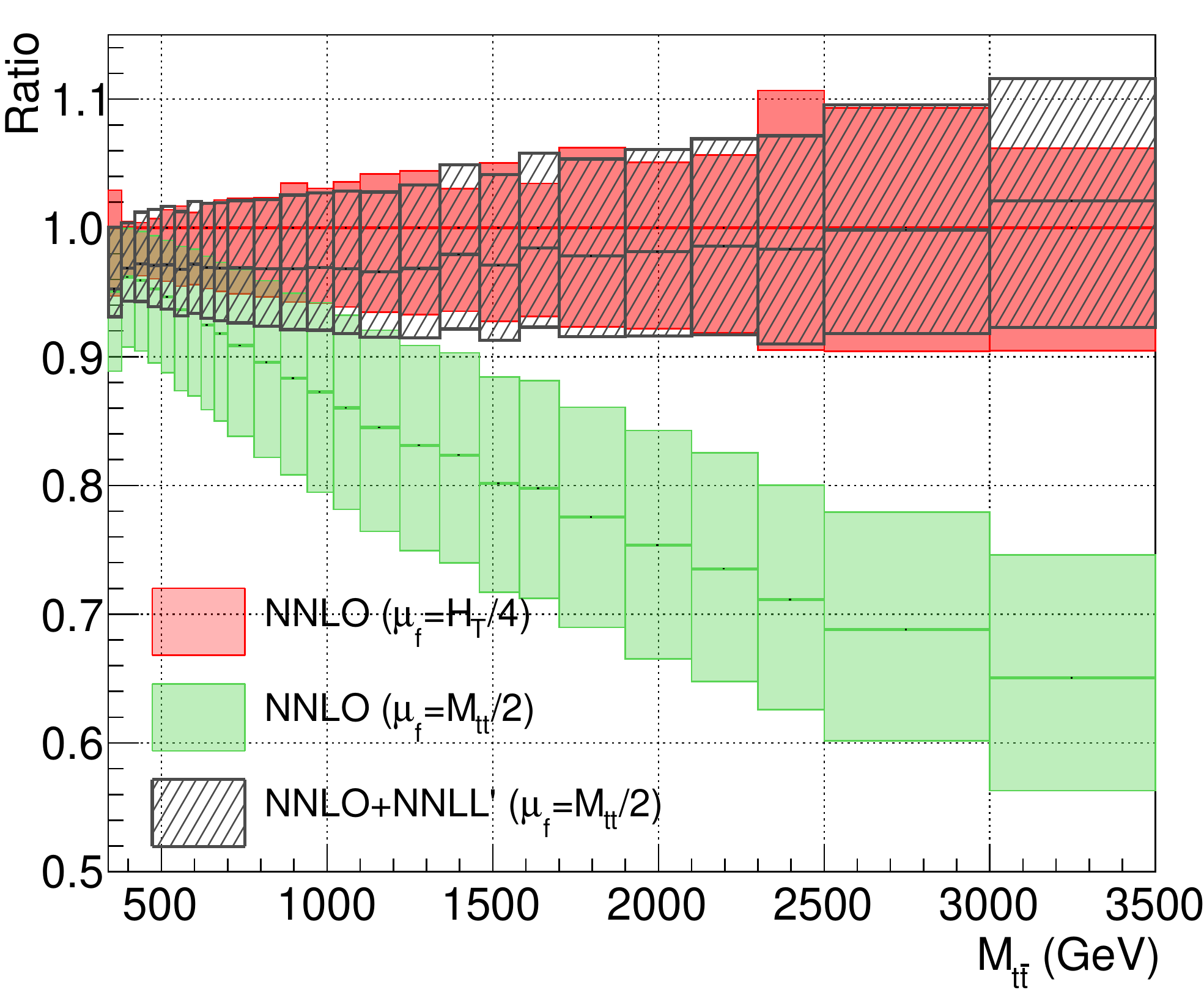}
\includegraphics[width=0.495\textwidth]{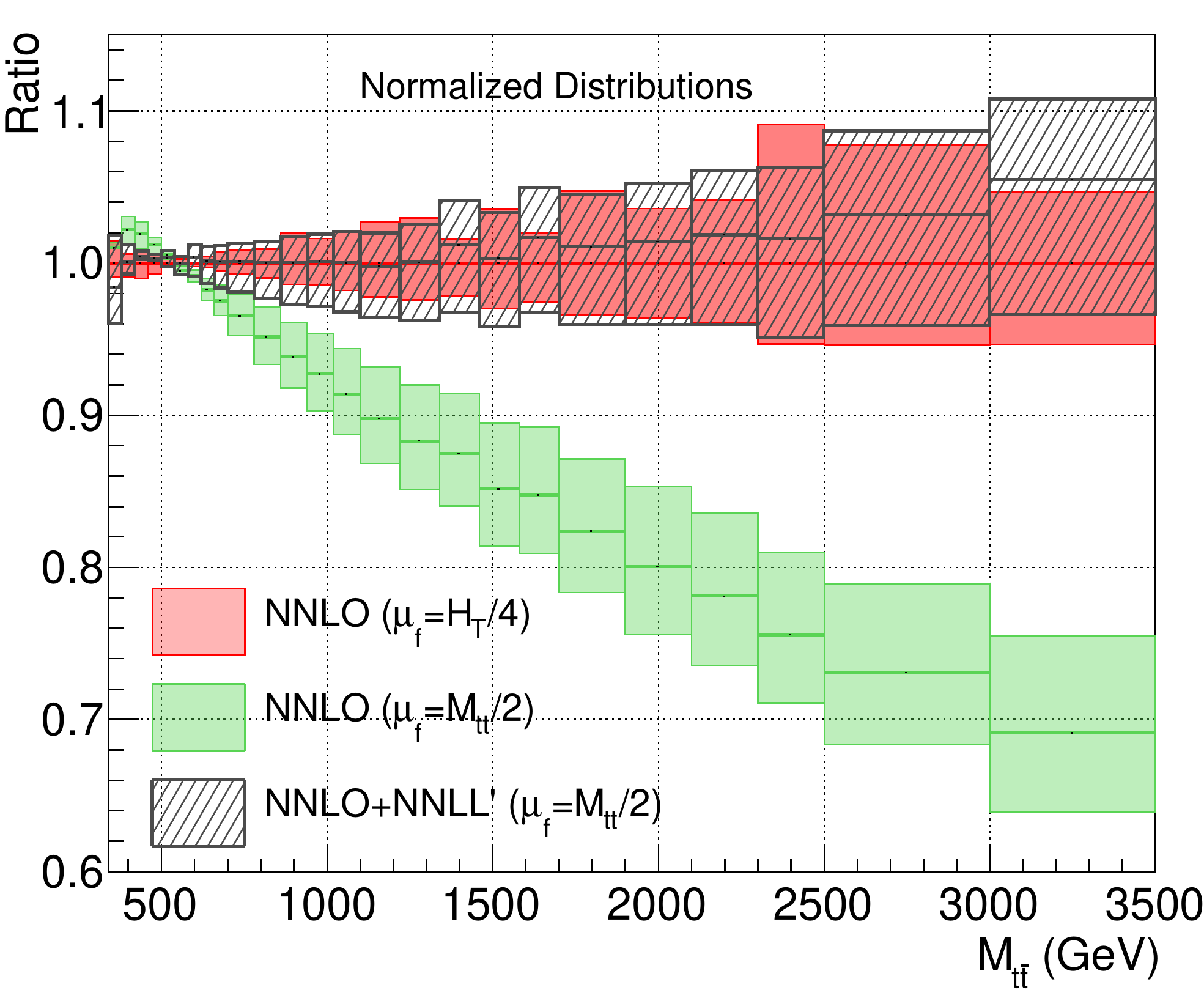}
\caption{\label{fig:mttDists_muf_compare} Results for the absolute (left) and normalized (right) 
top-pair invariant mass distribution at the LHC with $\sqrt{s}=13$~TeV as a ratio to the NNLO result evaluated using $\mu_f=H_T/4$. The uncertainty bands are obtained through scale 
variations as described at the beginning of section~\ref{sec:discussion} and in eqs.~(\ref{eq:unc_ref1}) and~(\ref{eq:unc_ref2}).}
\end{figure}

A remarkable feature of figure~\ref{fig:mttDists_final} is that the
NNLO+NNLL$'$ and NNLO results are in close agreement when $\mu_f =
H_T/4$ is chosen. To add context to this result, we compare in
figure~\ref{fig:mttDists_muf_compare} the ratio of the NNLO and
NNLO+NNLL$'$ results with $\mu_f = M_{t\bar{t}}/2$ to the NNLO result
with $\mu_f =H_T/4$, using the same set of matching scales and method
of estimating perturbative uncertainties as in
figure~\ref{fig:mttDists_final}. These two figures deliver a couple of
important messages. Firstly, the NNLO+NNLL$'$ result is rather stable
against switching the factorization scale between $H_T$-based and
$M_{t\bar{t}}$-based schemes. This implies that the even higher order
corrections to the NNLO+NNLL$'$ result are not so important. On the
other hand, the NNLO result changes drastically when switching the
schemes. In particular, higher order contributions beyond NNLO encoded
in the resummation produce a very large effect for the choice $\mu_f =
M_{t\bar{t}}/2$, as already forseen in \cite{Pecjak:2016nee}. Given
these observations, the close compatibility between the NNLO+NNLL$'$
result (with either scale choice) and the NNLO result with
$\mu_f=H_T/4$ is a highly non-trivial fact. This provides an important
confirmation of the result of \cite{Czakon:2016dgf}, which favors the
choice $\mu_f=H_T/4$ for the fixed-order calculation of the
$M_{t\bar{t}}$ distribution. The overall picture emerging from the
above analysis is that the perturbative description of the top-quark
pair invariant mass distribution is under good control.

\begin{figure}[t!]
\centering
\includegraphics[width=0.495\textwidth]{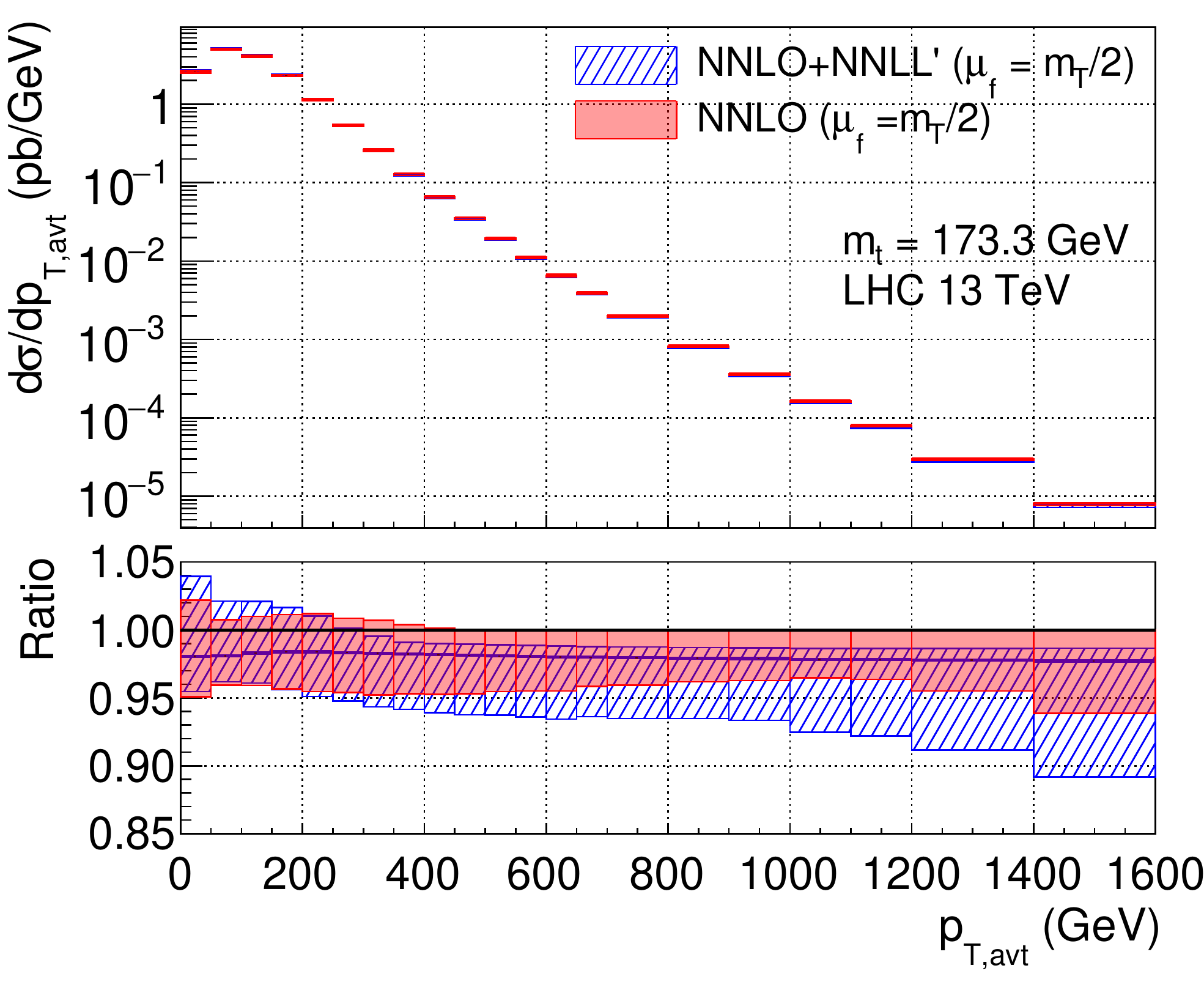}
\includegraphics[width=0.495\textwidth]{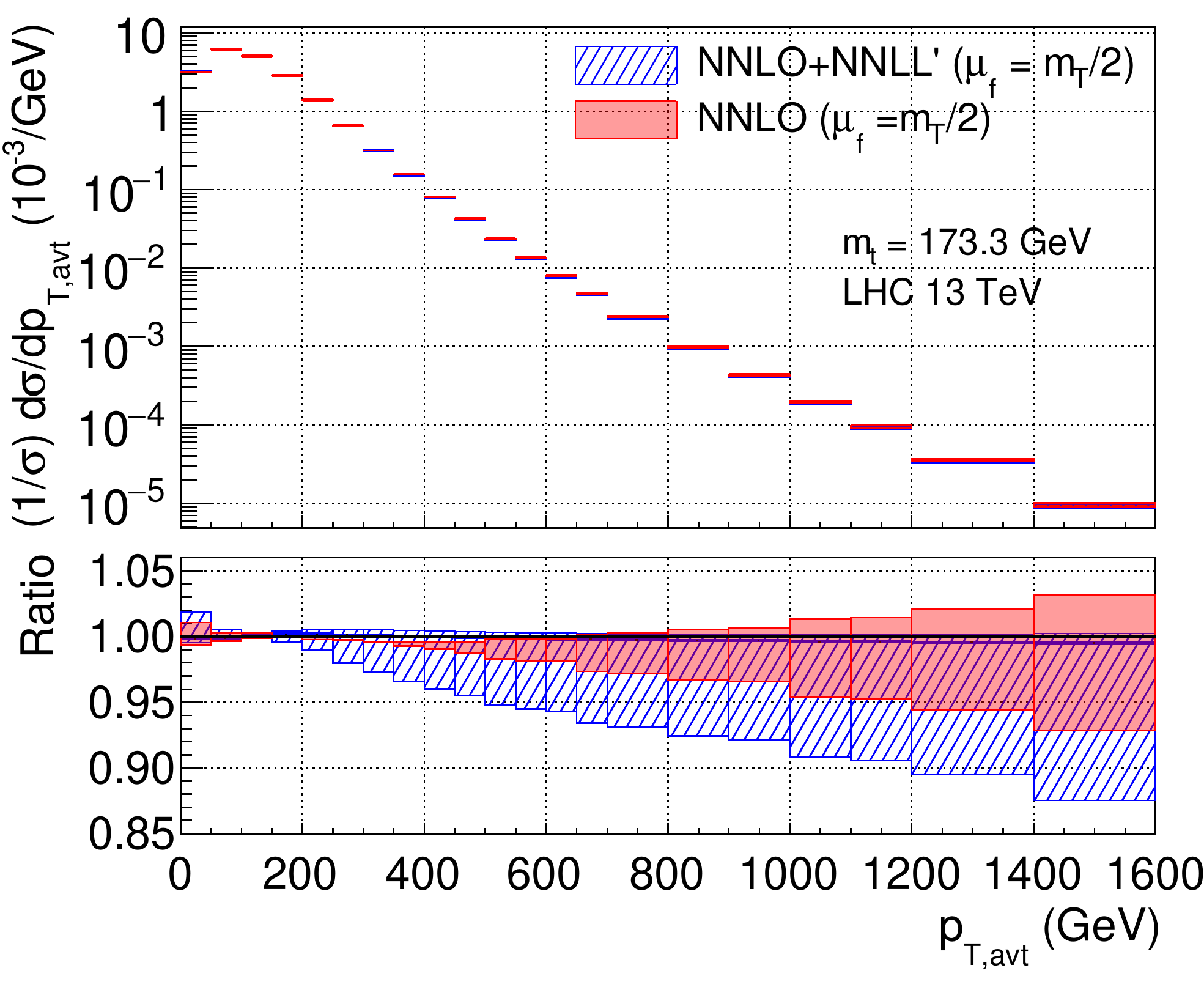}
\vspace{-1ex}
\caption{\label{fig:ptDists_final}Results for the absolute (left) and normalized (right) $p_{T,\text{avt}}$ distributions at the LHC with $\sqrt{s}=13$~TeV.
In all cases the ratio is to the NNLO result with $\mu_f = m_T/2$. Uncertainty bands are obtained in complete analogy to those in figure~\ref{fig:mttDists_final}.}
\end{figure}

Results for the absolute (normalized) average top/anti-top
($p_{T,\text{avt}}$) distribution at NNLO and NNLO+NNLL$'$ are shown
in the left (right) panel of figure~\ref{fig:ptDists_final}.  The NNLO
results (with which resummation is matched) have been calculated using
the definition
\begin{equation}
\label{eq:avt}
\frac{d \sigma}{d p_{T,\text{avt}}} = \frac{1}{2}\left(\frac{d \sigma}{d p_{T,t}} + \frac{d \sigma}{d p_{T,\bar{t}}}\right) \, ,
\end{equation}
where $p_{T,t}$ ($p_{T,\bar{t}}$) denotes the transverse momentum of
the top (anti-top) quark, and we have labeled the distributions in
figure~\ref{fig:ptDists_final} accordingly.  The $p_T$ distribution is
calculated using the scale choice $\mu_f = m_T/2$ (where $m_T$ refers
to the transverse mass of either the top or anti-top quark depending
on the distribution under consideration), which is favored by the
study \cite{Czakon:2016dgf}.  The resummed results use $\mu_h = m_T$
and $\mu_s = 2m_T/\bar{N}$ by default, as justified in the previous
section. The bands refer to perturbative uncertainties estimated
through scale variations using the same procedure as for the
$M_{t\bar{t}}$ distribution above. We see that the NNLO+NNLL$'$ result
is consistent with the NNLO one.  On the other hand, we show in
appendix~\ref{sec:comparisons} that upgrading matching with
fixed-order from NLO+NNLL$'$ to NNLO+NNLL$'$ is an important effect
for the $p_{T}$ distributions, especially in reducing the scale
uncertainties in the high $p_T$ region. This is an important fact to
keep in mind when using NLO-based Monte Carlo event generators to
model $p_T$ distributions.

\begin{figure}[t!]
\centering
\includegraphics[width=0.85\textwidth]{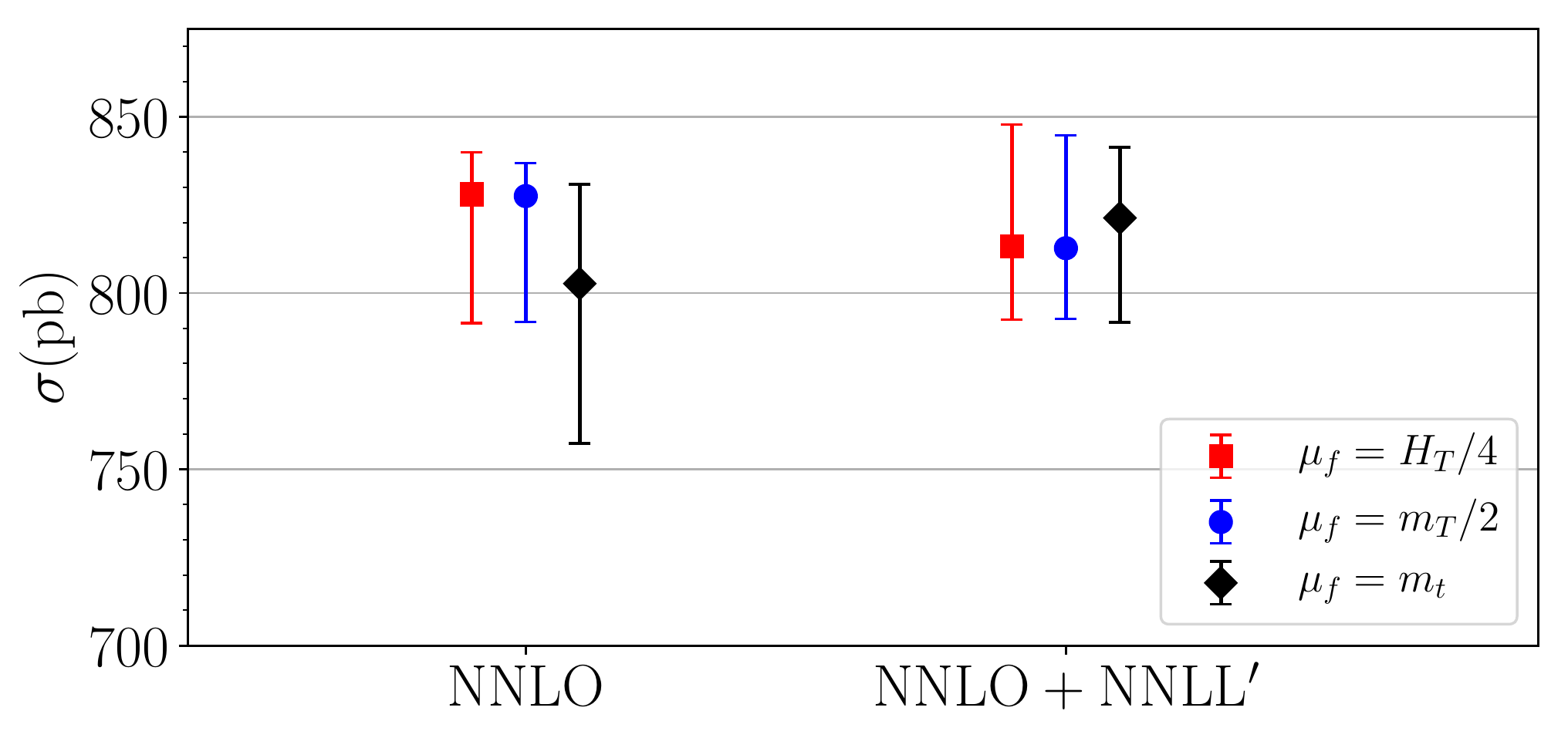}
\vspace{-1ex}
\caption{\label{fig:totXS13_final} Predictions for the total top-pair production cross section at the LHC with 
$\sqrt{s}=13$~TeV, where the error bars represent perturbative uncertainty estimates
through scale variations. The method for obtaining results and the uncertainty estimates 
at different values of $\mu_f$ is described in the second to last paragraph of section~\ref{sec:discussion}. } 
\end{figure}

Finally, in figure~\ref{fig:totXS13_final} we show results for the
total cross section, obtained in several different ways. The NNLO and
NNLO+NNLL$'$ results with $\mu_f = H_T/4$ are obtained by integrating
the top-pair invariant mass distribution in
figure~\ref{fig:mttDists_final}, while those with $\mu_f = m_T/2$ are
obtained by integrating the $p_T$ distribution in
figure~\ref{fig:ptDists_final}. In these results with dynamical
scales, perturbative uncertainties are estimated through the same
procedure of scale variations used for the distributions, and are
displayed as error bars in figure~\ref{fig:totXS13_final}. These are
compared to the ``standard'' results for the total cross section, which
are calculated using fixed scales with $\mu_f = \mu_r = m_t$ by
default. We obtain them from the {\tt Top++} program
\cite{Czakon:2011xx}, which implements both the NNLO results from
\cite{Czakon:2013goa} as well as a soft-gluon resummation in the
absolute threshold production limit $\beta_t\to 0$
\cite{Cacciari:2011hy}. In these fixed scale results, perturbative
uncertainties are estimated in both the NNLO and the NNLO+NNLL$'$
results by varying $\mu_f$ and $\mu_r$ up and down by a factor of two
using the seven-point method.  Evidently, while this resummation
result for the total cross section is also labelled NNLO+NNLL$'$ in
figure~\ref{fig:totXS13_final}, one should keep in mind that it uses a
different framework than the current work, including the treatment of
resummation scales and how they are varied as just described.

From Figure~\ref{fig:totXS13_final} we see that the integral of both
the NNLO+NNLL$'$ $M_{t\bar{t}}$ distribution with $\mu_f= H_T/4$ and
the NNLO+NNLL$'$ $p_T$ distribution with $\mu_f = m_T/2$ yield nearly
the same total cross section as the widely quoted result from the {\tt
  Top++} program. This shows that the results of this work not only
provide the most precise QCD results for the top pair invariant mass
distribution and the top quark transverse momentum distribution across
phase space, but also give the correct normalization for these
distributions.

\section{Conclusions}
\label{sec:conclusions}

In this paper we combined state-of-the-art results from soft-gluon
resummation (NNLL$'$) and fixed-order perturbation theory (NNLO) in
order to produce NNLO+NNLL$'$ predictions for the top-pair invariant
mass and the average top/anti-top quark $p_{T,\text{avt}}$ distributions at hadron colliders.  These
results represent the most complete QCD calculations of these
observables to date. They are also the first instance where an NNLO
calculation has been supplemented with resummation in a process where
the Born-level cross section contains four partons and thus has
non-trivial matrix structure in color space.

The resummation formalism used here contains several elements which
have not appeared in the literature so far.  Some of these involve the
details of weaving together three different kinds of calculations to
obtain results optimized throughout phase space. In this procedure, it
is crucial to avoid counting the same contribution more than once.  In
particular, in section~\ref{sec:prelim} we presented a matching
procedure which allows us to combine NNLO results in fixed order
\cite{Czakon:2016dgf}, NNLL$_b'$ results in a joint resummation of
overlapping soft and collinear logarithms \cite{Ferroglia:2012ku}, and
NNLL$_m$ results in pure soft-gluon resummation \cite{Ahrens:2010zv},
in order to achieve what we have called NNLO+NNLL$'$ accuracy.  All
ingredients required to implement this matching procedure, as well as
the resummation itself, carried out in Mellin space, were given in
section~\ref{sec:MellinRes} and appendix~\ref{sec:gis}.

Our analysis in section~\ref{sec:systematic} revealed some important
kinematic features of the $M_{t\bar{t}}$ and $p_T$
distributions. These can not only be understood analytically using
results from soft-gluon resummation, but also affect its
implementation, and are useful to keep in mind when interpreting the
numerical results for the $M_{t\bar{t}}$ and $p_T$ distributions given
in section~\ref{sec:discussion}.  In the case of the $M_{t\bar{t}}$
distribution, the analysis of section~\ref{sec:systematic}
demonstrates that in the boosted regime where $m_t \ll M_{t\bar{t}}$,
the most relevant hard scale is not $M_{t\bar{t}}$ itself but rather
$H_T$. This fact is related to the dynamical enhancement of the
forward and backward scattering regions due to the $t$- and
$u$-channel diagrams appearing in the Born level partonic cross
section in the $gg$ initiated production process. We used this feature
to identify a well-motivated set of matching scales for the
kinematics-dependent hard and soft functions appearing in the
resummation formalism, and also to argue that the high $M_{t\bar{t}}$
region is particularly amenable to soft gluon resummation.  On the
other hand, we observed that the high-energy region of the $p_{T}$
distribution is rather sensitive to hard emissions, so that matching
resummation with NNLO is an essential improvement.  This is certainly
the case for the analytic resummation performed here, but should also
be kept in mind when predicting the high-energy tail of the $p_{T}$
distribution with NLO-based event generators.

In section~\ref{sec:discussion} we presented numerical results for the
absolute and normalized $M_{t\bar{t}}$ and $p_{T,\text{avt}}$
distributions, as well as the total cross section, valid to
NNLO+NNLL$'$.  For the $p_{T,\text{avt}}$ distribution the resummation
effects are mild, especially at the scale $\mu_f = m_T/2$ favored by
the NNLO analysis of perturbative convergence in
\cite{Czakon:2016dgf}.  For the $M_{t\bar{t}}$ distribution, an
interesting outcome of our analysis is the stability of the
NNLO+NNLL$'$ results under parametric changes in $\mu_f$, as shown in
figures~\ref{fig:mttDists_final} and~\ref{fig:mttDists_muf_compare}.
Given the large shifts in the NNLO calculation under such $\mu_f$
changes, it is an important result that the NNLO+NNLL$'$ results
stabilize the differential cross section close to the NNLO prediction
with $\mu_f = H_T/4$, which is the setting favored by
\cite{Czakon:2016dgf} and currently being used in all NNLO
phenomenology. The consistency between the NNLO and NNLO+NNLL$'$
results gives us confidence that even higher-order corrections are
under good control.

\section*{Acknowledgements}
A.M.\ thanks the Department of Physics at Princeton University for hospitality during the completion of this work. This work was supported in part by the National Natural Science Foundation of China under Grant No.\ 11575004 and 11635001. D.J.S.\ has received support from an STFC Postgraduate Studentship and is supported under the ERC grant ERC-STG-2015-677323. The work of M.C.\ was supported in part by a grant of the BMBF.
The work of D.H.\ and A.M.\ is supported by the UK STFC grants ST/L002760/1 and ST/K004883/1 and by the European Research Council Consolidator Grant NNLOforLHC2. The work of A.F.\ is supported in part by the National Science Foundation under Grant No.\ PHY-1417354.

\appendix

\section{Perturbative stability across orders}
\label{sec:comparisons}

\begin{figure}[t!]
\centering
\includegraphics[width=0.83\textwidth]{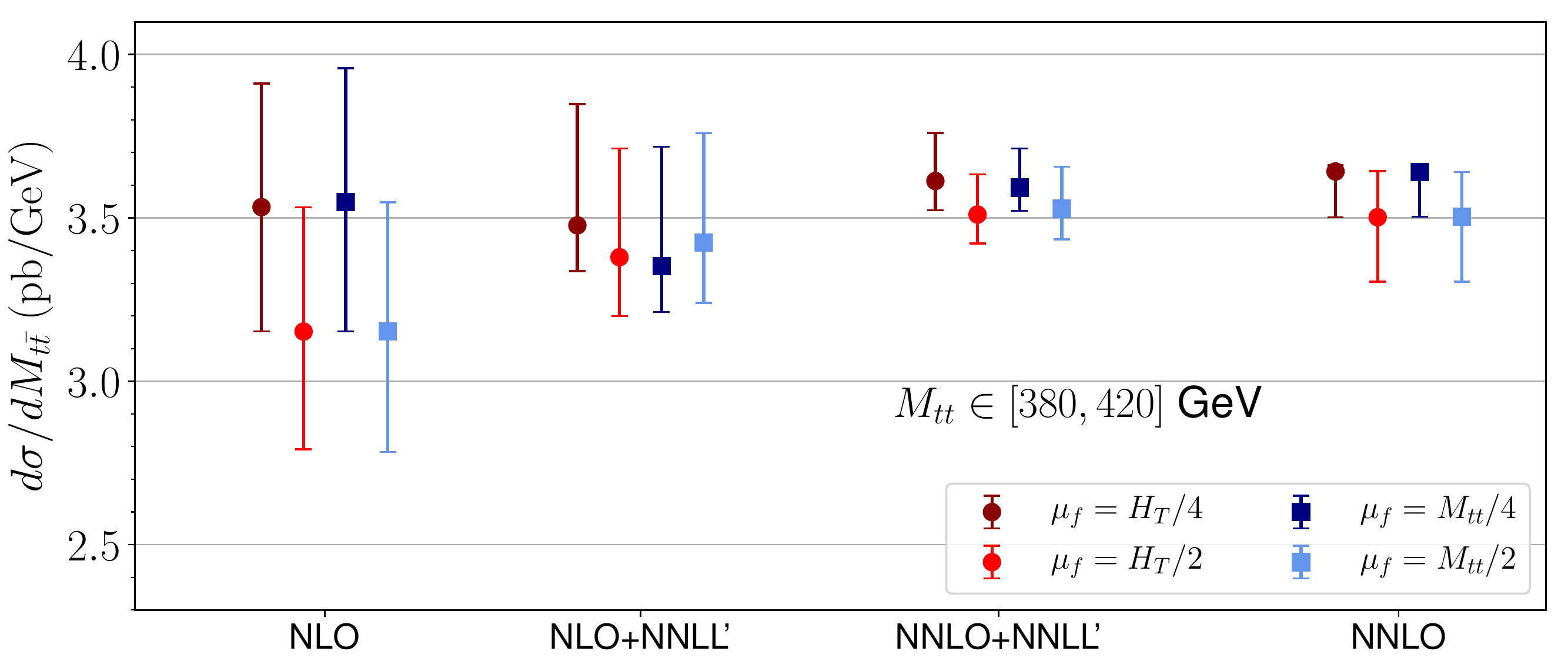}\\
\includegraphics[width=0.83\textwidth]{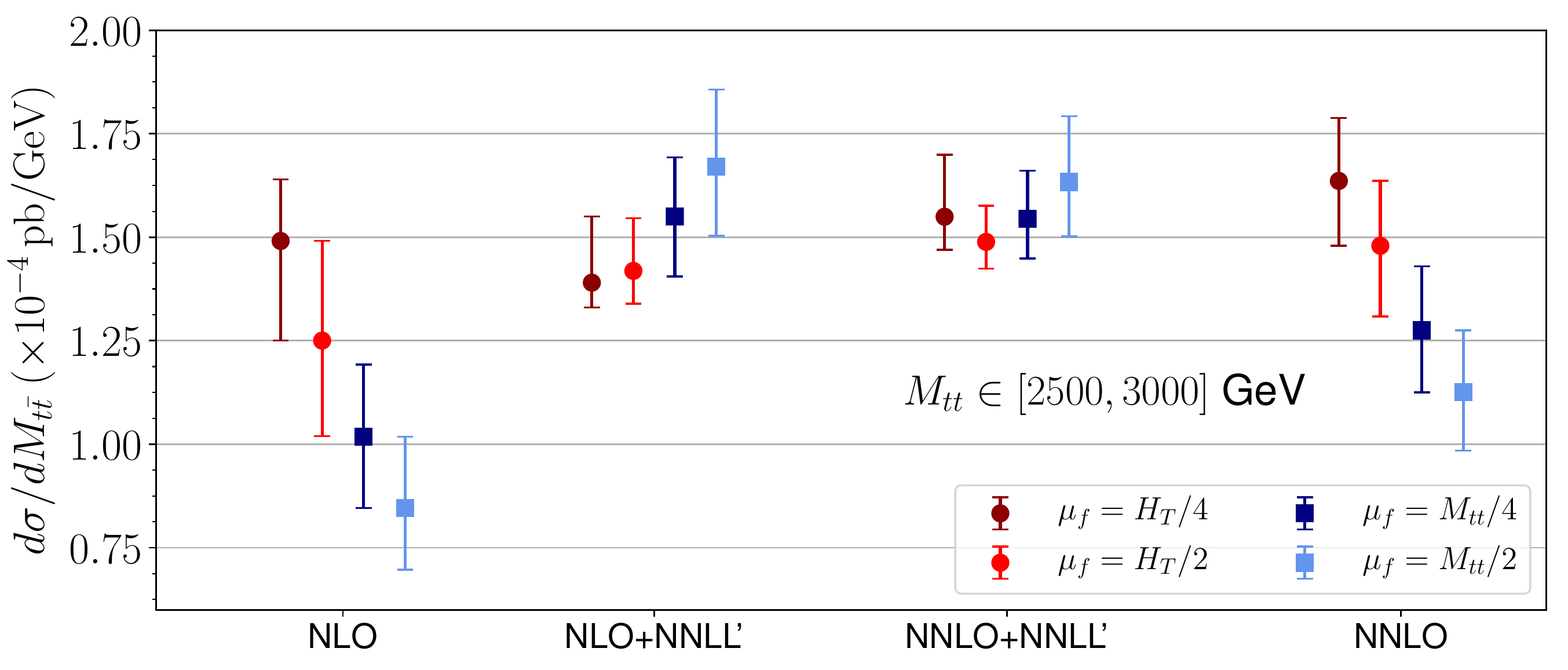}
\caption{\label{fig:mtt_muf_stability} Cross sections obtained for two sample bins, $M_{t\bar{t}}=[380-420] \, \GeV$ (upper plot) and $M_{t\bar{t}}=[2500-3000] \, \GeV$ (lower plot). The default value of $\mu_f$ is indicated explicitly, and the 
error bars represent perturbative uncertainties estimated through scale variations as described at the beginning of section~\ref{sec:discussion} and in eqs.~(\ref{eq:unc_ref1}) and ~(\ref{eq:unc_ref2}).}
\end{figure}

In this section we perform some comparisons of results for the
top-pair invariant mass and average top/anti-top-quark $p_T$ distribution across
different perturbative orders.

Figure~\ref{fig:mtt_muf_stability} displays results for the cross
section at (N)NLO and (N)NLO+NNLL$'$ in the same low-energy and
high-energy bins of $M_{t\bar{t}}$ considered in
section~\ref{sec:KinFeatures}.  The NLO results are generated using
NLO PDFs, while all other results are generated using NNLO PDFs (including
the NLO+NNLL$'$ ones).  The resummed results use the default matching
scales from section~\ref{sec:discussion}. The figure compares results
obtained with the default $\mu_f$ indicated explicitly in the figure legend, with
perturbative uncertainties estimated through scale variations and
displayed as error bars.  In both bins, adding resummation to the
fixed-order result is a clear improvement: the (N)NLO+NNLL$'$ results
are considerably more stable against $\mu_f$ variations than the
(N)NLO ones, especially in the high $M_{t\bar{t}}$ bin.  An
important message to be drawn from the figure is that the NNLO+NNLL$'$
results at different $\mu_f$ congregate near the NNLO one with $\mu_f
= H_T/4$.

\begin{figure}[t!]
\centering
\includegraphics[width=0.83\textwidth]{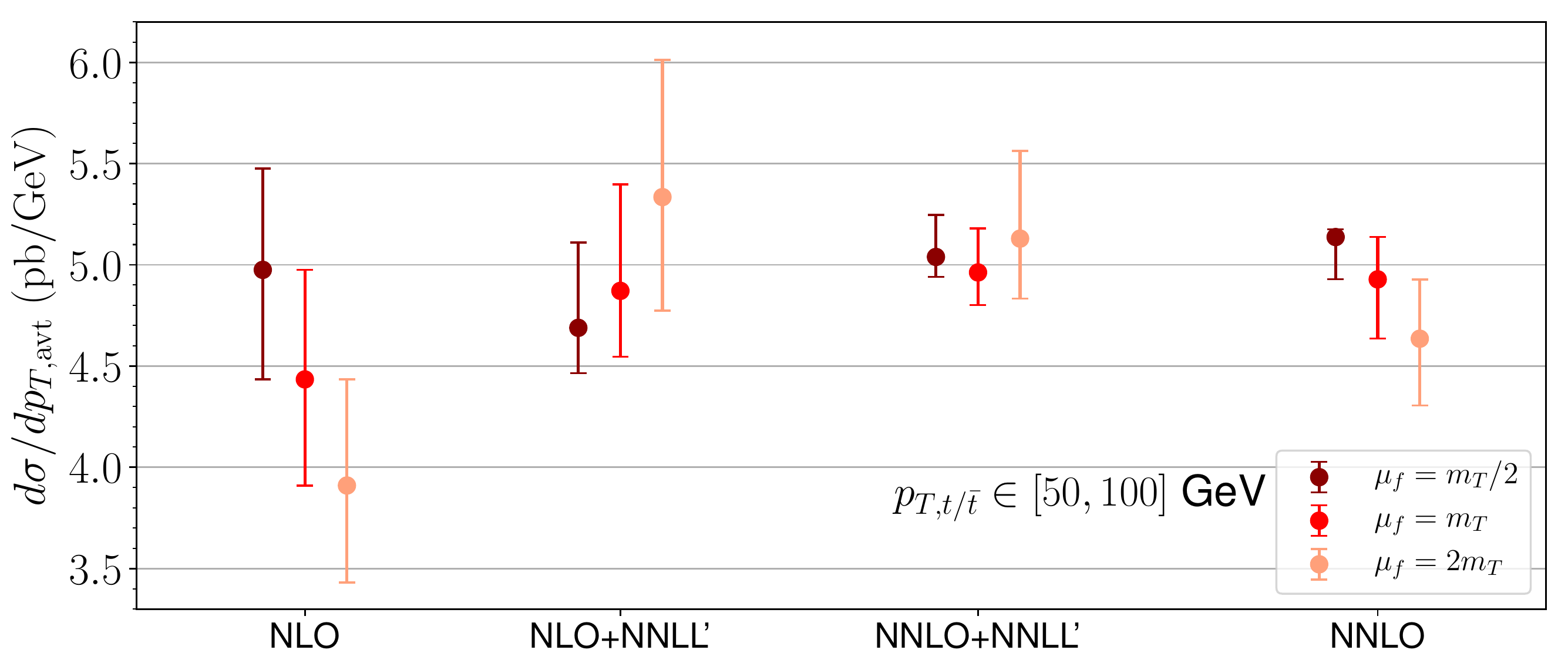}\\
\includegraphics[width=0.83\textwidth]{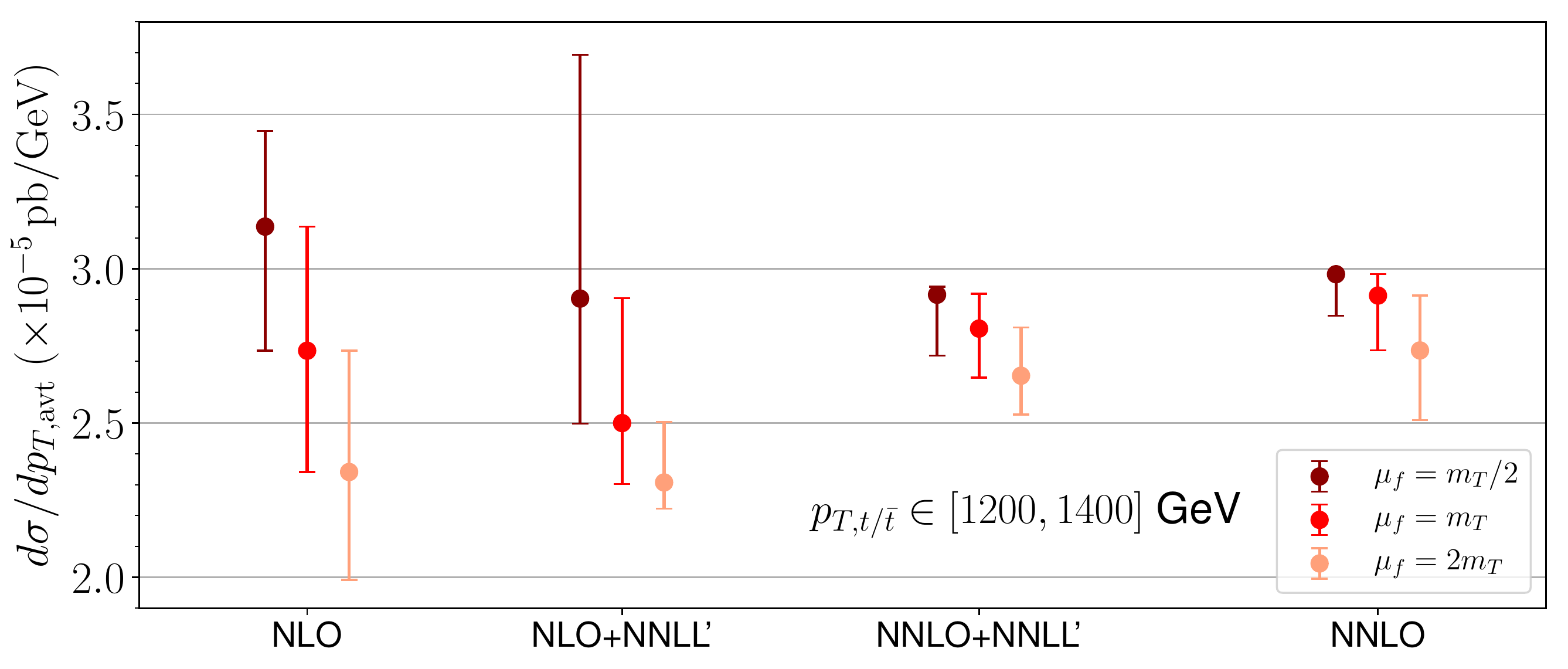}
\caption{\label{fig:pt_muiStability} Same as figure~\ref{fig:mtt_muf_stability} but for the $p_{T,\text{avt}}$ distribution.}
\end{figure}

Figure~\ref{fig:pt_muiStability} shows results for the $p_{T,\text{avt}}$ 
distribution at different perturbative orders in the two sample bins considered in
section~\ref{sec:KinFeatures}.  The resummed results use
the same default matching scales as in section~\ref{sec:discussion} and are 
completely analogous to those in figure~\ref{fig:mtt_muf_stability}.
Compared to the $M_{t\bar{t}}$ distribution, where the parametric difference between the $H_T$ and
$M_{t\bar{t}}$-based $\mu_f$ is large, there is no such hierarchy of scales to consider for
the $p_T$ distribution.  Therefore, we have shown results for three
different $m_T$-based choices, ranging from $\mu_f = m_T/2$ to $\mu_f
= 2m_T$ by default.  While the resummation is of some benefit in
stabilizing the (N)NLO results in the low-$p_T$ bin, the picture is
less clear in the high-$p_T$ bin. For instance, there is a dramatic
reduction in $\mu_f$ dependence in the NNLO+NNLL$'$ results compared
to the NLO+NNLL$'$ ones.  This is an indication that the high-$p_T$
region is more sensitive to hard radiation than the
high-$M_{t\bar{t}}$ region.  We have given some qualitative explanations for why
this should be the case when discussing the $R_T$ distribution in
section~\ref{sec:KinFeatures}.  Numerically, we have found that the
NLO results for the high-$p_T$ region of the distribution are quite
sensitive to both the $qg$ channel and the $R_T>1$ region, in a strongly 
$\mu_f$-dependent fashion. Soft-gluon resummation cannot stabilize such $\mu_f$ dependence,
which explains the importance of matching to NNLO in fixed order.
\begin{table}[t!]
\centering
\begin{tabular}{|c|c|r|}
\hline
                                & Bin    (GeV)          &\multicolumn{1}{c|}{ NNLO+NNLL$'$ (pb)}                                                                                 \\ \hline
\multirow{2}{*}{$M_{t\bar{t}}$} & $[380,420]$   & $3.60$ {\footnotesize $^{+4\%}_{-2\%}$}{\footnotesize $^{+1\%}_{-1\%}$}                \\ \cline{2-3} 
                                & $[2500,3000]$ & $(1.55 \times 10^{-4})$ {\footnotesize $^{+9\%}_{-4\%}$}{\footnotesize $^{+3\%}_{-4\%}$} \\ \hline
\multirow{2}{*}{$p_{T,\text{avt}}$}          & $[50,100]$    & $5.04$ {\footnotesize $^{+4\%}_{-2\%}$}{\footnotesize $^{+1\%}_{-1\%}$}              \\ \cline{2-3} 
                                & $[1200,1400]$ & $(2.92\times 10^{-5})$ {\footnotesize $^{+0\%}_{-7\%}$}{\footnotesize $^{+1\%}_{-1\%}$}  \\ \hline
\end{tabular}
\caption{\label{tab:errSource}Cross section in ``high'' and ``low'' energy bins. The first uncertainties refer to those from $\mu_f$ variation only while the second set are generated by the variation of the matching scales.}
\end{table}

The uncertainties associated with each of the distributions presented here result from a combination of the uncertainties generated from the variation of each scale in accordance with eqs.~(\ref{eq:unc_ref1}) and~(\ref{eq:unc_ref2}). It is interesting to decompose the source of these uncertainty bands in terms of the contributions which arise from varying the factorization scale $\mu_f$ compared with the other matching scales. We present such a sample decomposition in table~\ref{tab:errSource} for the two ``low'' and ``high'' energy bins used throughout section~\ref{sec:systematic}. We show the cross section in each bin with two sets of uncertainties, the first refers to those obtained through variations of $\mu_f$ alone while the second refers to the combined uncertainty generated by varying each of the matching scales. In most instances, the dominant contribution to the uncertainty arises from the variation of $\mu_f$. For the $M_{t\bar{t}}$ distribution at high $M_{t\bar{t}}$ the uncertainty from the matching scales is larger than at low $M_{t\bar{t}}$ while across the $p_T$ spectrum this source of uncertainty remains constant.

\begin{figure}[t!]
\centering
\includegraphics[width=0.495\textwidth]{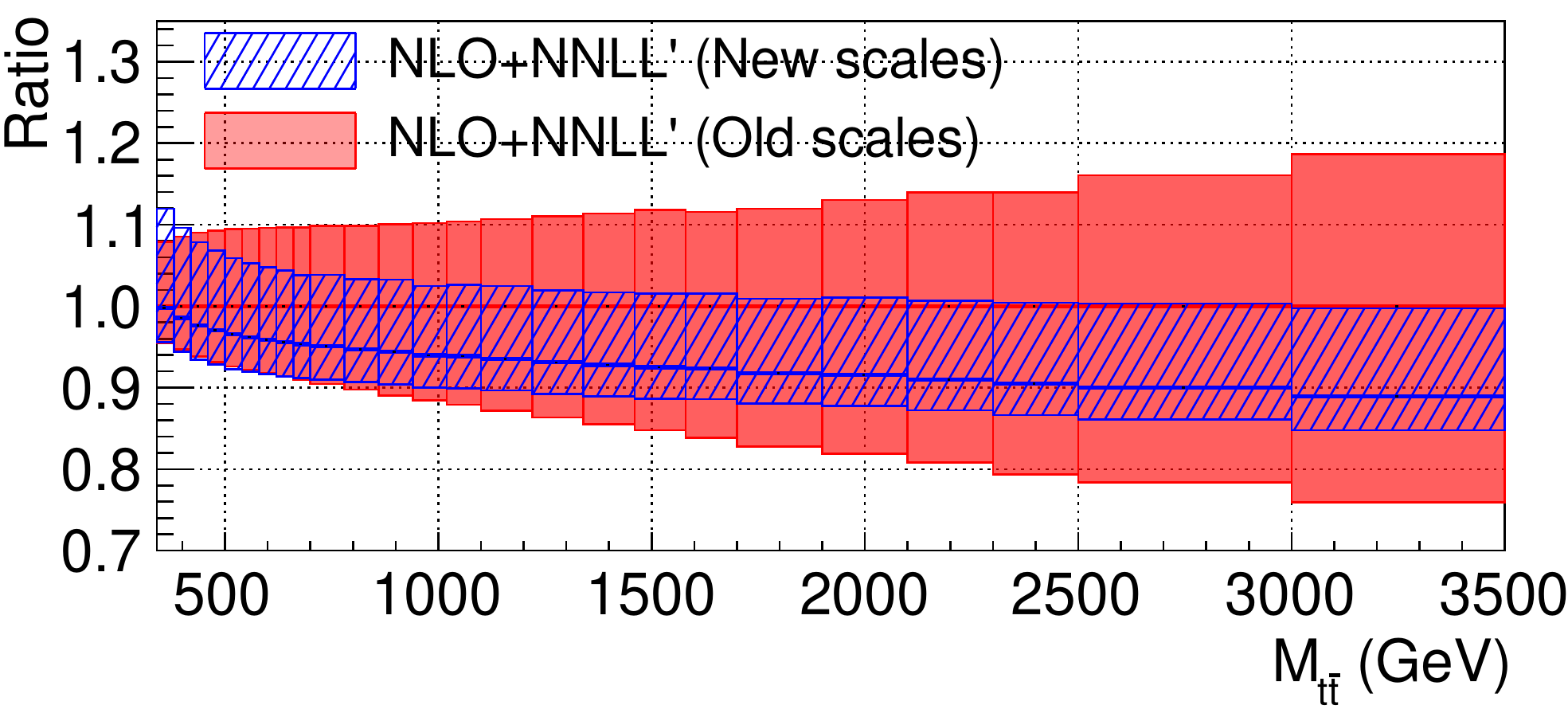}
\includegraphics[width=0.495\textwidth]{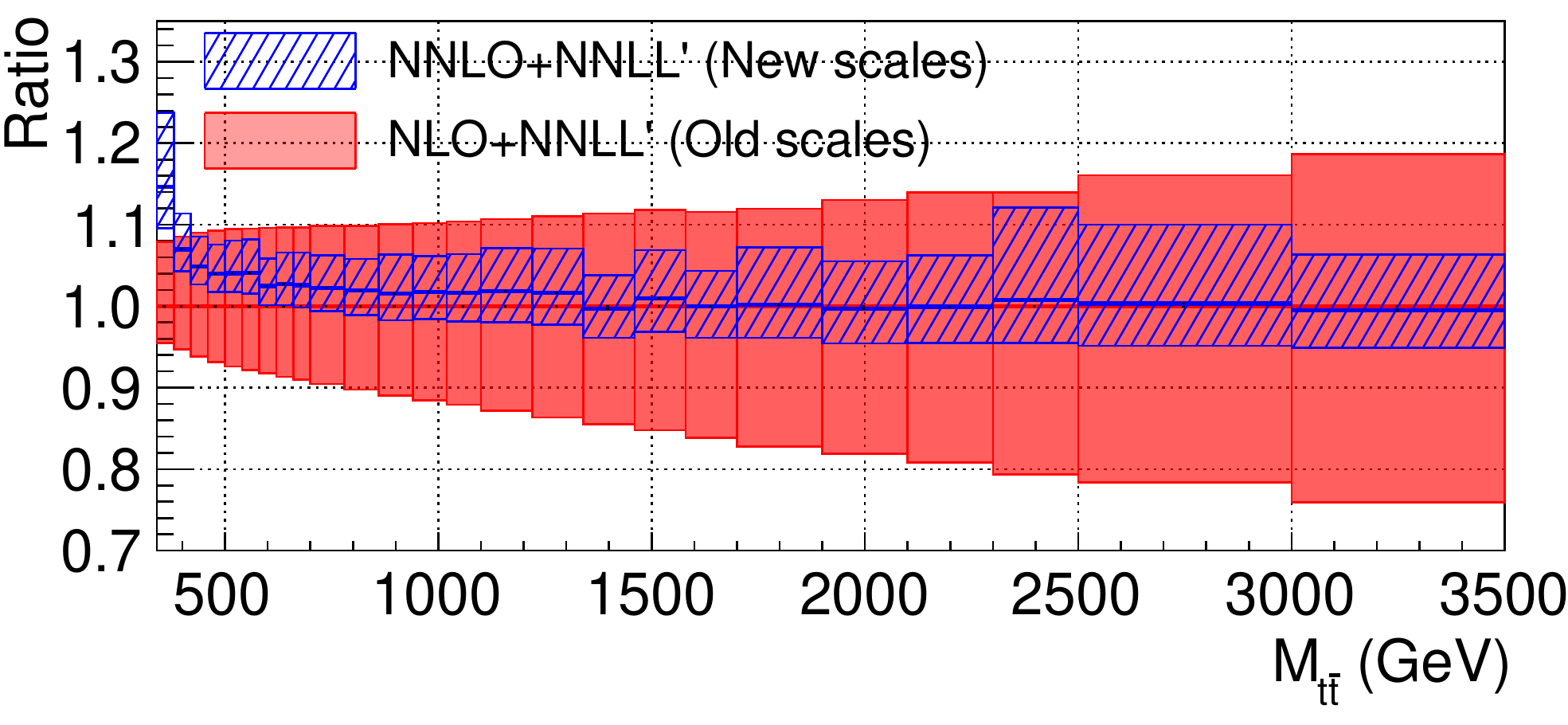}
\\
\includegraphics[width=0.495\textwidth]{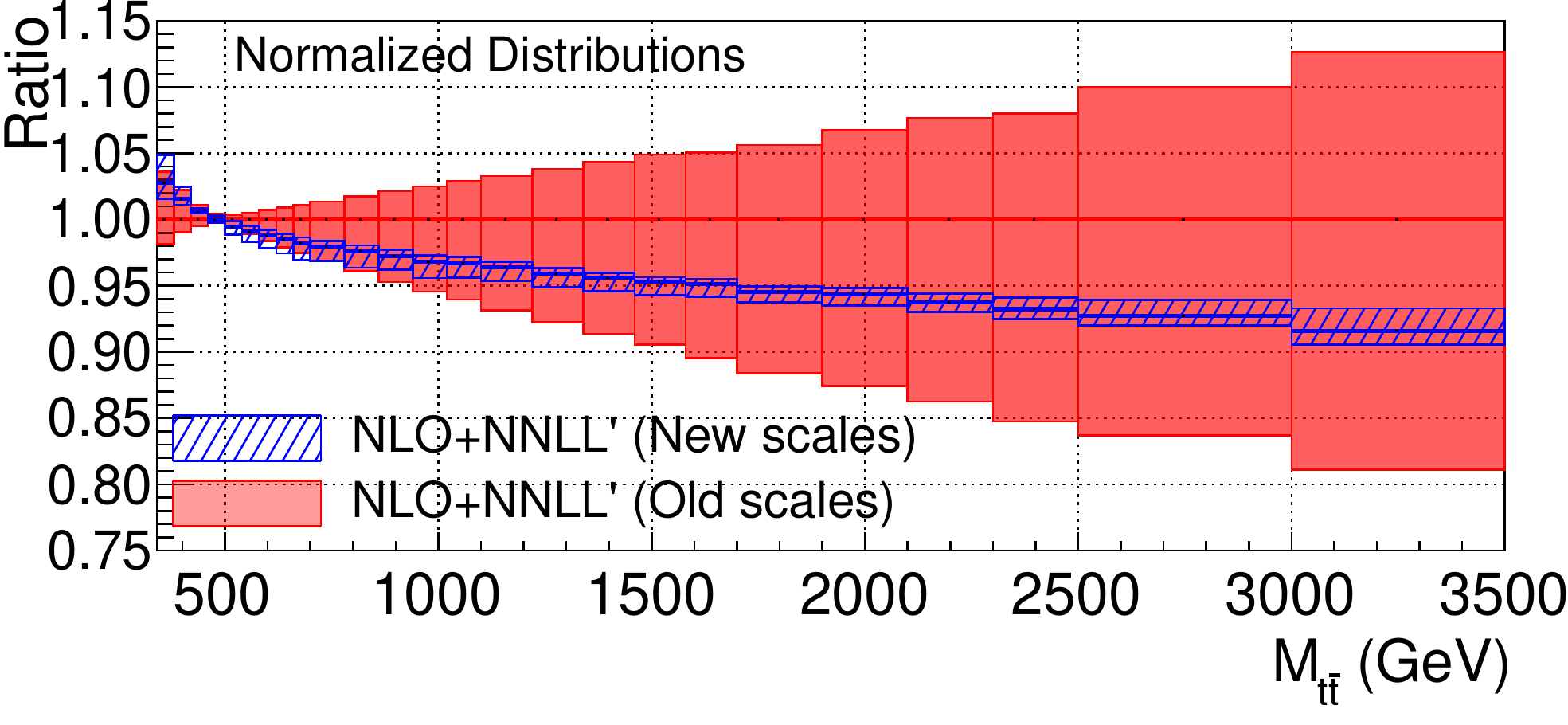}
\includegraphics[width=0.495\textwidth]{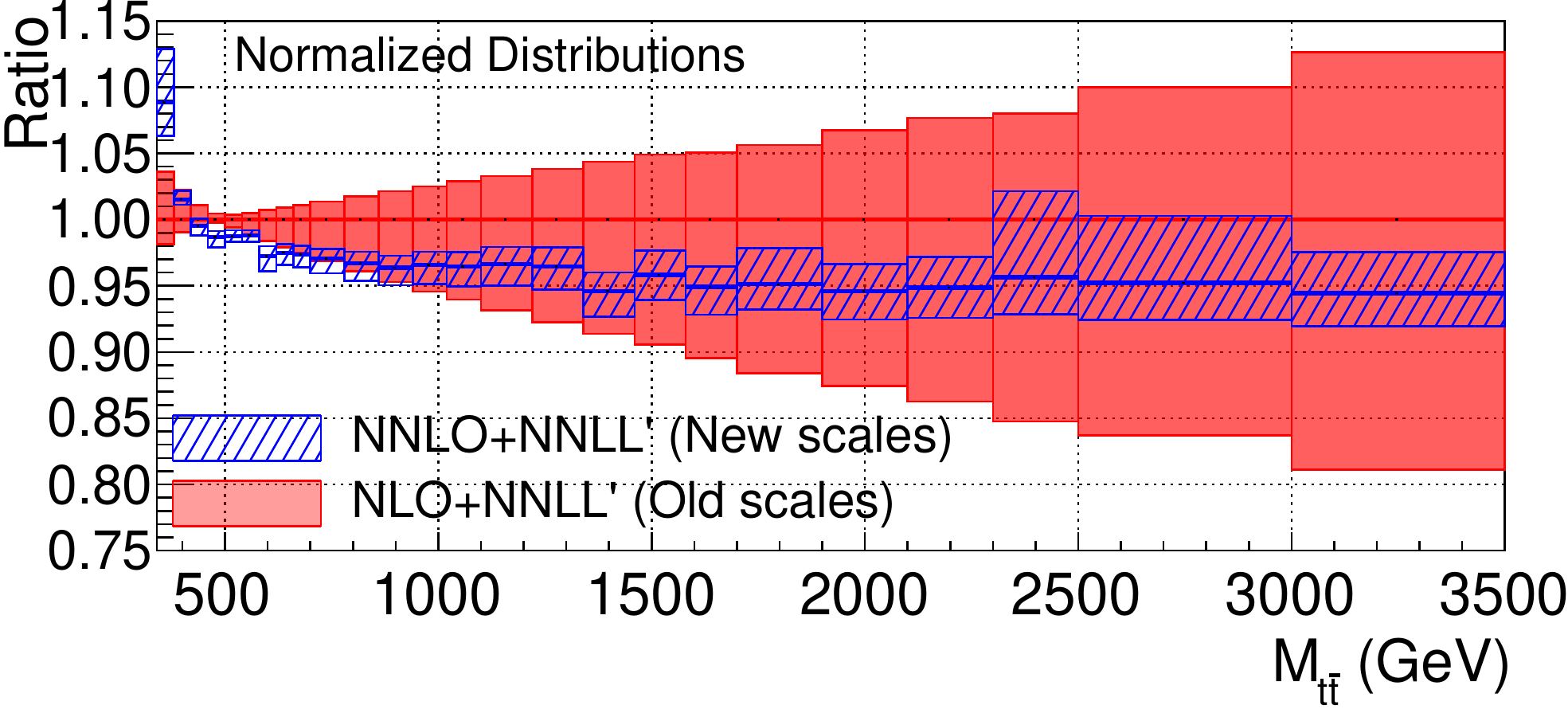}
\caption{\label{fig:old_vs_new_mtt} Comparison of the 3 results for the absolute (upper) and normalized (lower) $M_{t\bar{t}}$ distributions: 1) NLO+NNLL$'$ with old scales (red bands); 2) NLO+NNLL$'$ with new scales (blue bands on the left side); 3) NNLO+NNLL$'$ with new scales (blue bands on the right side). They are all normalized by the central values of 1).}
\end{figure}
Finally, it is interesting to compare the results of the present work with the earlier resummation results at NLO+NNLL$'$ accuracy in \cite{Pecjak:2016nee}. To this end it is important to note that the differences between the current work and the work in \cite{Pecjak:2016nee} are two-fold: 1) we have matched with the exact NNLO calculation here compared to an NLO matching in \cite{Pecjak:2016nee}; and 2) we have employed different settings of the factorization scale and the matching scales than \cite{Pecjak:2016nee}. Concerning the scale choices, we remind the reader that in this work we by default use $\mu_h=H_T/2$, $\mu_s=H_T/\bar{N}$, $\mu_f= H_T/4$ for the $M_{t\bar{t}}$ distribution, and $\mu_h=m_T$, $\mu_s=2m_T/\bar{N}$, $\mu_f=m_T/2$ for the $p_T$ distribution. We refer to this scale setting as the ``new scales''. On the other hand, \cite{Pecjak:2016nee} by default uses $\mu_h=M_{t\bar{t}}$, $\mu_s=M_{t\bar{t}}/\bar{N}$, and $\mu_f= M_{t\bar{t}}/2$ (for the $M_{t\bar{t}}$ distribution) or $\mu_f=m_T$ (for the $p_T$ distribution). We call this scale setting the ``old scales''.

\begin{figure}[t!]
\centering
\includegraphics[width=0.495\textwidth]{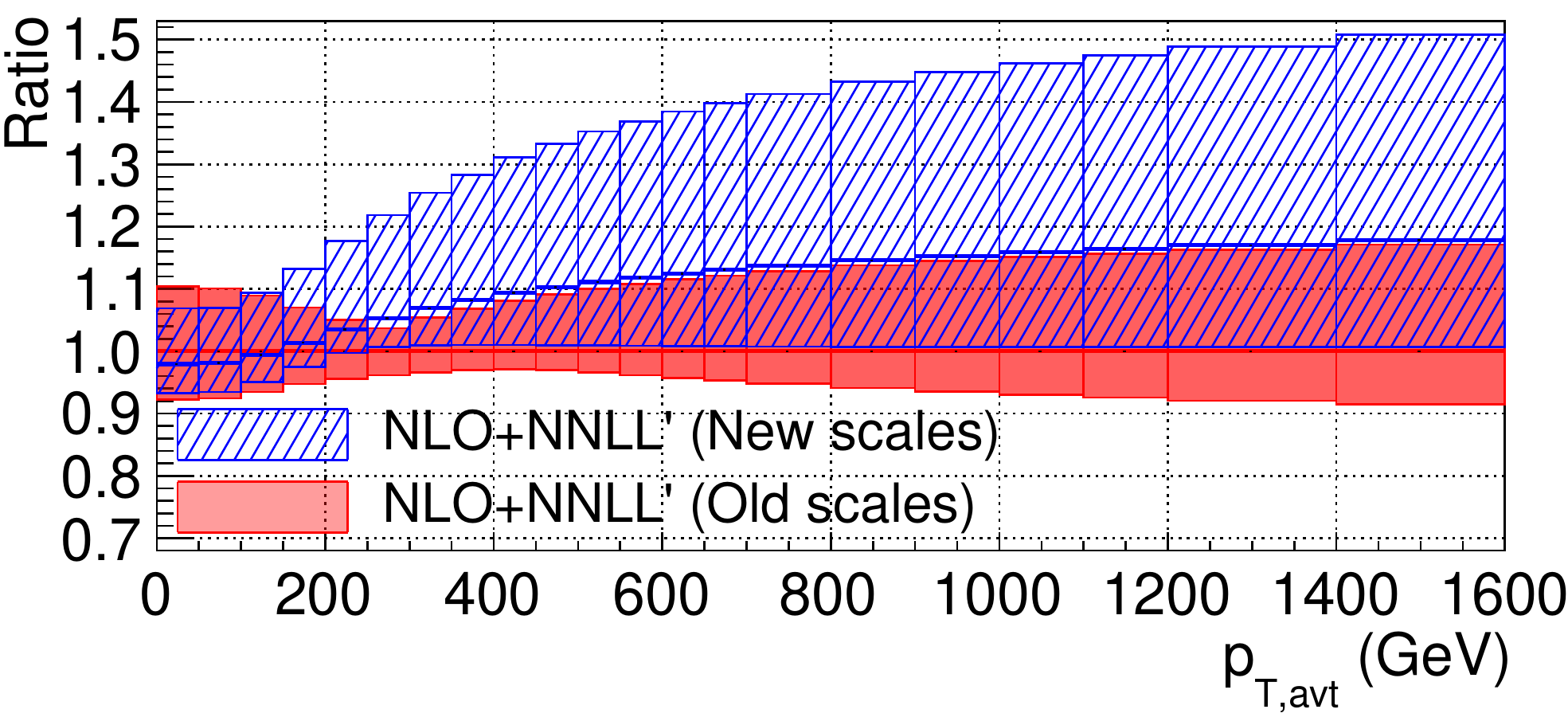}
\includegraphics[width=0.495\textwidth]{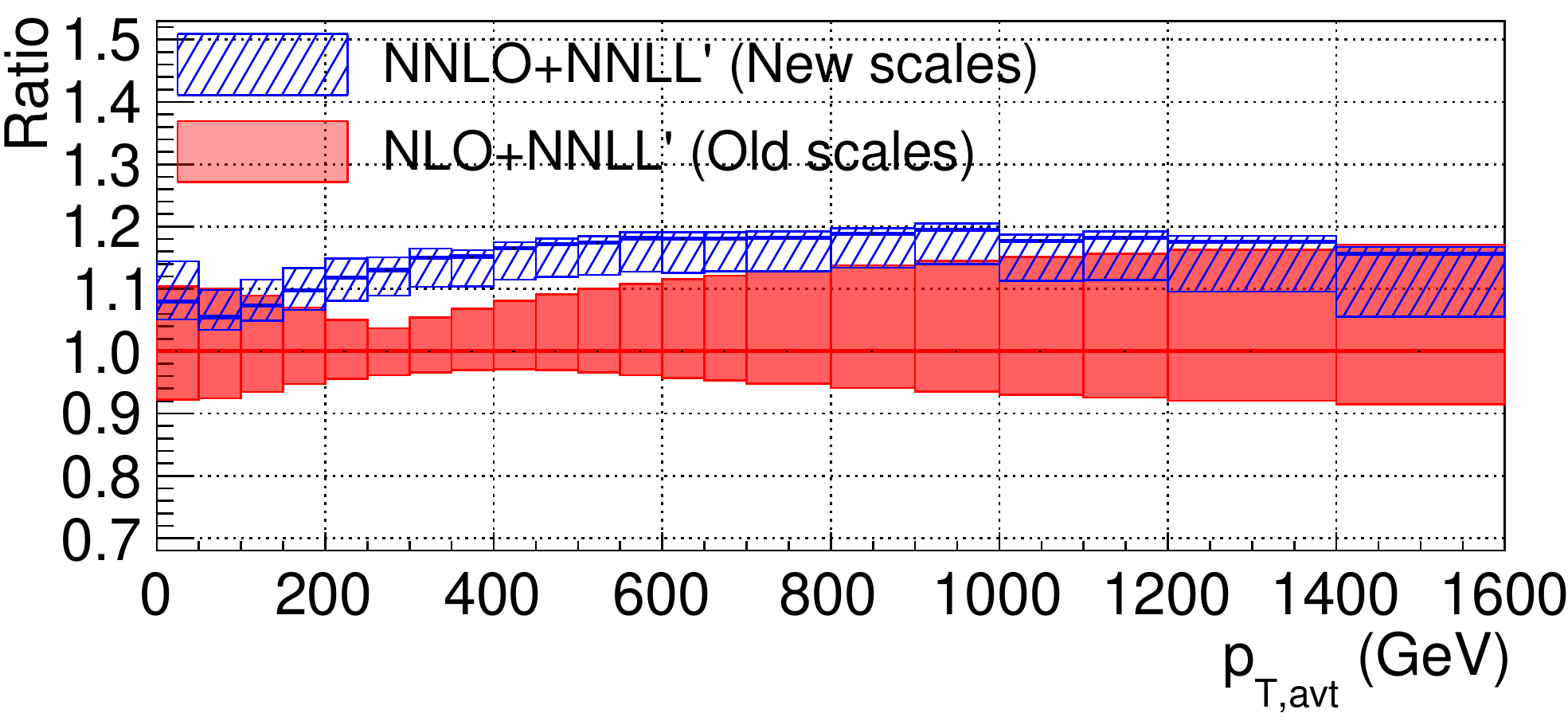}
\\
\includegraphics[width=0.495\textwidth]{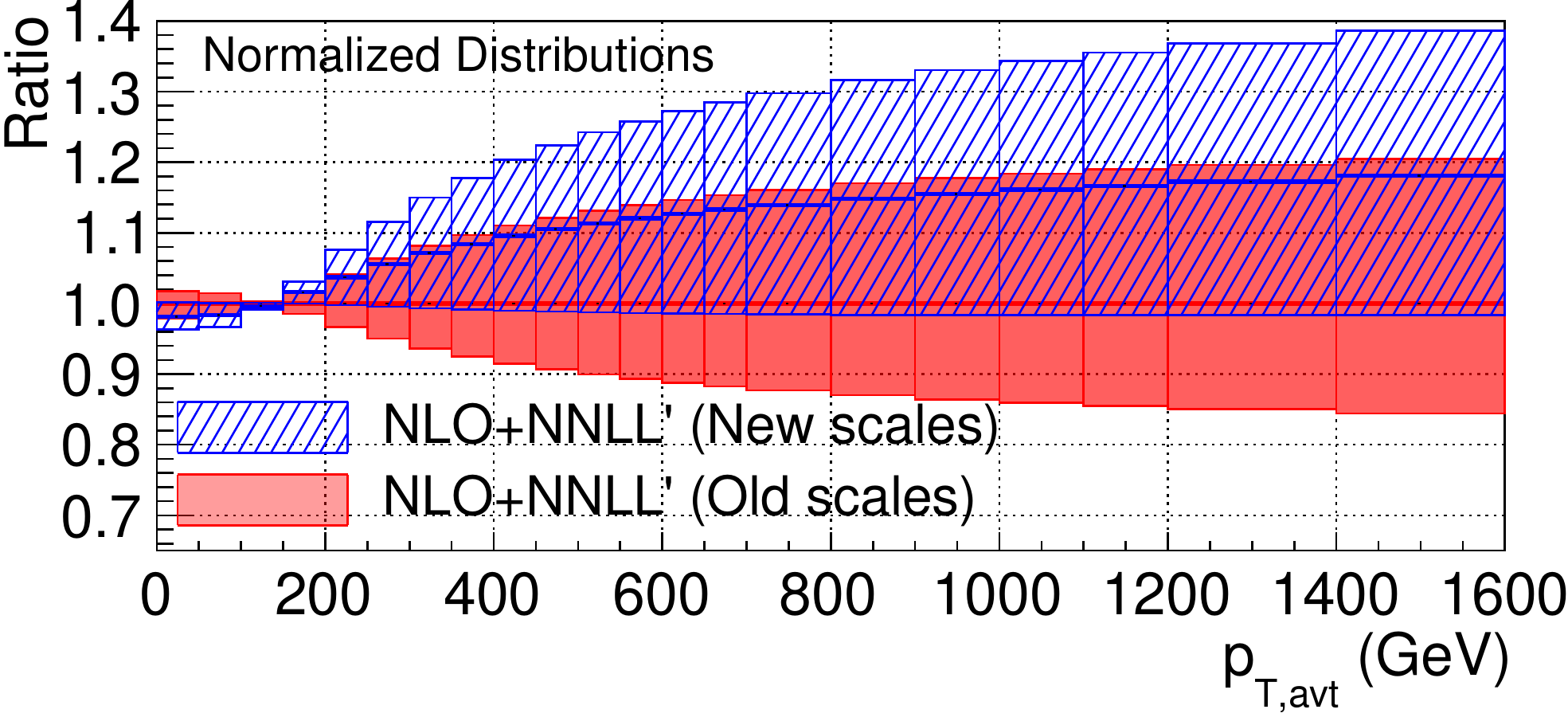}
\includegraphics[width=0.495\textwidth]{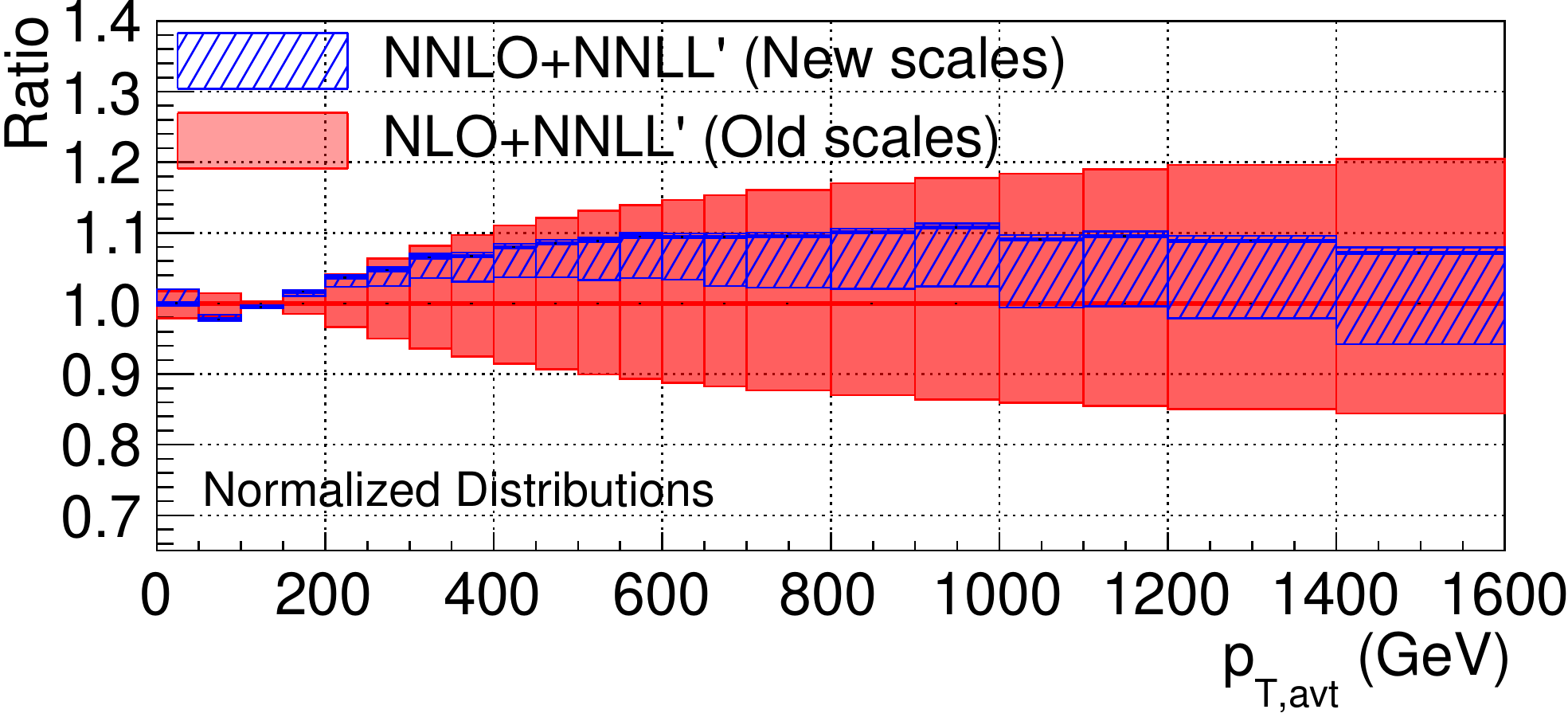}
\caption{\label{fig:old_vs_new_ptt} Same as figure~\ref{fig:old_vs_new_mtt} but for the $p_{T,\text{avt}}$ distribution.
}
\end{figure}

In order to quantify the effect of switching to the new scales and the effect of matching with NNLO, we compare the 3 kinds differential cross sections: 1) NLO+NNLL$'$ with old scales from \cite{Pecjak:2016nee}; 2) NLO+NNLL$'$ with new scales from the present work; and 3) NNLO+NNLL$'$ with new scales which are the best predictions of the present work. Such a comparison for the absolute as well as normalized $M_{t\bar{t}}$ distributions is shown in figure~\ref{fig:old_vs_new_mtt}. It can be seen from the plots on the left side that, by changing to the new scales, the apparent scale uncertainties are reduced. This is a hint that with the new scale choice the higher order corrections are indeed smaller. The effect of matching with the NNLO results is shown in the plots on the right side of figure~\ref{fig:old_vs_new_mtt}. We see that the matching changes the differential cross section most significantly in the low $M_{t\bar{t}}$ bins, which can be expected. A similar comparison in the case of $p_T$ distributions is shown in figure~\ref{fig:old_vs_new_ptt}.
 One can see that the change of default scales does not reduce the scale uncertainties (actually leading to larger uncertainties in the high $p_T$ bins). Matching with the NNLO is important here, which stabilizes the scale variation. Overall, the NNLO+NNLL$'$ results are consistent with the NLO+NNLL$'$ results when the new scales are used.

\section{Comparison with resummed PDFs}
\label{sec:resPDFs}

In the main part of this work, all computations have been carried out
using the (N)NLO NNPDF3.0 PDF set~\cite{Ball:2014uwa}. However, one
can also use threshold resummed PDFs~\cite{Bonvini:2015ira}, into
which the effects of soft gluon radiation are incorporated. Including
these contributions in the PDFs can produce non-negligible differences
in threshold resummed cross section predictions when compared to those
obtained with regular PDFs. An example of this can be found
in~\cite{Beenakker:2015rna}, which considered sparticle
pair production. In particular, for a large range of sparticle masses
considered it was found that while $K$-factors (defined as the ratio
of the resummed result to the fixed order one) obtained with regular
PDFs were almost always greater than unity, this was not always the
case when resummed PDFs were employed.

Although the resummed PDFs are obtained with a reduced data set and
are therefore not yet suitable for precision calculations, we consider
this an appropriate opportunity to explore their implications for
top-quark pair production at the LHC.
In order to compare results obtained with threshold resummed PDFs to
those with standard PDFs we compute the \Mtt\ and $p_{T,\text{avt}}$ distributions
using the same settings as in section~\ref{sec:discussion}. To enable
a fair comparison the NNPDF collaboration also provides PDFs which do
not include threshold resummation but which are compiled from the same
data set as the resummed ones. Specifically we use the
\texttt{NNPDF30\textunderscore nnlo\textunderscore disdytop} PDFs as
the benchmark for PDFs without threshold resummation and the
\texttt{NNPDF30\textunderscore nnll\textunderscore disdytop} PDFs for
those with threshold resummation.
\begin{figure}[t!]
\centering
\includegraphics[width=0.495\textwidth]{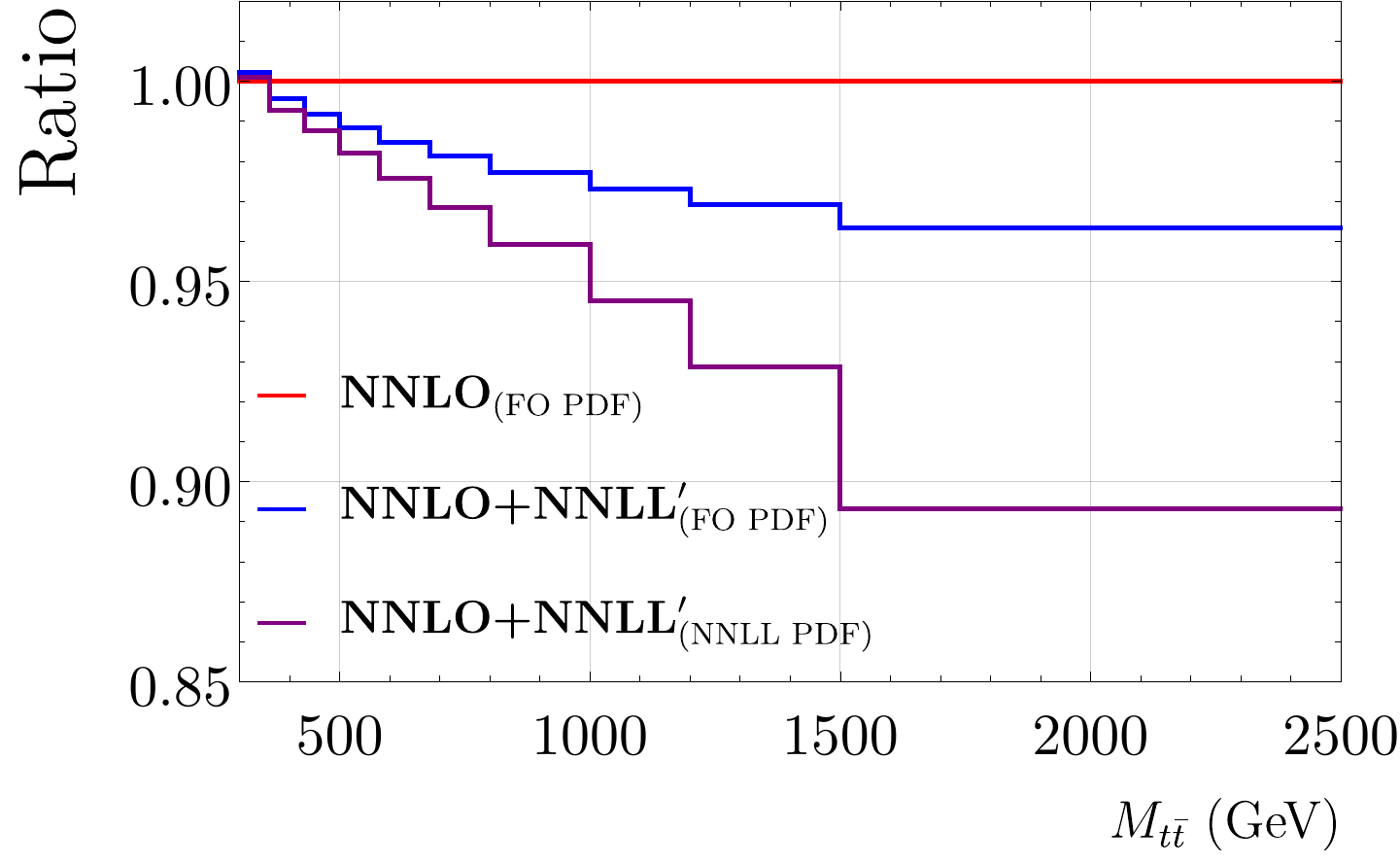}
\includegraphics[width=0.495\textwidth]{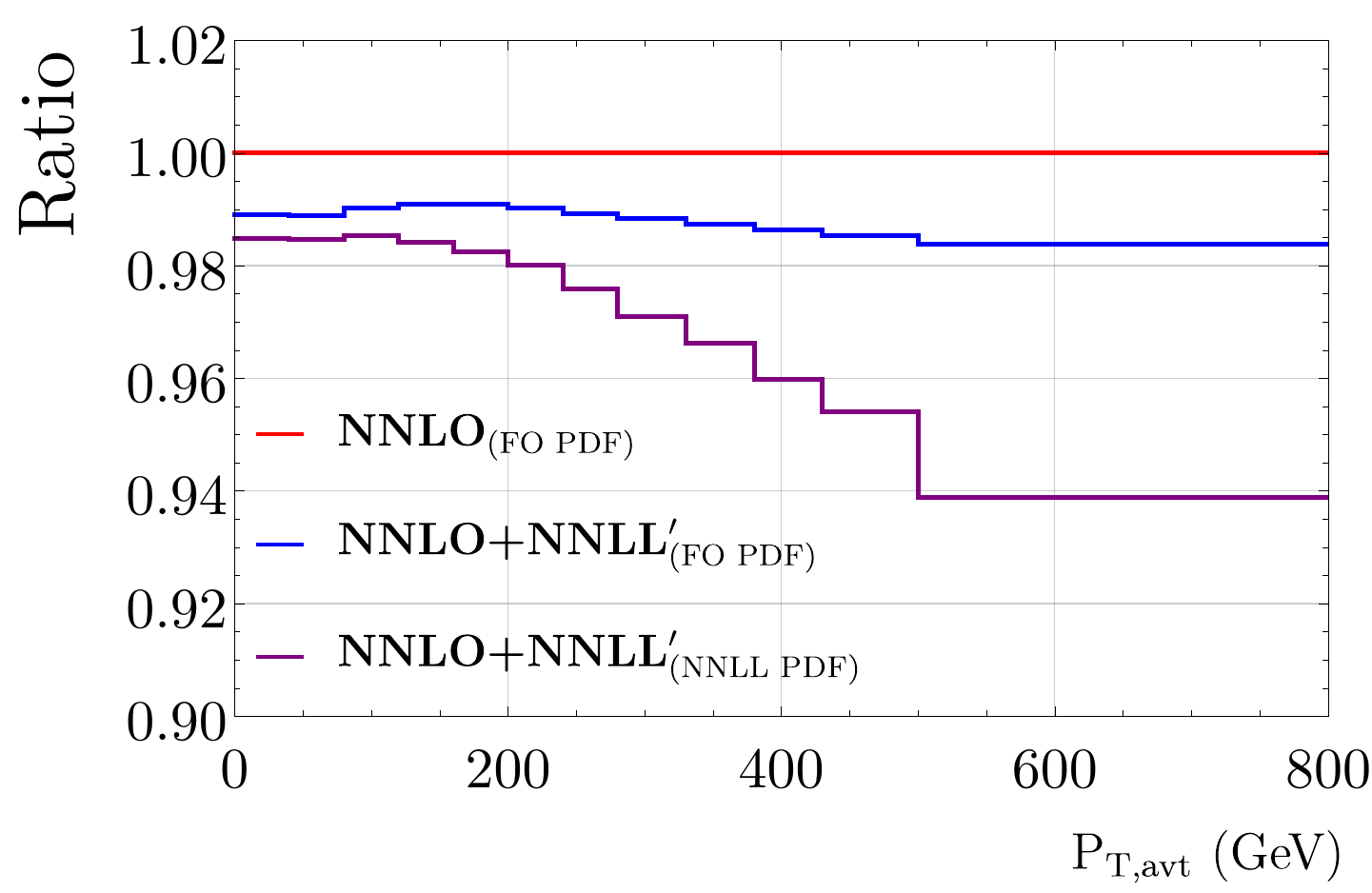}
\caption{\label{fig:resPDF} Comparison of predictions obtained for the \Mtt\ and $p_{T,\text{avt}}$ distributions with regular and resummed PDFs. Plotted are the NNLO prediction (red) using a fixed order PDF as well as the NNLO+NNLL$'$ predictions computed both with fixed order (blue) and NNLL threshold resummed PDFs (purple). Each distribution is normalized to the NNLO one.
}
\end{figure}

In figure~\ref{fig:resPDF} we compare the central values of the NNLO
and NNLO+NNLL$'$ predictions for the $p_{T,\text{avt}}$ and \Mtt\ distributions,
obtained using $\mu_f=m_T/2$ and and $\mu_f=H_T/4$, respectively.  The
NNLO predictions are calculated using the PDFs without resummation,
while the NNLO+NNLL$'$ predictions are computed using the PDFs with
(labelled \texttt{NNLL~PDF}) and without (labelled \texttt{FO PDF}) threshold
resummation. In this manner, the NNLO and NNLO+NNLL$'$ predictions
with the fixed order PDFs act as an equivalent set of predictions to
those in the main part of our paper, but now with a PDF set which can
be compared to those incorporating threshold resummation. The left
plot in figure~\ref{fig:resPDF} shows the \Mtt\ distribution, where
the effect of using resummed PDFs is significant. Here the use of
resummed PDFs produces a suppression in the cross section at high
\Mtt\ compared to predictions produced with regular fixed order
PDFs. The plot on the right, showing the $p_{T,\text{avt}}$ distribution, also
displays noticeable changes. Here again the resummed PDFs produce a greater
suppression of the cross section in the tail of the distribution than
when regular PDFs are used. It will be interesting to perform further studies using resummed PDFs in the
future as the fits improve and experimental measurements in the tails
of these distributions become more accurate.

\section{The RG exponents}
\label{sec:gis}

In this appendix we collect explicit expressions for the RG exponents appearing in eqs.~(\ref{eq:Umatgs}), (\ref{eq:Umatg-boosted}), and (\ref{eq:Umatg-boosted-d}). In order to ease notation we introduce the following shorthand
\begin{align*}
L_h = \ln\frac{M_{t\bar{t}}^2}{\mu_h^2} \, , \quad L_s = \ln\frac{M_{t\bar{t}}^2}{\bar{N}^2\mu_s^2} \, , \quad L_{dh} = \ln\frac{m_t^2}{\mu_{dh}^2} \, , \quad L_{ds} = \ln\frac{m_t^2}{\bar{N}^2\mu_{ds}^2} \, ,
\end{align*}
and remind the reader that
\begin{align*}
\lambda_i=\frac{\alpha_s(\mu_h)}{2 \pi} \beta_0 \ln \frac{\mu_h}{\mu_i} \, .
\end{align*}
Perturbative expansions of the anomalous dimensions and beta function are given by
\begin{align*}
\gamma(\alpha_s) = & \left(\frac{\alpha_s}{4 \pi}\right)\gamma_0 + \left(\frac{\alpha_s}{4 \pi}\right)^2\gamma_1+\left(\frac{\alpha_s}{4 \pi}\right)^3\gamma_2 + \ldots \, ,\\
\beta(\alpha_s) =& -2 \alpha_s \bigg[\left(\frac{\alpha_s}{4 \pi}\right) \beta_0 + \left(\frac{\alpha_s}{4 \pi}\right)^2 \beta_1 + \ldots \bigg] \, .
\end{align*}

Note that since the $g$-functions are derived from the part of the evolution functions in eqs.~(\ref{eq:Umat}) and (\ref{eq:Umat-massless}) which are proportional to the identity matrix in color space, the factors of $i\pi$ which appear there cancel out in the resummed cross section. As such, we do not retain these $i\pi$ factors in our expressions for the $g$-functions.

\subsection{Soft limit}
\label{app:gfn_massive}

First, we present the $g^m_i$ functions appearing in the evolution factor eq.~(\ref{eq:Umatgs}) for the threshold resummed result:
\begin{align}
g^m_1(\lambda_s,\lambda_f) &= \frac{\Gamma_0}{2 \beta_0^2} \Bigg[ \lambda_s+(1-\lambda_s)\ln(1-\lambda_s)+\lambda_s \ln(1-\lambda_f) \Bigg] \, ,
\\
g^m_2(\lambda_s,\lambda_f) &= \frac{\Gamma_0 \beta_1}{2 \beta_0^3} \Bigg[\ln(1-\lambda_s)+\frac{1}{2}\ln^2(1-\lambda_s)\Bigg]-\frac{\Gamma_1}{2 \beta_0^2}\ln(1-\lambda_s)+\frac{\gamma_0^\phi}{\beta_0}\ln \frac{1-\lambda_s}{1-\lambda_f}
 \nonumber \\ 
&+\frac{\Gamma_0}{2\beta_0} L_s \ln\frac{1-\lambda_s}{1-\lambda_f} + \frac{\Gamma_0}{2\beta_0} L_h \ln(1-\lambda_f) \nonumber \\
&+ \frac{1}{1-\lambda_f} \Bigg\{ \frac{\Gamma_0 \beta_1}{2 \beta_0^3} \lambda_s \, [1+\ln(1-\lambda_f)]-\frac{\Gamma_1}{2 \beta_0^2}\lambda_s \Bigg\} \, ,
\\
g^m_3(\lambda_s,\lambda_f) &= \frac{1}{1-\lambda_s}\Bigg\{\frac{\Gamma_0 \beta_1^2}{4\beta_0^4}\bigg[\lambda_s+2\lambda_s \ln(1-\lambda_s)+\ln^2(1-\lambda_s)\bigg]
\nonumber \\
&\hspace{4em} +\frac{\Gamma_0 \beta_2}{2 \beta_0^3}\bigg[\frac{\lambda_s}{2}+(1-\lambda_s)\ln(1-\lambda_s)\Bigg] -\frac{\Gamma_1 \beta_1}{2 \beta_0^3}\bigg[ \frac{3}{2}\lambda_s+\ln(1-\lambda_s) \bigg] 
\nonumber \\
&\hspace{4em} + \frac{\Gamma_2}{4 \beta_0^2}\lambda_s+\frac{\beta_1 \gamma_0^\phi}{\beta_0^2}\big[1+\ln(1-\lambda_s)\big]-\frac{\gamma_1^\phi}{\beta_0}
\nonumber \\
&\hspace{4em} +\frac{\Gamma_0 \beta_1}{2 \beta_0^2}\Big[ \big[1+\ln(1-\lambda_s)\big] L_s - (1-\lambda_s) L_h \Big] \!+\! \frac{\Gamma_1}{2 \beta_0}\big[(1-\lambda_s)L_h - L_s \big]\Bigg\}
\nonumber \\
&+ \frac{1}{1-\lambda_f}\Bigg\{-\frac{\Gamma_0 \beta_1^2}{2 \beta_0^4}\lambda_s + \frac{\Gamma_0 \beta_2}{2 \beta_0^3}\lambda_s-\frac{\gamma_0^\phi \beta_1}{\beta_0^2}\big[1+\ln(1-\lambda_f)\big] + \frac{\gamma_1^\phi}{\beta_0} 
\nonumber \\
&\hspace{4em} + \frac{\Gamma_1}{2 \beta_0}\big[L_s-L_h\big]
 + \frac{\Gamma_0 \beta_1}{2 \beta_0^2}\big[L_h-L_s\big]\big[1+\ln(1-\lambda_f)\big]\Bigg\}
\nonumber \\
&+ \frac{\lambda_s}{(1-\lambda_f)^2}\Bigg\{\!\frac{\Gamma_0 \beta_1^2}{4 \beta_0^4}\bigg[1-\ln^2(1-\lambda_f)\bigg] -\!\frac{\Gamma_0 \beta_2}{4 \beta_0^3}\! +\! \frac{\Gamma_1 \beta_1}{2 \beta_0^3}\bigg[\frac{1}{2}\!+\!\ln(1-\lambda_f)\bigg] - \frac{\Gamma_2}{4 \beta_0^2}\!
\Bigg\} \, . 
\end{align}

\subsection{Boosted-soft limit}
\label{app:gfn_massless}
Here we present the $g$-functions which appear in the evolution factors eqs.~(\ref{eq:Umatg-boosted}) and~(\ref{eq:Umatg-boosted-d}) for the boosted-soft resummation formula. The functions $g_i$ are simply given by their massive counterparts ($g_i^m$ as above) using the replacement,
\begin{align*}
\gamma^\phi & \rightarrow \gamma^\phi + \gamma^{\phi_q} \, ,
\end{align*}
for each instance of $\gamma^\phi$ in the $g_i^m$ in appendix~\ref{app:gfn_massive}. Each instance of $\Gamma_{\rm cusp}(\alpha_s)$ in $g_i^m$ is replaced with $A(\alpha_s)$ (this is due to the presence of $S_A$ and $a_A$ rather than $S_{\Gamma}$ and $a_{\Gamma}$ in eq.~(\ref{eq:Umat-massless})) with $A(\alpha_s)$ given by
\[
    A(\alpha_s) \rightarrow 
\begin{dcases}
    2 \Gamma_{\rm cusp}^q, & q\bar{q}\text{-channel} \\
    \Gamma_{\rm cusp}^q + \Gamma_{\rm cusp}^g, & gg\text{-channel} \, .
\end{dcases}
\]
We decompose each of the $g_i^D$, which are functions of three arguments into two two-argument functions $g^D_{i,dh}$ and $g^D_{i,ds}$ as follows
\begin{align*}
g^D_i(\lambda_{dh},\lambda_{ds},\lambda_f) = g^D_{i,dh}(\lambda_{dh},\lambda_f) + g^D_{i,ds}(\lambda_{ds},\lambda_f) \, .
\end{align*}
Using this decomposition, we present below the functions as used in this work.

\begin{align}
g^D_{1,dh}(\lambda_{dh},\lambda_f) &= \frac{\Gamma_0}{2 \beta_0^2}\Bigg[\ln(1-\lambda_{dh})+\lambda_{dh}\left[1-\ln\left( \frac{1-\lambda_{dh}}{1-\lambda_f}\right)\right]\Bigg] \, ,
\\
g^D_{1,ds}(\lambda_{ds},\lambda_f) &= -\frac{\Gamma_0}{2 \beta_0^2}\Bigg[\ln(1-\lambda_{ds})+\lambda_{ds}\left[1-\ln\left( \frac{1-\lambda_{ds}}{1-\lambda_f}\right)\right]\Bigg] \, ,
\\
g^D_{2,dh}(\lambda_{dh},\lambda_f) &= \frac{\beta_1 \Gamma_0}{2 \beta_0^3}\Bigg[ \left[1+\frac{1}{2}\ln(1-\lambda_{dh})\right]\ln(1-\lambda_{dh})\Bigg] - \frac{\Gamma_1}{2 \beta_0^2}\ln(1-\lambda_{dh})
\nonumber \\
& -\frac{\gamma_0^S}{\beta_0}\ln(1-\lambda_{dh})
+ \frac{\Gamma_0}{2 \beta_0} L_{dh} \ln(1-\lambda_{dh})-\frac{\gamma_0^{\phi_q}}{\beta_0}\ln\left(\frac{1-\lambda_{dh}}{1-\lambda_f}\right)
\nonumber \\
& +\frac{1}{1-\lambda_f}\Bigg\{\frac{\beta_1 \Gamma_0}{2 \beta_0^3}\lambda_{dh} \big[1+\ln(1-\lambda_f) \big] -\frac{\Gamma_1}{2 \beta_0^2}\lambda_{dh}\Bigg\} \, ,
\\
g^D_{2,ds}(\lambda_{ds},\lambda_f) &= -\frac{\beta_1 \Gamma_0}{2 \beta_0^3}\Bigg[ \left[1+\frac{1}{2}\ln(1-\lambda_{ds})\right]\ln(1-\lambda_{ds})\Bigg] + \frac{\Gamma_1}{2 \beta_0^2}\ln(1-\lambda_{ds})
\nonumber \\
&+\frac{\gamma_0^S}{\beta_0}\ln(1-\lambda_{ds})- \frac{\Gamma_0}{2 \beta_0}L_{dh}\ln(1-\lambda_{ds})-\frac{\Gamma_0}{2 \beta_0}\big[L_{ds}
 - L_{dh}\big]\ln\left(\frac{1-\lambda_{ds}}{1-\lambda_f}\right)
\nonumber \\
&+ \frac{1}{1-\lambda_f}\Bigg\{-\frac{\beta_1 \Gamma_0}{2 \beta_0^3}\lambda_{ds} \big[1+\ln(1-\lambda_f)\big]+\frac{\Gamma_1}{2 \beta_0^2}\lambda_{ds}\Bigg\} \, ,
\\
g^D_{3,dh}(\lambda_{dh},\lambda_f) &= -\frac{\beta_1^2\ \Gamma_0}{2 \beta_0^4}\ln(1 - \lambda_{dh}) + \frac{\beta_2 \Gamma_0}{2 \beta_0^3}\ln(1-\lambda_{dh})
\nonumber \\
&+ \frac{1}{1-\lambda_{dh}}\Bigg\{\frac{\beta_1^2 \Gamma_0}{4 \beta_0^4}\big[1+\ln(1-\lambda_{dh})\big]^2 + \frac{\beta_2 \Gamma_0}{4 \beta_0^3}-\frac{\beta_1 \Gamma_1}{2 \beta_0^3}\bigg[\frac{3}{2}+\ln(1-\lambda_{dh})\bigg]
\nonumber \\
&\hspace{4em} + \frac{\Gamma_2}{4 \beta_0^2}-\frac{\beta_1}{\beta_0^2}\big(\gamma_0^{\phi_q}+\gamma_0^S\big)\big[1+\ln(1-\lambda_{dh})\big] +\frac{1}{\beta_0}\big(\gamma_1^{\phi_q}+\gamma_1^S\big)
\nonumber \\
&\hspace{4em} + \frac{\beta_1 \Gamma_0}{2 \beta_0^2}\big[1+\ln(1-\lambda_{dh})\big]L_{dh}-\frac{\Gamma_1}{2 \beta_0}L_{dh} \Bigg\}
\nonumber \\
&+\frac{1}{1-\lambda_f}\Bigg\{-\frac{\beta_1^2 \Gamma_0}{2\beta_0^4}\lambda_{dh} + \frac{\beta_2 \Gamma_0}{2 \beta_0^3}\lambda_{dh}
+\frac{\beta_1}{2 \beta_0^2}\gamma_0^{\phi_q}\big[1+\ln(1-\lambda_f)\big]-\frac{\gamma_1^{\phi_q}}{2 \beta_0}
\nonumber \\
&\hspace{4em} +\frac{\beta_1 \Gamma_0}{4 \beta_0^2}\big[L_{ds}-L_{dh}\big]\big[1+\ln(1-\lambda_f)\big]
- \frac{\Gamma_1}{4 \beta_0}\big[L_{ds}-L_{dh}\big]\Bigg\}
\nonumber \\
&+\frac{1}{(1-\lambda_f)^2}\Bigg\{\frac{\beta_1^2 \Gamma_0}{4 \beta_0^4}\lambda_{dh}\big[1-\ln^2(1-\lambda_f)\big]
-\frac{\beta_2 \Gamma_0}{4 \beta_0^3}\lambda_{dh}
\nonumber \\
&\hspace{4em} +\frac{\beta_1 \Gamma_1}{2 \beta_0^3}\lambda_{dh}\bigg[\frac{1}{2}+\ln(1-\lambda_f)\bigg]
-\frac{\Gamma_2}{4 \beta_0^2}\lambda_{dh}\Bigg\} \, , 
\\
g^D_{3,ds}(\lambda_{ds},\lambda_f) &= \frac{\beta_1^2\ \Gamma_0}{2 \beta_0^4}\ln(1 - \lambda_{ds}) -\frac{\beta_2 \Gamma_0}{2 \beta_0^3}\ln(1-\lambda_{ds})
\nonumber \\
&+\frac{1}{1-\lambda_{ds}}\Bigg\{-\frac{\beta_1^2 \Gamma_0}{4 \beta_0^4}\big[1+\ln(1-\lambda_{ds})\big]^2
-\frac{\beta_2 \Gamma_0}{4 \beta_0^3}+\frac{\beta_1 \Gamma_1}{2 \beta_0^3}\bigg[\frac{3}{2}+\ln(1-\lambda_{ds})\bigg]
\nonumber \\
&\hspace{4em}- \frac{\Gamma_2}{4 \beta_0^2}
 +\frac{\beta_1}{\beta_0^2} \gamma_0^S \big[1+\ln(1-\lambda_{ds})\big]-\frac{\gamma_1^S}{\beta_0}
\nonumber \\
&\hspace{4em}- \frac{\beta_1 \Gamma_0}{2 \beta_0^2}\big[1+\ln(1-\lambda_{ds})\big]L_{ds}+\frac{\Gamma_1}{2 \beta_0}L_{ds}\Bigg\}
\nonumber \\
&+\frac{1}{1-\lambda_f}\Bigg\{ \frac{\beta_1^2 \Gamma_0}{2\beta_0^4}\lambda_{ds} - \frac{\beta_2 \Gamma_0}{2 \beta_0^3}\lambda_{ds}
+\frac{\beta_1}{2 \beta_0^2}\gamma_0^{\phi_q} \big[1+\ln(1-\lambda_f)\big]-\frac{\gamma_1^{\phi_q}}{2 \beta_0}
\nonumber \\
&\hspace{4em} +\frac{\beta_1 \Gamma_0}{4 \beta_0^2}\big[L_{ds}-L_{dh}\big]\big[1+\ln(1-\lambda_f)\big]
- \frac{\Gamma_1}{4 \beta_0}\big[L_{ds}-L_{dh}\big]\Bigg\}
\nonumber \\
&+\frac{1}{(1-\lambda_f)^2}\Bigg\{-\frac{\beta_1^2 \Gamma_0}{4 \beta_0^4}\lambda_{ds}\big[1-\ln^2(1-\lambda_f)\big]
+\frac{\beta_2 \Gamma_0}{4 \beta_0^3}\lambda_{ds}
\nonumber \\
&\hspace{4em} -\frac{\beta_1 \Gamma_1}{2 \beta_0^3}\lambda_{ds}\bigg[\frac{1}{2}+\ln(1-\lambda_f)\bigg]
+\frac{\Gamma_2}{4 \beta_0^2}\lambda_{ds}\Bigg\} \, .
\end{align}

\end{document}